\documentclass[11pt]{article}
\usepackage{fleqn}
\usepackage{epsfig}

\normalbaselines
\newcommand{\be}{\begin{eqnarray}}
\newcommand{\ee}{\end{eqnarray}}

\def\ll#1{\left#1}
\def\r#1{\right#1}
\def\fr{\frac{1}{2}}

\def\mref#1{(\ref{#1})}
\def\zero{\setcounter{equation}{0}}

\def\bd{\begin{displaymath}}
\def\ed{\end{displaymath}}

\def\ba#1{\begin{array}{#1}}
\def\ea{\end{array}}
\def\nn{\nonumber}
\newfont{\Bbb}{msbm10 scaled 1200}

\renewcommand{\theequation}{\thesection.\arabic{equation}}

\begin{document}

\pagestyle{empty}

\begin{center}

{\LARGE\bf Exactness of Conventional and Supersymmetric JWKB Formulae \\ 
and \\
Global Symmetries of Stokes Graphs }

\vskip 12pt

{\large {\bf  Piotr Milczarski \footnote{Supported by
the {\L}\'od\'z University Grant No 580 }
\footnote{ Ph.D. thesis written under supervision of Prof. Stefan 
Giller } }}

\vskip 6pt

Theoretical Physics Department II, University of {\L}\'od\'z,\\
Pomorska 149/153, 90-236 {\L}\'od\'z, Poland\\ 
e-mail: jezykmil@krysia.uni.lodz.pl
\end{center}
\vspace{18pt} 
\thispagestyle{empty}
\begin{abstract}
It has been shown that the cases of the JWKB formulae in 1--dim
QM quantizing the energy levels exactly are results of
essentially one global symmetry of both potentials and their
corresponding Stokes graphs. Namely, this is the invariance of
the latter on translations in the complex plain of the space
variable i.e. the potentials and the Stokes graphs have to be
periodic. A proliferation of turning points in the basic period
strips (parallelograms) is another limitation for the exactness
of the JWKB formulae. A systematic analyses of a single-well
class of potentials satisfying suitable conditions has been
performed.  Only ten potentials (with one or two real parameters) 
quantized exactly by the JWKB formulae have been found all of
them coinciding (or being equivalent to) with the well-known
ones found previously. It was shown also that the exactness of
the supersymmetric JWKB formulae is a consequence of the
corresponding exactness of the conventional ones and vice versa.
Because of the latter two exactly JWKB quantized potentials have
been additionally established. These results show that the exact
SUSY JWKB formulae choose the Comtet {\underline{at al}} \cite{7} form of
them independently of whether the supersymmetry is broken or
not. A close relation between the shape
invariance property of potentials considered and their
meromorphic structure on the $x$-plane is also demonstrated.

\end{abstract}

\vskip 6pt
\begin{tabular}{l}
{\small PACS number(s): 03.65.-W , 03.65.Sq , 02.30.Lt , 02.30.Mv} \\[6pt]
\end{tabular}

\newpage

\pagestyle{plain}

\setcounter{page}{1}
        
\section{Introduction\label{s1}}

\hskip+2em It has long been known that for some number of potentials in
1--dim quantization problems (and also in these cases of $n$-dim
problems which can be reduced to 1--dim ones) their
corresponding JWKB quantization formulae for energy levels (or
some their generalizations \cite{1} or modifications \cite{6})
are exact \cite{2,3,4,5}, whilst in most solvable cases
(i.e. in those for which their corresponding energy spectrum is
known by other means) the JWKB quantization appears to be only
approximate.

Evidently, the same solvable potentials which are quantized
accurately by the JWKB formulae are also quantized exactly when the
supersymmetric (SUSY) modification of the JWKB method is used
\cite{7,8,9}. But in general the SUSY JWKB quantization 
formulae similarly to the
standard JWKB ones do not provide accurate quantization
conditions for energy levels in most solvable cases of
potentials \cite{10}.

Of course there have been attempts of proving that these known
cases of both the conventional \cite{4} and SUSY JWKB
quantization formulae \cite{11} had to be accurate but from our
point of view both these attempts suffered from arbitrary and
erroneous conventions used to sum divergent series of necessary
phases and one can easily convince oneself that some relevant
parts of proofs in the thesiss mentioned are certainly incorrect
(see also Sec.\ref{s3} and Appendix 1).

A way to treat this problem properly is to use the Weierstrass
infinite product representations for mero- and holomorphic
functions which the quantized potentials really are.  But then
it appears that it is necessary to take into account properly
not only the phases coming from the infinite proliferations of
(complex) turning points but also the phases coming from other
(exponential) factors present in the corresponding Weierstrass
products (see Appendix 1).

Also a clear understanding of the exactness of the known (both
conventional and supersymmetric) JWKB formulae seems to be still
missing and they are considered to some extent as accidental. In
particular, unknown are (possibly simple) criteria (different
from a trivial direct comparison between JWKB results and exact
energy spectrum known by other means) allowing us to conclude
for which potentials the corresponding JWKB quantization
formulae could be exact and for which certainly not. Having such
criteria would be very important for applications of the JWKB
method since it would relax us from appealing to other exact
methods.

The following are the aims of this thesis:
\begin{enumerate}
\item To prove rigorously the exactness of the known JWKB
formulae; \item To clarify why the known JWKB quantization
formulae are exact; \item To clarify the observed relation
between the simultaneous (in-)accuracy of the JWKB and the SUSY
JWKB quantization formulae; and \item To provide criteria
allowing us to judge whether a given JWKB quantization formula
can be accurate or not.
\end{enumerate}       

For the beginning let us note however that in general there is
no a common meaning of what a JWKB approximation really is.

One of a typical way of considering it is just to get some
standard semiclassical solution provided by the Schr\"odinger
equation (SE) \cite{12} and valid in the considered domain.
Next to truncate the corresponding infinite series on a given
term. Such a truncation is then considered as an approximate (to
a given order in $\hbar$) solution to SE in the domain chosen.
In this way one gets (up to a $\hbar$-dependent constant) two
types of the approximate semiclassical solutions (ASS) to SE
correspondingly to two different signs of the classical momenta
generating them. Then correspondingly to the domain considered
the dominate ASS is chosen. If none of the two ASS's is
donaminating a linear combination of both is considered.

Next, the ASS's constructed in the above way in different
domains are smoothly joined and the ASS obtained in this way and
covering the total domain of importance is then considered as a
given order semiclassical solution to the problem considered.
Taking the lowest order of this approximation we get what is
then called the JWKB approximation to the problem.

The above way of constructing of ASS has been described by Berry
and Mount \cite{25} for the 1--dim cases and by Maslov and
Fedoriuk \cite{26} for an arbitrary but finite number of
dimensions. In particular the first paper discusses the
difficult (and unsolved satisfactorily) problem of joining
smoothly the ASS's across the so called Stokes lines on which
they change their dominating character into a subdominant one
(the so called 'connection problem').

From our point of view the above way of constructing ASS's
suffers on a complete ignorance of the exact solutions to SE
which are approximated in this way and which in fact are
unknown. This causes many confuses in applications of the method
since together with the exact solutions also corresponding
boundary conditions the solutions have to satisfy are ignored,
too.

Because of that and similarly to that it has been done in earlier papers 
\cite{1,13,14,15} we shall be following here rather oppositely
first starting with a particular set of well defined solutions
to SE satisfying a sufficient number of boundary conditions and
having well defined semiclassical asymptotics (SA) as well.
Since the latter property depends on domains the exact solutions
are defined in we shall choose as such a set of them these
having this property in a maximal way.

It has been shown \cite{1,13,14,15} that
such a set of the solutions to SE really exists and is known as
a set of fundamental solutions (FS). In this thesis the
descriptions 'ASS' and 'JWKB approximation' shall be understood
only just as the corresponding approximations to the fundamental
solutions. This assumption has serious consequences for the form
of the JWKB approximations which can differ seriously from the
one described above. In particular, the presence of simple and
second order poles in a considered potentials generate
unavoidably changes in the corresponding JWKB formulae.

A set of the FS's are accompanied by the so called Stokes graph
(SG). Making use of the latter we get a uniform and systematic
way of solving any interesting 1--dim problem both exactly and
in the semiclassical limit \cite{1,13,14,15}. The
main property of SG is that it takes into account {\it global}
features of a given problem considered in the complex planes of
variables entered the problem (i.e. a position variable, energy,
the Planck constant, some potential parameter(s), etc.). It is
just these global features determining global structures of
corresponding Stokes graphs which allows us to justify all the
known cases of exact JWKB formulae as well as to get an insight
what decides that a given JWKB formula can be exact or not.

The fundamental solutions which play the main role in our
approach have the following three basic properties:
\begin{enumerate}
\item They satisfy some definite and desired boundary conditions 
\cite{13,14};
\item They possess well defined semiclassical limits
\cite{13,14}; and
\item The corresponding semiclassical series are Borel summed to
the solutions themselves \cite{16}.
\end{enumerate}

In fact FS's are the unique solutions to a given SE which have
the above three properties altogether \cite{15}.

The thesis is organized in the following way.

In the next section we summarized essentialities related to
Stokes graphs and fundamental solutions.

In Sec.\ref{s3} we establish necessary symmetry conditions for
potentials and their SG's to ensure the corresponding JWKB
quantization formulae to be exact. Using these conditions we
perform a systematic analyses showing that among all potentials
which satisfy them only eight of them provide us with the exact
JWKB quantization formulae.

In Sec.\ref{s4} we describe the way of generalization of the
results of the previous section invoking some our earlier
results.

In Sec.\ref{s5} we argue that supersymmetric JWKB formulae
cannot be exact if the conventional ones are not as such too and
demonstrate why in all cases of the potentials of Sec.\ref{s3}
quantized exactly by the corresponding JWKB formulae the SUSY
forms of the latter have to be also exact. By direct
calculations we find in all these cases the validity of the SUSY
JWKB quantization formula in the form of Comtet \underline{et
al} \cite{7} independent of that whether the considered
superpotentials satisfy or break the supersymmetry conditions.
The latter result is not however in a contradiction with that of
Inomata \underline{et al} \cite{27} since our result concerns
the {\it exact} JWKB quantization whilst that of the last
authors is only the JWKB approximation.

We discuss also in this section the result of Dutt \underline{et
al} \cite{31} and Barclay \underline{et al} \cite{32} that the
SUSY JWKB formulae are exact for the shape invariant potentials
\cite{30} and notice that all of
them known as being shaped invariant under translational
transformation are also quantized exactly by the JWKB formulae.
According to that two more general theorems on the exactness of
the SUSY and conventional JWKB quantizations of the shaped
invariant potentials are formulated and proved.

In Sec.\ref{s6} we summarize our results and draw our
conclusions.

\section{Stokes graphs, fundamental solutions and quantization \label{s2}}
\zero

\hskip+2em We shall resume here the basic facts about Stokes graphs and
fundamental solutions
\cite{1,13,14,17}.

\subsection{ Stokes graphs \label{s2.1}}

Consider the SE written in the following form:
\be\label{w2.1}
& \Psi''(x,E,\lambda)-\lambda^2q(x,E,\lambda)\Psi(x,E,\lambda)=0 &
\ee
where: $\lambda^=2m\hbar^{-2},\; q(x,E,\lambda)=V(x,\lambda)-E$
and a potential $V(x,\lambda)$ is assumed to be a meromorphic function
of $x$ and $\lambda$\ with the following asymptotic behaviour for
$\lambda\to+\infty\;\; (\hbar\to 0)$:
\be\label{w2.2}
& V(x,\lambda)\sim
V_0(x)+\frac{1}{\lambda}V_1(x)+\frac{1}{\lambda^2}V_2(x)+\ldots &
\ee

Together with q(x,E,$\lambda$) we shall consider a function
$\tilde{q}(x,E,\lambda)\equiv q(x,E,\lambda)+\delta(x,E,\lambda)/\lambda^2$
where $\delta(x,E,\lambda)$\ behaves according to \mref{w2.2}
when $\lambda\to+\infty$. The precise form of
$\delta(x,E,\lambda)$\ depends on types of singularities of
$q(x,E,\lambda)$\ in particular on whether the latter possesses
simple or second order poles (see a discussion below).

Let E be real and let $x_1, x_2,\ldots,$\ be roots of
$\tilde{q}(x,E,\lambda)$\ and $y_1, y_2,\ldots$\ be its simple
poles. 

Some of them can therefore be real but the rest ones are complex
and conjugated pairwise.

For each point $x_i, y_i, i=1,2,\ldots$, let us construct
actions:
\be\label{w2.3}
W^r_i(x,E,\lambda)&=&\int\limits^x_{x_i}\sqrt{\tilde{q}(y,E,\lambda)}dy\nn\\
&{\rm and}&\\
W_i^p(x,E,\lambda)&=&\int\limits^x_{y_i}\sqrt{\tilde{q}(y,E,\lambda)}dy\nn
\ee
and associate with them a system of lines defined by the
conditions:
\be\label{w2.4}
\Re W_i^{r,p}(x,E,\lambda)&=&0
\ee

These are Stokes lines (SL). To be a little bit more precise we
call a Stokes line each connected set of points of the $x$-plane
satisfying the conditions \mref{w2.4}. A collection of all
Stokes lines is called a Stokes graph. 

If the $\Re$-operation in \mref{w2.4} is substituted by the
$\Im$-one then the corresponding set of lines are called
anti-Stokes lines (ASL). The two sets of lines are orthogonal to
each other at all the points except the roots or poles of
$\tilde{q}(x,E,\lambda)$.

Let $z_k, k=1,2,\ldots$, be infinite points of the actions
\mref{w2.3} i.e. the points where the integrals
\mref{w2.3} diverge to infinity. They are created by poles 
of $\tilde{q}(x,E,\lambda)$\ (including these at infinities of
the $x$-plane) and, therefore, their positions coincides with
these of the poles. A total number of infinite points is assumed
to be even infinite.

The roots of $\tilde{q}(x,E,\lambda$) and its simple poles are
starting points for SL's. Due to our assumption about the root
multiplicity only three Stokes lines can emanate from each
$x_i$. On the other hand only one SL can emerge from each simple
pole. Each of SL's starting from some root or simple pole:
$1^0$\ can end at some other root or simple pole, 
$2^0$\ can end at the same root forming a loop around a second order pole, 
or
$3^0$\ runs to an infinity point $z_k$. 
     
A domain $S_k$\ containing a point $z_k$\ and bounded by some
Stokes lines emenating from the roots of
$\tilde{q}(x,E,\lambda)$\ is called a sector of SG. Therefore,
there are at least as many sectors as the infinite points $z_k$.
Typically the points $z_k$\ collect a number of sectors which
depends on a rate of increasing of the action when it approaches
$z_k$'s. Sectors corresponding to finite $z_k$'s are also
finite. The remaining ones extend to infinities of the $x$-plane.
There are no roots $x_i$\ and no simple poles $y_i$ inside $S_k$, but there 
are some at its boundary.

For the purposes of the thesis it is enough to know only the
proper topology of relevant SG's i.e. their precise metric
structures can be ignored. Therefore to draw the corresponding
SG's it is sufficient for all the cases considered here to apply
the following rules:
\begin{description}
\item[$1^o$] From each root of $\tilde{q}(x,E,\lambda)$\ (all
roots are assumed to be simple) emanate three SL's and three
ASL's; \item[$2^0$] From each simple pole of
$\tilde{q}(x,E,\lambda)$\ emanates only one SL and only one ASL;
\item[$3^0$] From each second order pole of $\tilde{q}(x,E,\lambda)$\ 
emanate only SL's or only ASL's
depending on whether the pole coefficient is real negative or
real positive, respectively; \item[$4^0$] From each higher order
pole $(n>2)$\ of $\tilde{q}(x,E,\lambda)$\ emanate $n-2$
directions to which SL's are tangent asymptotically;
\item[$5^0$] Any SL and any ASL can have only single common point;
\end{description}

The above rules together with the asymptotic properties of the
actions \mref{w2.3} for $x\to \infty$\ as well as an analytic
behaviour of SG's on the parameters of $\tilde{q}(x,E,\lambda)$\
the latter function can depend on allows us to draw any such SG
considered in the thesis.

\subsection{ Fundamental solutions \label{s2.2}} 

\hskip+2em  To any given SG
we can attach to each of its sectors $S_k$\ a solution $\Psi_k$\
to SE called a fundamental solution and having the following
structure \cite{2,13,14,17}:
\be\label{w2.5}
\Psi_k(x)&=&\tilde{q}^{-\frac{1}{4}}(x)e^{\sigma_k\lambda W_i(x)}\chi_k(x)
\ee
where:
\be\label{w2.6}
\chi_k(x)=1+\sum\limits_{n\geq 1}\left[-\frac{\sigma_k}{2\lambda}\right]^n
\int\limits^x_{z_k}dy_1\int\limits^{y_1}_{z_k}dy_2\ldots 
\int\limits^{y_{n-1}}_{z_k}dy_n\omega(y_1)\omega(y_2)\ldots\omega(y_n)
\cdot\nn\\
\cdot\left(1-e^{-2\sigma_k\lambda(W_i(x)-W_i(y_1))}\right)
\left(1-e^{-2\sigma_k\lambda(W_i(y_1)-W_i(y_2))}\right)\ldots\\
\ldots\left(1-e^{-2\sigma_k\lambda(W_i(y_{n-1})-W_i(y_n))}\right)\nn
\ee
with
\be\label{w2.7}
\omega(y)&=&\frac{\delta(y)}{\tilde{q}^\fr(y)}-\frac{1}{4}
\frac{\tilde{q}''(y)}{\tilde{q}^\frac{3}{2}(y)}+\frac{5}{16}
\frac{\tilde{q}'^2(y)}{\tilde{q}^\frac{5}{2}(y)}
\ee

In the above formulae $x_i$\ is some of the roots lying at the
boundary of $S_k$\ and $\sigma_k=\pm1$\ is chosen each time
so as to ensure a sign of $\Re(\sigma_k W_i(x))$\ to be negative
for the whole sector $S_k$.

One of the conditions determining the function
$\delta(x,E,\lambda)$\ introduced earlier is to make all the
multiple integrals in \mref{w2.6} convergent at their lower
limits $z_k$. It appears that for the last reason this function
has to be defined as non zero only if $z_k$\ is a second order
pole of the potential considered. The second reason appears when
the FS is to be continued to a point being a simple or double
pole for the potential. Namely, for both these cases we have to
correct the potential always by the same term
$\delta(x,E,\lambda)=(2(x-z_k))^{-2}$\ at {\it each} simple or
double pole of the potential. Of course, in the case of infinite
number of these singularities the arising infinite series has to
be sum into some function having them as its own simple and
double poles. The $\delta$--terms correcting the potentials
considered in the above way we shall call the Langer
corrections.

We would like to stress at this moment that introducing the
Langer corrections are {\it unavoidable} part of the FS
constructions whenever it is necessary to take into account the
presence of the simple and double poles in the potential.

However, there are still another reasons for which particular
forms of $\delta(x,E,\lambda)$\ have to be considered (see the
next sections).

If $x\in S_k$\ then all the integrations in \mref{w2.6} can be
performed along so called canonical paths for which the
condition $\Re(\sigma_kW_i(y_j)-\sigma_kW_i(y_{j+1}))\leq 0$\ is
fulfilled for any two successive integration variables. 

If $x$ is any point such that the integrations in \mref{w2.6} can
be performed along some canonical paths then it is called a
canonical point. A collection of all canonical points
corresponding to the solution $\Psi_k$\ is called a canonical
domain. We denote the latter by $D_k$.

\indent  $\Psi_k$'s have the following two properties in their 
corresponding $D_k$'s:
\begin{description}       
\item[a.] Their series \mref{w2.6} are uniformly convergent;       
\item[b.] Their asymptotic expansions when $\lambda\to+\infty$\  are 
dominated by the first two factors in
\mref{w2.5} whilst the third one approaches then unity.
\item[] Additionally we have:
\item[c.] Every $\Psi_k$\  is Borel summable in some $B_k (S_k\subset 
B_k\subset D_k)$\ and $\Psi_k$'s are the only solutions
to SE with this property.
\end{description}

The two first factors mentioned in the property b. above are the
ones which just constitute our JWKB approximation to $\Psi_k$\
we have talked about in the Introduction. But this approximation
is valid only in the canonical domain $D_k$\ of $\Psi_k$.

All the fundamental solutions are pairwise independent. But
since they are the solutions of the second order linear ODE
\mref{w2.1} each three of them are linearly dependent. If for some of
such a triad, say ,$\Psi_i,\; \Psi_j,\; \Psi_k$, canonical
domains corresponding to them have common points pairwise then
coefficients of a respective linear relation connected them can
all be calculated in the following way \cite{13,14,17}:
\be\label{w2.8}
\Psi_i(x)&=&\alpha_{\frac{i}{j}\to k}\Psi_j(x)+\alpha_{\frac{i}{k}\to j}
\Psi_k(x)
\ee
where
\be\label{w2.9}
\alpha_{\frac{i}{j}\to k}&=&\lim\limits_{x\to 
z_k}\frac{\Psi_i(x)}{\Psi_j(x)},
\;\;\;\ldots etc
\ee
and $x$ runs to $z_k$\ or $z_j$\ along canonical paths.  The
latter calculations allows us to get immediately corresponding
JWKB approximations for the coefficients $\alpha_{i/j\to k}$.

\subsection{Singularities of fundamental solutions\label{s2.3}}

\hskip+2em  Loci of singularities of the fundamental solutions in the
$x$-plane coincide with the ones of $q(x,E,\lambda)$\ and their
nature is governed by the general rules (see for example
\cite{18}). Therefore
if $q(x,E,\lambda)$\ is holomorphic then such is each
fundamental solution.  The singularities at zeros of
$q(x,E,\lambda)$\ as provided by the representations \mref{w2.5}
- \mref{w2.7} are therefore only apparent i.e they mutually
cancel when the corresponding sums are performed. Nevertheless,
they are real singularities in each factor in \mref{w2.5} as
well as in each integral in \mref{w2.6} where they can cause
troubles with taking the integrals (see the next section).

Each simple pole $y_k$\ of $q(x,E,\lambda)$\ is a source of a
logarithmic branch point singularity for each fundamental
solution which close to $y_k$ behaves as
$a(x-y_k)+b(x-y_k)\ln(x-y_k)$\ with $b\not=0$.

Each second order pole $z_k$ of $q(x,E,\lambda)$\ generates in
each solution to SE~\mref{w2.1} a branch point of the form
$a(x-z_k)^{\alpha}+b(x-z_k)^{\beta}$\ if $\alpha-\beta$\ is not
an integer or the branch point of the form
$a(x-z_k)^\alpha+b(x-z_k)^\beta\ln(x-z_k)$\ in the opposite
case. In the case of these fundamental solutions which have to
vanish at $z_k$\ a real part of at least one of the numbers
$\alpha,\beta$\ has to be positive (one of the coefficients $a$ and
$b$ can vanish in the case when the real part of the corresponding
$\alpha$\ or $\beta$\ is not positive). 

Each higher order pole $z_k$\ of $q(x,E,\lambda)$ generates a
branch point which is simultaneously an essential singularity
for each fundamental solution. The fundamental solutions which
are defined in the sectors containing this $z_k$\ have to vanish
at $z_k$\ in the corresponding sectors.

It is easy to verify that the behaviour of the FS's (corrected
if necessary by the Langer terms) around singular points of a
potential considered is exactly such as described above sometimes being
determined totally by their first two JWKB factors. However, let
us stress it once again, that at the simple and second order poles
this behaviour appears as a result of the Langer corrections.

On all the figures of SG's drawn for the purposes of the thesis
only cuts corresponding to the real branch points of FS's are
marked whilst the branch points which follows from the F--F
representation of FS's as given by \mref{w2.5}--\mref{w2.7} are
completely ignored.

A set of all zeros $x_k$, simple poles $y_k$\ and other poles
$z_k$\ of $q(x,E,\lambda)$\ determine for a real $E$ (and
$\lambda$) in a unique way a possibility of constructing a
corresponding set of fundamental solutions among which there
should be these two of them which determine the corresponding
problem of quantization. If some (or both) of these two solutions cannot be
constructed because of some specific properties of
$q(x,E,\lambda)$\ then the corresponding quantization problem
cannot be formulated i.e. supposed bound states do not exist.
The latter possibility is uniquely related to a particular
pattern of SG which should be drawn in such cases.

\subsection{Quantization\label{s2.4}}

\hskip+2em  A quantization of 1--dim quantum systems with the help of the
fundamental solutions has been described in many earlier
papers \cite{1,13,14,16}. Here we
sketch only the procedure for the case of two real turning
points $x_1,\; x_2$\ whilst the rest of them are complex and
conjugated pairwise (we assume $\tilde{q}(x,E,\lambda)$\ and $E$
to be real). We assume also that our physical problem is limited
to a segment $z_1\leq x\leq z_2$\ at the ends of which the
potential has poles. In particular we can push any of $z_{1,2}$\
(or both of them) to $\mp\infty$\ respectively.

To write the corresponding quantization condition for energy $E$
and to handle simultaneously the cases of second and higher
order poles we assume $z_1$\ to be the second order pole and
$z_2$\ to be the higher ones.

It is also necessary to fix to some extent the closest
environment of the real axis of the $x$--plane to draw a piece of
SG sufficient to write the quantization condition. To this end
we assume $x_3$\ and $\bar{x}_3$\ as well as $\bar{x}_4$\ and $\bar{x}_4$\ 
to be another four turning points and $z_3$\ and $\bar{z}_3$\ another two
second order poles of $V(x,\lambda)$\ closest the real axis.
Then a possible piece of SG can look as in Fig.1.

\begin{tabular}{c}
\psfig{figure=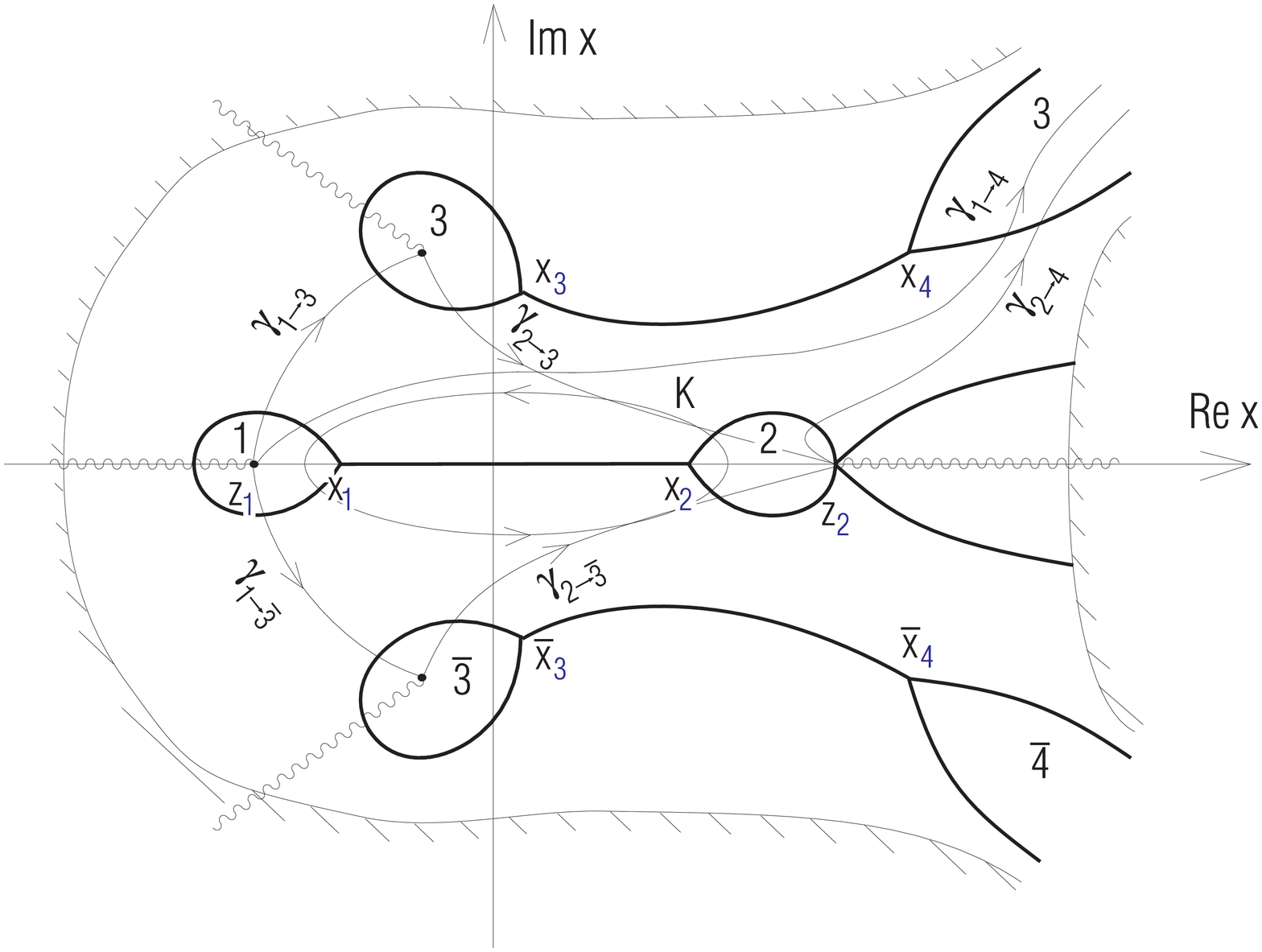, width=12cm} \\
Fig.1  The SG corresponding to general quantization rule \mref{w2.10}
\end{tabular}

There is no a unique way of writing the quantization condition
corresponding to the figure. Some possible three forms of this
condition can be written as \cite{16}:
\be\label{w2.10}
exp\ll[-\lambda\oint\limits_K\tilde{q}^\fr(x,\lambda,E))dx\r]=
-\frac{\chi_{1\to 3}(\lambda,E)\chi_{2\to
\bar{3}}(\lambda,E)}{\chi_{1\to
\bar{3}}(\lambda,E)\chi_{2\to 3}(\lambda,E)}=\nn\\
=-\frac{\chi_{1\to 4}(\lambda,E)\chi_{2\to
\bar{3}}(\lambda,E)}{\chi_{1\to\bar{3}}(\lambda,E)\chi_{2\to 4}(\lambda,E)}
\ee
and $\chi_{k\to j}(\lambda,E) k,j=1,2,3,4$\ are the coefficients
\mref{w2.6} calculated for $x\to z_j$. The closed integration path $K$ 
is shown in Fig.1.
In the figure the paths $\gamma_{1\to 3},\; \gamma_{2\to 3}$,
etc., are the integration paths in the formula \mref{w2.6}
whilst the wavy lines designate corresponding cuts of the
$x$--Riemann surface on which all the FS are defined. The same
conventions in designations are maintained on the remaining
figures 2--17.

The above three forms \mref{w2.10} of the quantization condition
are equivalent and can be substituted by another equivalent
forms if the latter are still admitted by the SG considered. It
means that in general we can choose between different forms of
the conditions \mref{w2.10} according to our needs. In
particular depending on the form of SG there are forms of
\mref{w2.10} completely
deprived of the JWKB phase factor on its LHS i.e. such forms are
composed only from the $\chi$-factors of FS's.

The condition \mref{w2.10} is {\it exact}. Its LHS has just the JWKB
form. If we substitude each $\chi_{k\to j}(\lambda,E)$\ in
\mref{w2.10} by unity (which these coefficients approach when 
$\lambda\to+\infty$) we obtain the
well--known JWKB quantization rule. But in this way the latter
is in general only an approximation to \mref{w2.10}. It is not
in the following three cases only:
\begin{description}
\item[$1^0$] All $\chi_{k\to j}(\lambda,E)$'s in \mref{w2.10} 
are really equal to (identical with) 1;
\item[$2^0$] They all cancel out mutually by some reasons;
\item[$3^0$] Both the above cases take place i.e. some of 
$\chi_{k\to j}(\lambda,E)$'s satisfy $1^0$\ 
and some $2^0$.
\end{description}

The first case is very rare and the only known example of it is
just the harmonic oscillator potential (see below). It needs in
fact for a given $\chi_{k\to j}(\lambda,E)$\ to have a
possibility to deform its integration path properly to make all
the integrations in \mref{w2.6} vanishing i.e. this condition
demands some particular topology of turning points on the
$x$--plane to happen.

The next one if not happens accidentally can take place due to a
possible reality of the coefficients entering the formula
\mref{w2.10} (where the coefficients can appear in pairs with their
complex conjugations dividing them) or due to some possible
symmetry of the potential $V(x,\lambda)$\ relating the
$\chi$--coefficients present in the formulae. We shall show in
the next sections that the latter case is the main reason for
all the known cases of the JWKB formulae which provide us with
the exact quantization conditions. In fact the symmetry
properties of the potential as well as a particular topology of
its turning point distribution cooperating together are the most
frequent way of the realization of the JWKB formula exactness.

\section{Global symmetries of Stokes graphs and quantization  \label{s3}}
\zero

\hskip+2em  The case when all $\chi_{j\to k}$\ in the condition \mref{w2.10}
reduce to unity is exceptional and within the holomorphic
potentials can contain only a single potential namely the
harmonic one. (Note that for the holomorphic potentials we can
always put $\delta(x,E,\lambda)\equiv0)$.

To see this we note that for the case to happen it is necessary
to have possibilities to deform the integration paths in the
formula \mref{w2.6} for $\chi$'s so as to make all the
integrations vanishing. The latter property can happen if the
integration paths can be pushed out to infinity i.e. none of
roots of $q(x,E,\lambda)$ (which are branch points for the
integrands in \mref{w2.6}) can prevent such a deformation. It
means that these roots cannot extend to infinity and that the
corresponding SG for the case should look as in Fig.2 with the
blob containing all the roots of the case.  Secondly, a number
of sectors has to be limited to four as it is shown in Fig.2
since only then the condition \mref{w2.10} can contain only
$\chi$'s with vanishing integrations. Next, a number of roots
inside the blob has to be finite since $q(x,E,\lambda)$\ would
vanish identically in the opposite case.  Therefore,
$q(x,E,\lambda)$ has to be polynomial. It is just a harmonic one
since for the polynomial potentials a number of sectors exeeds
by two a degree of a potential.
\begin{tabular}{c}
\psfig{figure=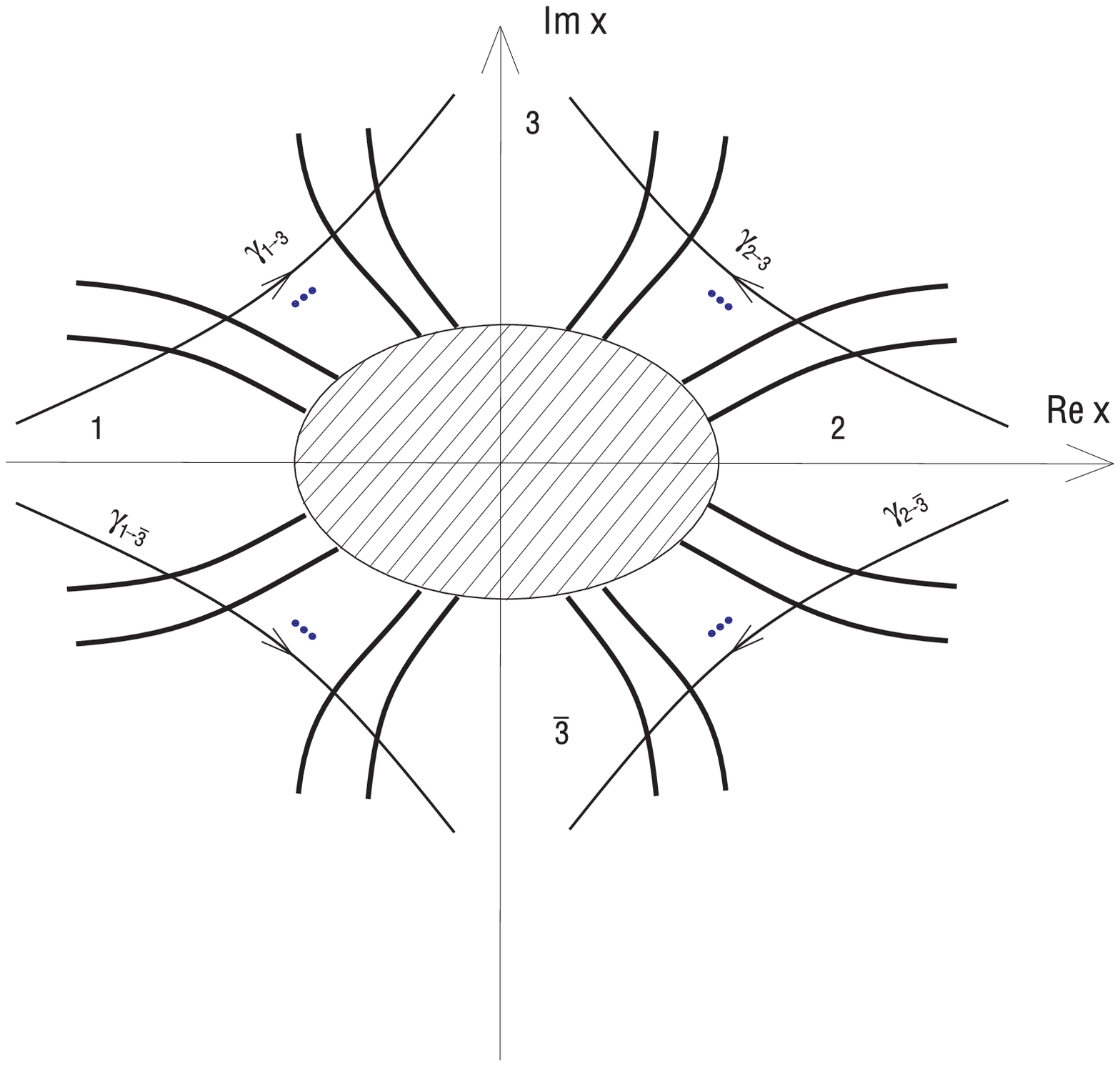,width=12cm} \\
Fig.2  The case of SG satisfied only by the harmonic \\
oscillator potential among the holomorphic ones
\end{tabular}

Therefore, other possibilities for the condition \mref{w2.10} to
be the pure JWKB one are related to possibilities for $\chi$'s
in \mref{w2.10} to cancell mutually. Such cases to happen if not
accidentall have to be related to some symmetries of
corresponding SG's the latter being determind by relevant
symmetries of underlying potentials.

Suppose therefore $q(x,E,\lambda)$\ to satisfy the following
symmetry relation:
\be\label{w3.1}
q(y(x),E,\lambda)&=&q(x,E,\lambda)
\ee
for $x\to y(x)$. We shall assume that it is always possible to
find (if necessary) $\delta(x,E,\lambda)$\ such that
\mref{w3.1} is satisfied by $\tilde{q}(x,E,\lambda)$\ ÿÿ
as well under the same transformation. If the corresponding SG
is also to be invariant under such a transformation then the
full set of the actions \mref{w2.3} which define this SG has to
be invariant too, up to multiplicative constants. The latter
freedom follows from the conditions \mref{w2.4} defining SL's.
But according to \mref{w3.1} we have:
\be\label{w3.2}
\int\limits^x_{x_k}\sqrt{\tilde{q}(\xi,E,\lambda)}d\xi=
\int\limits^x_{x_k}\sqrt{\tilde{q}(y(\xi),\lambda,E)}d\xi
=\int\limits^{y(x)}_{y(x_k)}\sqrt{\tilde{q}(\xi,E,\lambda)}
\frac{dx}{dy}(\xi)d\xi
\ee
from which we can conclude that the mentioned action set
invariance is achieved if $y'(x)=C$, where $C$ is real.

Therefore the allowed transformations $y(x)$ are linear. Since
they constitute a group then it is easy to see that if
$|C|\not=1$\ then $\tilde{q}(x,E,\lambda)$\ has to have common
accumulation points of their roots and poles. Because of that we
shall limit our further considerations to less singular cases of
$\tilde{q}(x,E,\lambda)$\ what means that we shall put $C=\pm
1$. The latter limitation leaves us with only two types of the
allowed symmetry transformations: the one which is essentially a
reflection $x\to -x$\ and the other being a complex translation
of the $x$--plane.

Therefore the two resulting classes of potentials remaining
invariant under the above two symmetry transformations are: a
class of even potentials and a class of periodic potentials. Of
course both the classes are not necessarily disjoint. It is,
however, rather clear that the evenness of a potential alone is
too week to ensure overall cancellations in \mref{w2.10} and it
is just rather a periodicity of it which can work effectively to
cause the relevant cancellations in \mref{w2.10} to happen if it
is possible at all. We shall show this below.

\subsection{Periodic holomorphic (entire) potentials\label{s3.1}}

\hskip+2em  In general $q(x,E,\lambda)$\ as a meromorphic function of
complex $x$ can be periodic with at most two independent (in
general complex) periods \cite{19}. However, in the case of
being {\it holomorphic} $q(x,E,\lambda)$\ can have only one
period (being a constant in the presence of the second one).
Further, since $q(x,E,\lambda)$\ is assumed to be real for its
real arguments then its period can be only real or only pure
imaginary. For the obvious reason we shall consider only the
last case assuming for simplicity the period to be equal to
$2\pi i$. In this case $q(x,E,\lambda)$\ can be expanded into
the following Fourier series \cite{19}:
\be\label{w3.3}
q(x,E,\lambda)&=&\sum\limits^\infty_{x=-\infty}q_n(E,\lambda)e^{nx}
\ee

If the behaviour of $q(x,E,\lambda)$\ at the $x$--infinity is to
be of a finite type the series \mref{w3.3} has to be abbreviated
providing us with a finite sum. The latter should contain at
least three terms if we want $q(x,E,\lambda)$\ to possess bound
states. Let $k$ and $l (k>l+1)$\ be therefore the upper and lower
limits of this abbreviation respectively. We shall consider just
below in details a few cases of such abbreviated $q$'s for which
$k-l=2,3,4$. By this we shall convince ourselves that the
remaining cases of $q(x,E,\lambda)$\ cannot provide us with
examples of the exactly JWKB quantized potentials.

In our investigations we shall make intensive use of the
Weierstrass product representation for the abbreviated series
\mref{w3.3} in order to perform necessary calculation of phases
of $q(x,E,\lambda)$\ alone as well as its functions. This
represention is considered in Appendix 1. We have calculated
there also explicitly, in order to provide us with an example of
such calculations, the relevant total phases of
$q(x,E,\lambda)$\ for the case $k=2,\; l=0$ considered just
below.
 
\underline{case: $k=2, l=0$}

We can write $q(x,E,\lambda)$ in this case as:
\be\label{w3.4}
q(x,E,\lambda)&=&\alpha(E,\lambda)e^{2x}-
2\beta(E,\lambda)e^x+\gamma(E,\lambda)
\ee
where $\alpha(E,\lambda),\;\beta(E,\lambda)$\ and
$\gamma(E,\lambda)$\ are known functions of $E$ and $\lambda$. In
particular, for $\alpha\equiv\beta\equiv1$ and $\gamma\equiv-E$\
we get the well--known Morse potential \cite{20}.

\begin{tabular}{c}
\psfig{figure=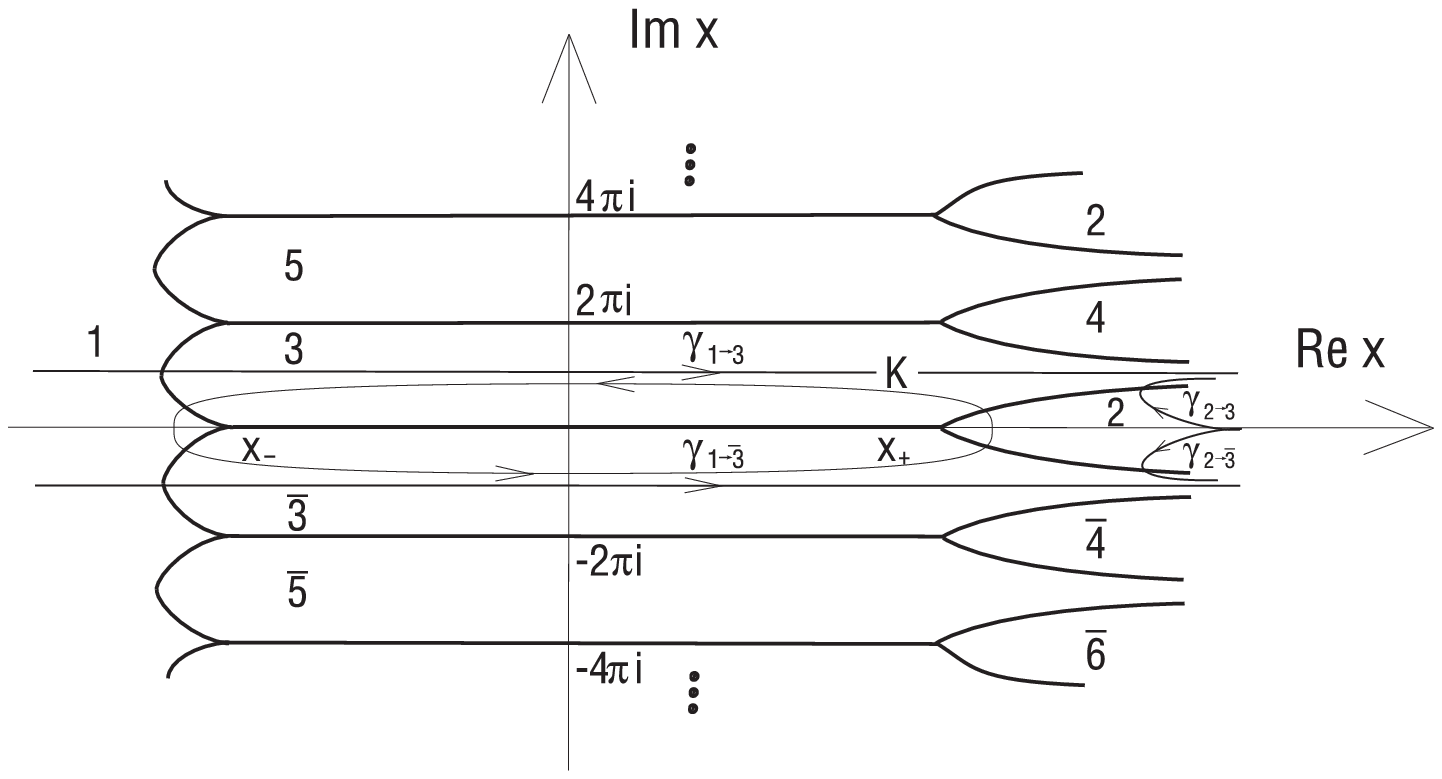,width=12cm} \\
Fig.3  The SG for the Morse type potential \mref{w3.4}
\end{tabular}

With $\alpha,\beta,\gamma>0$\ and $\beta^2>\alpha\gamma$\ we get
for $q(x,E,\lambda)=0$\ two real roots (modulo $2\pi i$) and the
corresponding SG shown in Fig.3 where $x_\pm=\ln(\beta
\pm\sqrt{(\beta^2-\alpha\gamma)})$. The quantization condition
\mref{w2.10} according to the figure looks now as follows:
\be\label{w3.5}
exp \left[-\lambda\oint\limits_K q^\fr(x,\lambda,E)dx\right]&=&
-\frac{\chi_{1\to 3}(\lambda,E)\chi_{2\to\bar{3}}(\lambda,E)}
{\chi_{1\to\bar{3}}(\lambda,E)\chi_{2\to 3}(\lambda,E)}
\ee

It follows from the figure that $\chi_{2\to 3}=\chi_{2\to\bar{3}}\equiv1$\ 
and $\chi_{1\to 3}\equiv\chi_{1\to \bar{3}}$. The first of these identities is
satisfied because both the paths $\gamma_{2\to 3}$\ and
$\chi_{2\to\bar{3}}$\ can be pushed out to infinities whilst the
second because of the periodicity of the corresponding
integrands in the formulae \mref{w2.6} for $\chi_{1\to 3}$\ and
$\chi_{1\to \bar{3}}$. Therefore, we are left finally with the
JWKB formula which gives $exact$ energy levels in this case.

It should be noticed, however, that the equality of the
coefficients $\chi_{1\to 3}$\ and $\chi_{1\to\bar{3}}$\ is not
immediate i.e. it does not follow as a direct result of the
periodicity of $q(x,E,\lambda)$. First we have to define the
total phase of $q(x,E,\lambda)$\ according to the prescriptions
of Appendix 1 in order to define uniquely its square roots
present in the coefficients $\chi_{1\to 3}$\ and
$\chi_{1\to\bar{3}}$. For the case just considered it has been
done in Appendix 1 where we have found that the phases of
$q(x,E,\lambda)$\ on the integration paths of $\chi_{1\to 3}$\
and $\chi_{1\to\bar{3}}$\ differ exactly by $4\pi$\ i.e. by the
period of the square roots of $q(x,E,\lambda)$\ just mentioned.

It should be stressed at this moment that to get the last result
the phases of the exponential factor in the corresponding
Weierstrass product (WP) have had to be taken into account i.e.
counting the relevant phases providing by the roots of
$q(x,E,\lambda)$\ alone would give us incorrect result.

\underline{case $k=-l=1$}

In this case $q(x,E,\lambda)$\ is given as:
\be\label{w3.6}
q(x,E,\lambda)=&\alpha e^x+\gamma e^{-x}-2\beta
\ee
where $\alpha,\beta,\gamma$\ all depend on E and $\lambda$\ and
are positive so that $q(x,E,\lambda)$\ represents the infinite
well.

\begin{tabular}{c}
\psfig{figure=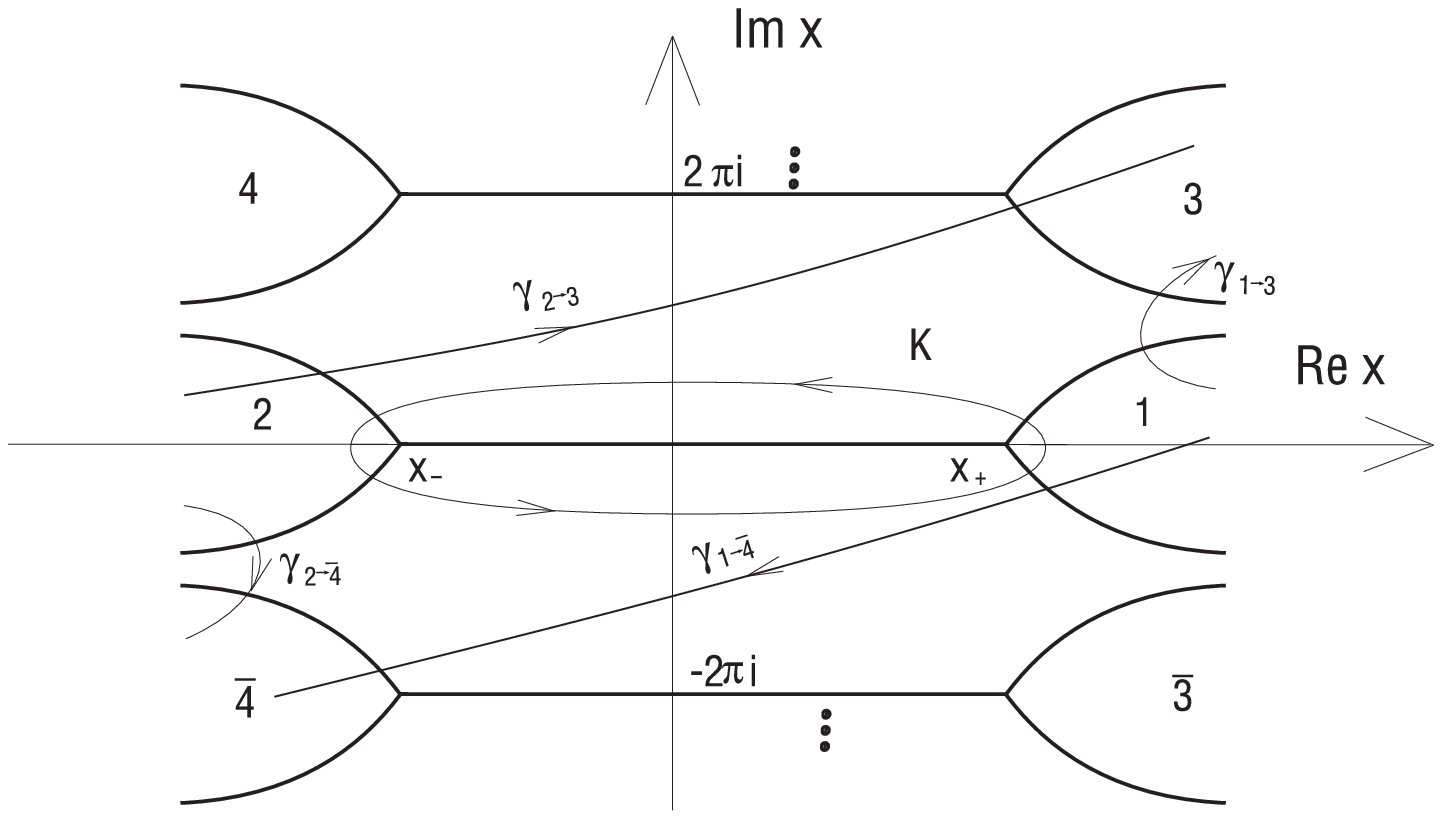,width=12cm} \\
Fig.4  The SG for the infinite potential well \mref{w3.6}
\end{tabular}

For $\beta^2>\alpha\gamma(>0)\; q(x,E,\lambda)=0$\ has again two
real roots (modulo $2\pi i$) and the corresponding SG looks as
in Fig.4. For the corresponding quantization condition we can
get the following two equivalent forms of it:

\be\label{w3.7}
exp\ll[-\lambda\oint\limits_kq^\fr(x,\lambda,E)dx\r]&=&-\frac{\chi_{2\to
3}(\lambda,E)\chi_{1\to\bar{4}}(\lambda,E)}{\chi_{2\to
\bar{4}}(\lambda,E)\chi_{1\to 3}(\lambda,E)}\nn\\
&{\rm or}&\\
\chi_{2\to 3}(\lambda,E)\chi_{1\to 4}(\lambda,E)&=&\chi_{1\to 3}(\lambda,E)
\chi_{2\to 4}(\lambda,E)\nn
\ee

It follows from Fig.4 that
${\chi_{1\to\bar{3}}(E,\lambda)=\chi_{2\to\bar{4}}(E,\lambda)=
\bar{\chi}_{2\to 4}(E,\lambda)\equiv 1}$,
$\chi_{1\to 4}(E,\lambda)=\bar{\chi}_{1\to\bar{4}}(E,\lambda)$\ and
$\chi_{2\to 3}(E,\lambda)=\chi_{\bar{4}\to 1}(E,\lambda)$. The
latter equality follows from the fact the phases of points of
the corresponding integration paths differ by $2\pi$\ (by
periodicity of $q(x,E,\lambda)$) what causes the square roots of
$q(x,E,\lambda)$\ to differ by their signs on the paths
compensated however by the opposite signatures of the
coefficients considered. (Note that this time the corresponding
phase difference calculations can take only into account the
phase contributions of roots of \mref{w3.6} alone since the
exponential factor is now absent in WP corresponding to
\mref{w3.6}).

Taking all these into account as well as the following general
equality \cite{13}:
\be
\chi_{j\to k}(E,\lambda)&=&\chi_{k\to j}(E,\lambda)\nn
\ee
we get finally for the quantization condition of the considered
potential:
\be\label{w3.8}
exp\ll[-\lambda\oint\limits_Kq^\fr(x,\lambda,E)dx\r]&=&
- \chi^2_{2\to 3}(\lambda,E)
\ee
together with: $|\chi_{2\to 3}(\lambda,E)|=1$.

We have to conclude therefore that in this case the exact
condition \mref{w3.7} cannot be reduced to the exact pure JWKB
ones by the pure symmetry arguments only.

The last potential, however, has been concluded initially by
Rosenzweig and Krieger \cite{4} as being exactly JWKB quantized
and next corrected by Krieger \cite{5} as to be not. The main
argument of the last author to support his conclusion was an
observation of the unvanishing first order correction to the
pure JWKB condition which follows from \mref{w3.7}. It is easy to
see that in terms of the coefficient $\chi_{2\to 3}(E,\lambda)$\
this argument means that the first integral in its
representation
\mref{w2.6} does not vanish i.e. the coefficient has to differ from unity.

It is still worth to note that the potential \mref{w3.6} can be
obtained from \mref{w3.4} just by multiplying the latter by
$e^{-x}$, so that if the coefficients $\alpha,\beta,\gamma$\ in
both of them are chosen to be the same then their sets of roots
coincide and their WP representations differ exactly by the
above factor $e^{-x}$. And this is just this factor which
introduces the dramatic difference between the corresponding
Stokes graphs of figures 3 and 4 and the corresponding
quantization conditions.

\underline{case $k=3,l=0$}

\begin{tabular}{c}
\psfig{figure=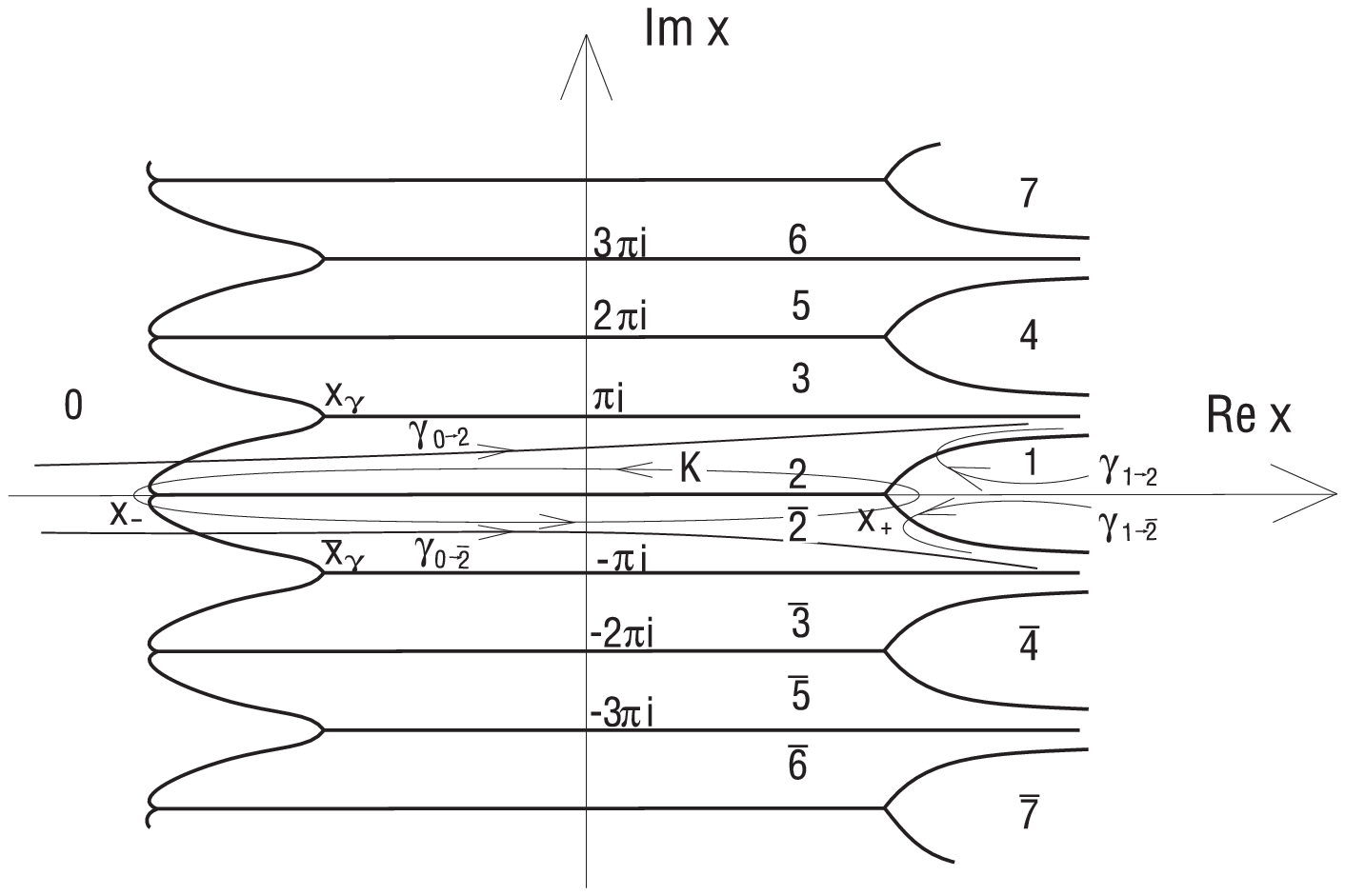,width=12cm} \\
Fig.5  The SG for the potential \mref{w3.9}
\end{tabular}

In order to satisfy the demand of two real turning points and
the reality condition $q(x,E,\lambda)$\ has to have now the
following form :
\be\label{w3.9}
q(x,E,\lambda)&=&\alpha(e^x-\beta_+)(e^x-\beta_-)(e^x+\gamma)
\ee
with $\alpha,\beta_\pm,\gamma>0$\ and all depending on $E$ and
$\lambda$\ as usually. The corresponding SG is shown in Fig.5
where $x_\pm=\ln(\beta_\pm)$\ and $x_\gamma=\ln(-\gamma)$. The
quantization condition is given therefore by:
\be\label{w3.10}
exp\ll[-\lambda\oint\limits_Kq^\fr(x,E,\lambda)dx\r]=-\frac{\chi_{0\to
2}(E,\lambda)\chi_{1\to\bar{2}}(E,\lambda)}{\chi_{0\to\bar{2}}
(E,\lambda)\chi_{1\to 2}(E,\lambda)}
=-\frac{\chi_{0\to 2}(E,\lambda)}{\bar{\chi}_{0\to 2}(E,\lambda)}
\ee
where the equality of $\chi_{1\to 2}$\ and $\chi_{1\to\bar{2}}$\
to unity which follow from Fig.5 have already been used.
Compairing further the phases of $q(x,E,\lambda)$\ on the lines
$\Im x=+\pi$\ and $\Im x=-\pi$\ we find that they differ by
$6\pi$.

The condition \mref{w3.10} shows therefore that if there are not
some accidental cancellations then the JWKB formula can be only
an approximation in this case i.e. it cannot be exact.

\underline{case $k=2, l=-1$}

This is the last case which we consider in details. The
corresponding $q(x,E,\lambda)$\ differs from \mref{w3.9} by the
last factor in which $e^x$\ is to be substituted by $e^{-x}$.
Thus $q(x,E,\lambda)$\ represents now an infinite potential well
with the corresponding SG shown in Fig.6. In comparison with
Fig.5 the horizontal SL emerging from $x_\gamma$\ and its
periodic distribution reverse their directions into the opposite
ones. This change of SG does not however allow all the $\chi$'s
to cancel in the corresponding quantization condition:
\be\label{w3.11}
exp\ll[-\lambda\oint\limits_Kq^\fr(x,E,\lambda)dx\r]&=&-\frac{\chi_{2\to
3}(E,\lambda)}{\bar{\chi}_{2\to 3}(E,\lambda)}
\ee
so that also in this case the JWKB formula cannot be exact.

\begin{tabular}{c}
\psfig{figure=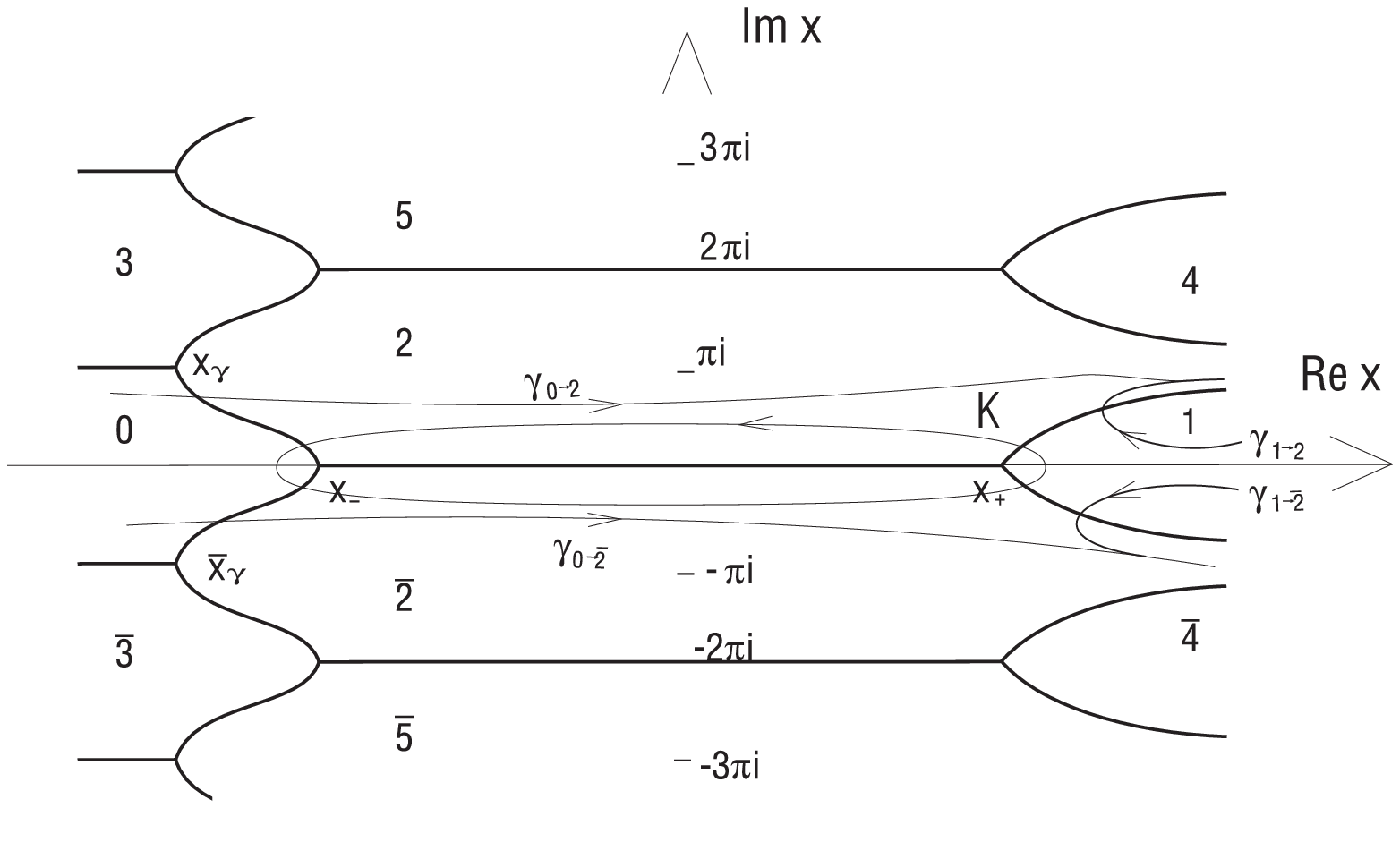,width=12cm} \\
Fig.6  The SG for the infinite potential well:\\ 
$V(x,\lambda)=\alpha(e^x - \beta_+)(e^x-\beta_-)
(e^{-x}+\gamma)-\alpha\beta_+\beta_-\gamma$ \\
corresponding to the quantization formula \mref{w3.11}
\end{tabular}

Considering higher values of $k-l\; (>3)$ it is easy to note
that all the conditions used as far (two real turning points,
reality and periodicity) are not sufficient to course the full
cancellations of the $\chi$'s in the corresponding quantization
conditions. The main reason for that is that for higher values
of $k+l$\ an increasing number of complex turning points in a
basic strip of periodicity causes the $\chi$'s entering the
corresponding quantization conditions not to be related any
longer by the condition of periodicity, because all the relevant
integrations are performed {\it inside} the same basic strip.

The above situation does not change even if we make
$q(x,E,\lambda)$\ to be additionally an even function of $x$.

\subsection{ Aperiodic exactly JWKB-quantized potentials \label{s3.2}}

\hskip+2em  The periodic potential \mref{w3.4} which provides us with the
exact JWKB quantization formulae \mref{w3.5} can serve also as
the source of aperiodic exactly JWKB-quantized potentials. The
latter can be obtained from the former by a trivial
change-of-variable procedure $x\to y(x)$\ resulting in the
following potential transformations:
\be\label{w3.12}
q(x,E,\lambda)\to\ll[\frac{q(x,E,\lambda)}{y'^2(x)}-\frac{1}{\lambda^2}
\ll[\frac{3}{4}\frac{y''^2(x)}{y'^4(x)}-\fr\frac{y'''(x)}{y'^3(x)}
\r]\r]_{x=x(y)}
\ee
   
The only necessary demand for a relevant change is to provide
by it in the resulting potentials a free (i.e. $x$--independent)
term which can play a role of the energy parameter (see
\cite{28}, for example).

The latter demand when applied to the potential \mref{w3.4}
permits the following two possibilities: $1^0\;\; e^{x/2}\to x$\
and $2^0 e^x\to x$. Adjusting properly $\alpha,\beta$\ and
$\gamma$\ in \mref{w3.4} we get in this way the 1--dim harmonic
oscillator potential and the radial parts of the 3--dim
homogeneous harmonic oscillator potential in the first case and
the Coulomb potential in the second one.

The same method can be applied to the potentials which are {\it
not} exactly JWKB quantized providing us with aperiodic
potentials with the same property i.e. the method allows us not to
make any further estimations of the resulting potentials for
their being not quantized exactly by the corresponding JWKB
formulae (what on their own could not be a simple task).

Applying the method to the potential well \mref{w3.6} for
example only the substitution $e^{x/2}\to x$\ is allowed
providing us with $q(x,E,\lambda)=\alpha x^{-4}-\beta x^{-2}-E$\
with $\alpha,\beta>0$\ and $-\beta^2/(4\alpha)<E<0$. Of course,
this potential can be considered as a radial part of a
spherically symmetric potential with the repelling term
$x^{-4}$\ and the attractive one $x^{-2}$\ the latter being
affected by the centrifugal term contribution. It is seen that
in this model a number of levels is finite being limited by the
increasing of the angular momentum. According to our earlier
result this potential cannot be JWKB exactly quantized.

It is worthwhile to note that the above substitution when
considered as the inverse transformation is nothing but the
well known Langer change--of--variable procedure applied to the
cases of the Coulomb and harmonic 3--dim potentials to obtain
the exact JWKB formulae for the latter potentials. It is clear
that in view of the above discussion Langer's substitutions are
just what is {\it necessary} to be done in order to achieve the
latter goal.
       
\subsection{Periodic meromorphic potentials\label{s3.3}}

\hskip+2em  The reality condition demanded for $q(x,E,\lambda)$\ allows us
to choose for its two possible basic periods the one being pure
real, and the second - pure imaginary.

Within this class of potentials we can ignore obviously
$q(x,E,\lambda)$\ with a real period but without real poles. We
have therefore to consider the following possibilities for
$q(x,E,\lambda)$:
\begin{description}
\item[a.] It is holomorphic in some vicinity of the real axis but 
meromorphic 
outside it and being periodic with its unique imaginary period equal to
$2\pi i$;
\item[b.] It is meromorphic on the real axis with the only imaginary period 
equal to $2\pi i$;
\item[c.] It is meromorphic on the real axis with the only real period equal 
to $2\pi$;
\item[d.] It is meromorphic on the real axis with two periods: a real one 
equal to $2\pi$\ and a pure imaginary one equal to $i\omega$\ 
with $\omega$\ being any positive real number.
\end{description}
\vskip 12pt
\underline{       case a.}
\vskip 12pt

In this case $q(x,E,\lambda)$\ is assumed again to have (for
some range of $E$) two real roots in its basic period strip
defined by $-\pi<\Im x\leq+\pi$. Let us assume also for a while
that it has only four complex pairwise conjugated poles in this
strip all lying on the imaginary axis (this position can always
be achieved by a simple translation). If the order of these
poles amounts to $n$ then $q(x,E,\lambda)$\ has to have the
following form:
\be\label{w3.13}
q(x,E,\lambda)=\frac{q_1(x,E\lambda)}
{\ll\{\sinh\fr(x-ia)\sinh\fr(x+ia)\r\}^n}+q_2(x,E,\lambda)
\ee
where $q_i(x,E,\lambda),\; i=1,2$, are holomorphic and periodic
with their period equal to $2\pi i$\ and $0<a\leq\pi$. Therefore
roots of $q(x,E,\lambda)$\ are given by the equation:
\be\label{w3.14}
q_1(x,E,\lambda)+\ll\{\sinh\fr(x-ia)\sinh\fr(x+ia)\r\}^n\cdot
q_2(x,E,\lambda)=0
\ee
with $q_i$'s having forms of the abbreviated series \mref{w3.3}.
According to our earlier observations that more than two roots
in the basic period streap prevent as a rule the JWKB formula to
be exact we have to put in \mref{w3.14}: $n=1,\;
q_1(x,E,\lambda)=\alpha e^x+\beta e^{-x} +\gamma$\ and
$q_2(x,E,\lambda)=const$.

However, we have to consider the cases $a\not=\pi$\ and $a=\pi$\
separately because their Stokes graphs differ in their
structures. In particular, whilst in the first case
$q(x,E,\lambda)$\ will have simple poles at the points $x=\pm ia$
($mod$ $2\pi i$)\ in the second case the second order poles are
generated.

Satisfying finally the assumption of two real roots of
$q(x,E,\lambda)$\ we get the following two potentials:
\be\label{w3.15}
&&V_1(x)=\frac{\alpha_1
e^x+\beta_1}{2\sinh\fr(x-ia)\sinh\fr(x+ia)}=\frac{\alpha_1e^x+\beta_1}{\cosh
x-\cosh a}\nn\\ &&V_2(x)=\frac{\alpha_2 e^x+\beta_2}{\cosh^2\fr
x}
\ee
the second of which is essentially the Rosen-Morse one
\cite{21}.

To obtain from \mref{w3.15} the potentials which would have
bound states some conditions on their parameters have to be
satisfied. For the first of them they are:
\be\label{w3.16}
\alpha_1>0>\beta_1
\ee
with the quantized energy $E$ varying in the following range:
\be\label{w3.17}
-\frac{2x(\alpha_1-x)^2}{(|\alpha_1-x)-|y|)^2+2|y||\alpha_1-x|(1-\cos
a)}<E<0\nn\\ x=\sqrt{\alpha_1^2+\beta_1^2+2\alpha_1\beta_1\cos
a},\;\;\;\;y=\beta_1+2\alpha_1\cos a
\ee
   
For the second potential in \mref{w3.15} we can put
$\alpha_2>0>\beta_2$\ not loosing its generality so that the
corresponding energy range is:
\be\label{w3.18}
-\frac{\beta^2_2}{\alpha_2-\beta_2}<E<0
\ee

\begin{tabular}{c}
\psfig{figure=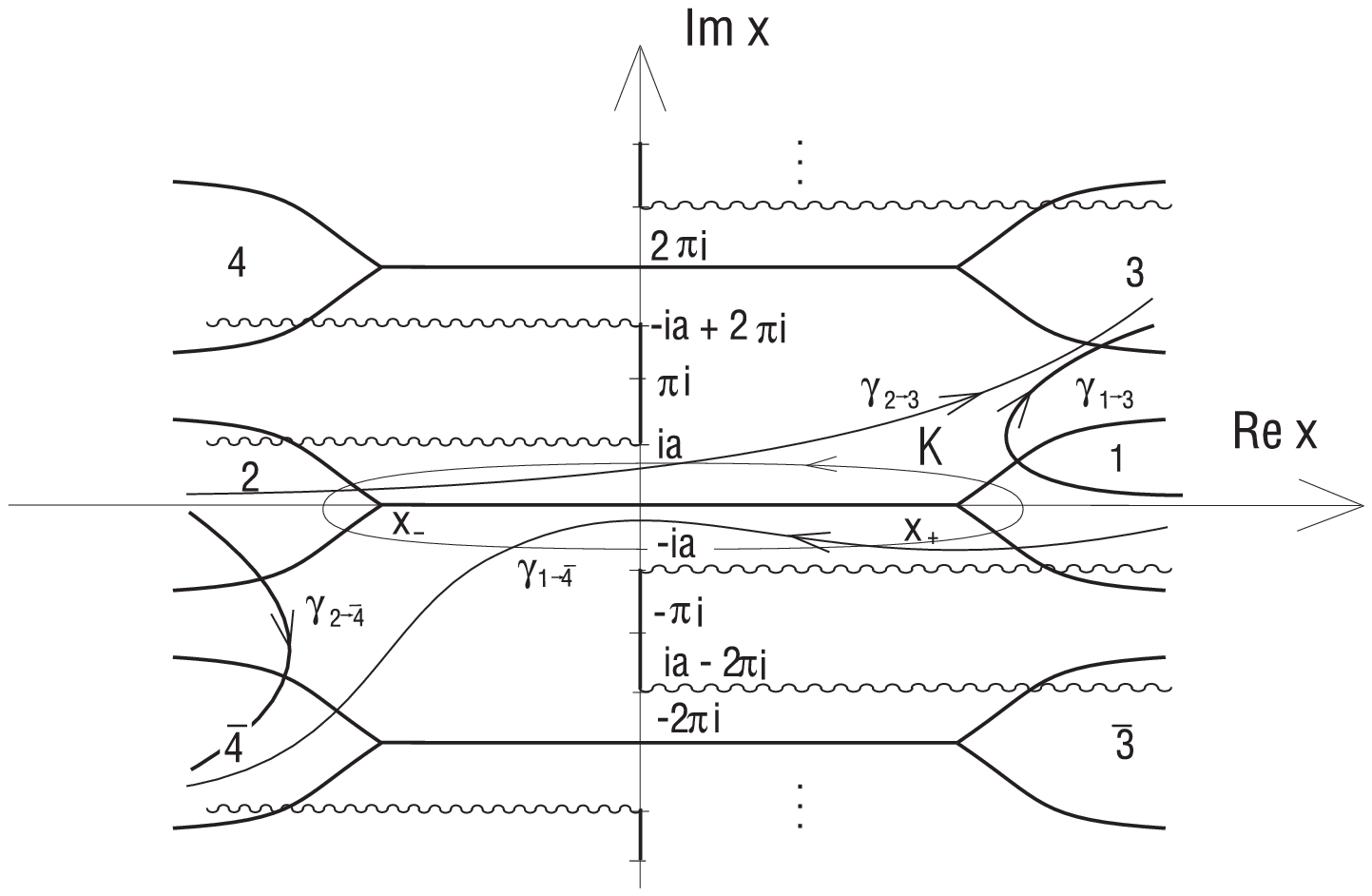,width=12cm} \\
Fig.7  The SG for the first of the potentials \mref{w3.15}
\end{tabular}

A relevant Stokes graph corresponding to the first of the
potentials \mref{w3.15} is shown in Fig.7. It follows
immediately from the figure that the JWKB formula:
\be\label{w3.19}
exp\ll[-\lambda\oint\limits_K\ll[\frac{\alpha_1e^x+\beta_1}{\cosh
x-\cosh a}-E\r]^\fr dx\r]=-1
\ee
cannot be exact in this case since the coefficients $\chi_{1\to
4}$\ and $\chi_{2\to 3}$\ are not related by periodicity and do
not cancel in the exact condition.

The SG for the second of the potential \mref{w3.15} is shown on
Fig.8a. In order however to continue the relevant solutions
corresponding to Sectors $1$ and $2$ to Sectors $3$ and $\bar{3}$ (the
latter two containing the second order poles at $x=\pm \pi i$\
respectively) we have to choose properly the $\delta$--piece of
$\omega$\ as defined by \mref{w2.7} to admit the integrals in
\mref{w2.6} to converge at the poles. One can
easily convince oneself that the choice
$\delta=[4\cosh(x/2)]^{-2}$\ is sufficient for such a goal
leaving simultaneously the original form of SG of Fig.8a.
unchanged i.e. it redefines the coefficient $\beta$\ into
$\beta'(=\beta-1/(4\lambda)^2)$\ only. Then Fig.8a provides us
with the following quantization condition for the case:
\be\label{w3.20}
exp\ll[-\lambda\oint\limits_K\sqrt{\frac{\alpha_2e^x+\beta_2 - 
\frac{1}{16\lambda^2}}{\cosh^2\fr
x}-E}dx\r]=-\frac{\chi_{1\to\bar{3}}(\lambda,E)\chi_{2\to 3}(\lambda,E)}
{\chi_{1\to 3}(\lambda,E)\chi_{2\to\bar{3}}(\lambda,E)}
\ee

\begin{tabular}{cc}
\psfig{figure=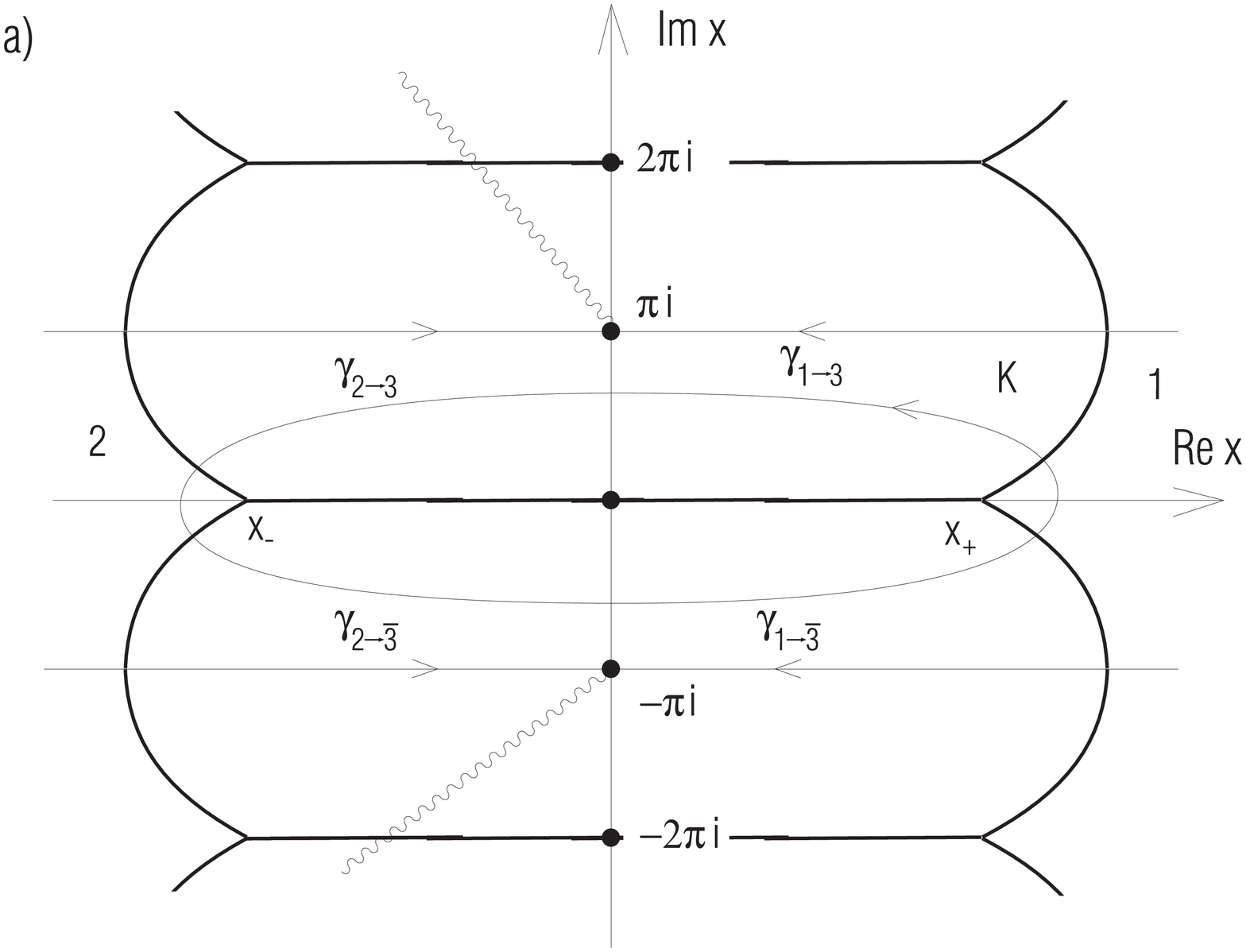,width=7cm}
&
\psfig{figure=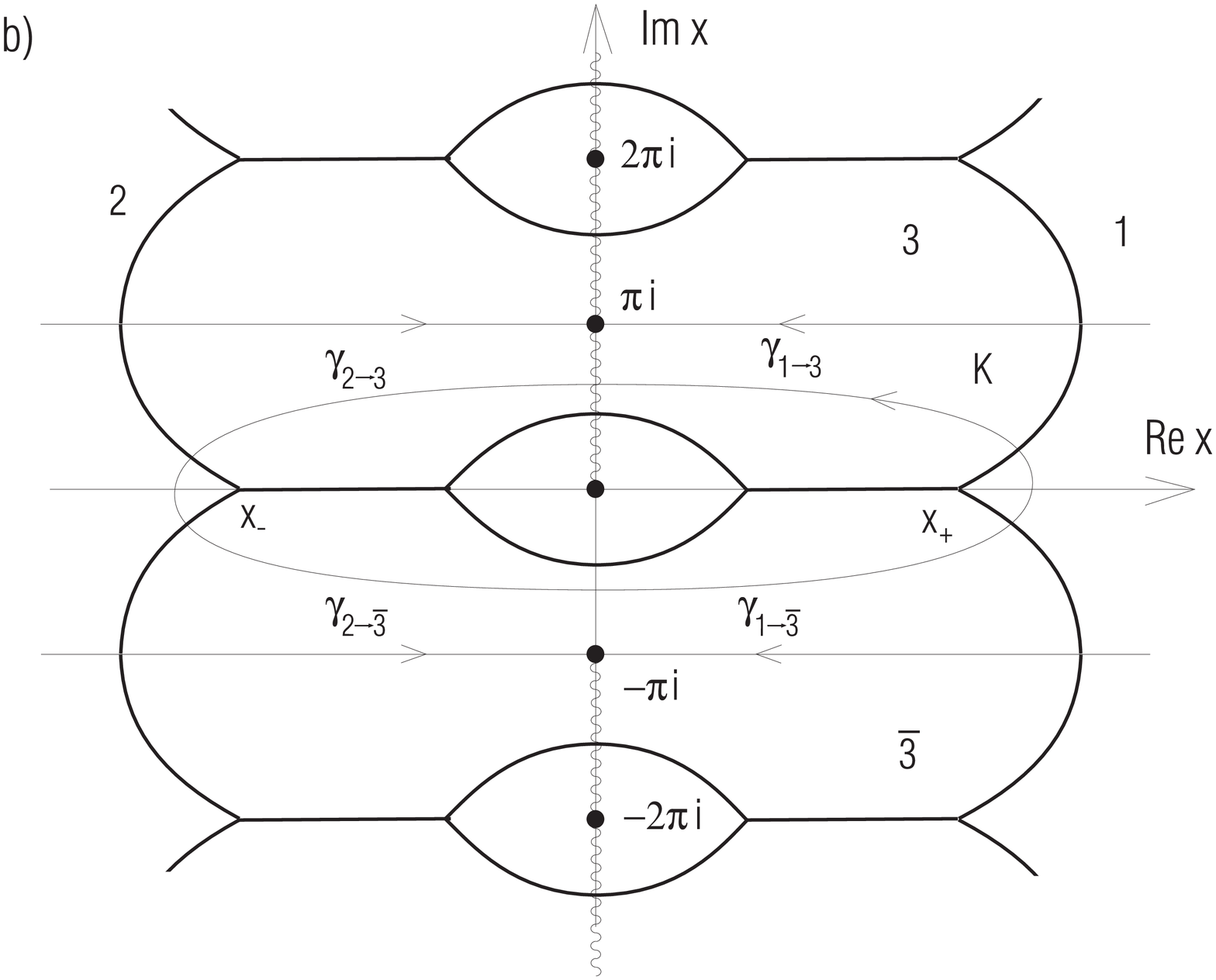,width=7cm} \\
\parbox{7cm}{Fig.8  The SG's corresponding to the second of the 
potentials \mref{w3.15}  and the quantization formulae  
\mref{w3.20} (Fig.8a) and \mref{w3.21} (Fig.8b) } &
\end{tabular}

According to Appendix 1 the total change of the phase of
$\tilde{q}(x,E,\lambda)$\ in \mref{w3.20} is determined only by
the distributions of zeros of its nominator as well as by the
corresponding zeros of $\cosh^2(x/2)$\ being the denominator.
Calculated (according to the rules of Appendix 1) with respect
to the points of the lines $\Im x=\pi$\ and $\Im x=-\pi$\
(shifted by the period $2\pi i$) the nominator phase change
amounts to $2\pi$\ what is exactly the same as the total phase
change of the denominator. Therefore, the total phase change of
$\tilde{q}(x,E\lambda)$\ is exactly equal to zero in this case
what is enough for $\chi_{1\to 3}$\ and $\chi_{1\to\bar{3}}$\ as
well as for $\chi_{2\to 3}$\ and $\chi_{2\to\bar{3}}$\ to
coincide. It means that the RHS of \mref{w3.20} is equal to $-1$
and the JWKB formula corresponding to \mref{w3.20} is {\it
exact} in the case considered.

It is interesting to note that the success of the JWKB formula
corresponding to \mref{w3.20} to be exact depends completely on
the total phase change of $\tilde{q}(x,E\lambda)$\ to be equal
to an integer multiple of $4\pi$. This condition permits also to
produce the forms of the exact JWKB formulae different than the
one discussed above. They can be obtained for example by
doubling the number of turning points in the basic period strip.
This can be done in many ways by choosing $\delta$\ properly.
One of such choices is $\delta=[4\cosh(x/2)]^{-2} +
a\lambda2\sinh^{-2}(x/2)$\ with real but arbitrary $a\not=0$.
This choice introduces to the basic period strip the second
order pole at $x=0$ but also two additional zeros (lying close to
the pole for small $a$). Both the zeros are real for $a>0$\ or
pure imaginary for $a<0$. Let us note further that the pole at
$x=0$ does {\it not} produce a singularity at this point for
$\Psi_1(x)$\ and $\Psi_2(x)$\ but it can do it for their
corresponding $\chi$--factors (since it does it for the
corresponding JWKB factors). However, the latter possibility does
not affect the procedure of writing the corresponding
quantization condition nor estimating periods corresponding to
the coefficients $\chi_1(x)$\ and $\chi_2(x)$. Choosing
therefore for definiteness $a$ to be positive we get the
corresponding SG in the form shown in Fig.8b and the following
quantization condition:

\be\label{w3.21}
exp\ll[-\lambda\oint\limits_K\ll[\frac{\alpha_2e^x+\beta_2-
\frac{1}{16\lambda^2}}{\cosh^2\fr x}
-\frac{a}{\sinh^2\fr x}-E\r]^\fr dx\r]=\nn\\
=-\frac{\chi_{1\to\bar{3}}(\lambda,E)\chi_{2\to
3}(\lambda,E)}{\chi_{1\to
3}(\lambda,E)\chi_{2\to\bar{3}}(\lambda,E)}
\ee
   
Note that $\chi_{1\to 3}$\ and $\chi_{2\to 3}$\ in \mref{w3.21}
are calculated on different sides of the (possible) cut
emanating from the point $x=0$.

Now it is easy to see that the $\chi$'s in the RHS of
\mref{w3.21} cancel mutually pairwise by
periodicity and we obtain from \mref{w3.21} the {\it exact} JWKB
quantization formula with $a$ as an {\it arbitrary} real
parameter. In particular we can put $a=0$ in the formula getting
it again in the 'standard' Bailey's form \cite{3}.

It follows from the above considerations that it is rather
hopeless to look for other potentials of the case considered
which could provide us with corresponding exact JWKB formulae.
The proliferation of roots unavoidable for these potentials
should prevent effectively the exact JWKB quantization
conditions to appear closing the relevant quantizations {\it
inside} the single period strip. 
\vskip 12pt
\underline{       case b.}
\vskip 12pt

Assuming the presence of simple or second order poles in
$q(x,E,\lambda)$\ it is clear that we can allow only one such a
pole in the main period strip. We assume its localization at
$x=0$. In this way the problem of quantization has to be reduced
to a half of the real axis which we choose not loosing a
generality to be the right one. The allowed classes of
potentials should have therefore the forms:
\be\label{w3.22}
V_1(x,\lambda)&=&\frac{\alpha_1e^x+\beta_1}{\sinh x}\nn\\
V_2(x,\lambda)&=&\frac{\alpha_2e^x+\beta_2}{\sinh^2\fr x}
\ee

If bound states are to exist for the first potential in
\mref{w3.22} its parameters have to satisfy
the following relations:
\be\label{w3.23}
\beta_1<0,&-\beta_1<\alpha_1,&\alpha_1+\sqrt{\alpha_1^2
\beta_1^2}<E<2\alpha_1
\ee

\begin{tabular}{cc}
\psfig{figure=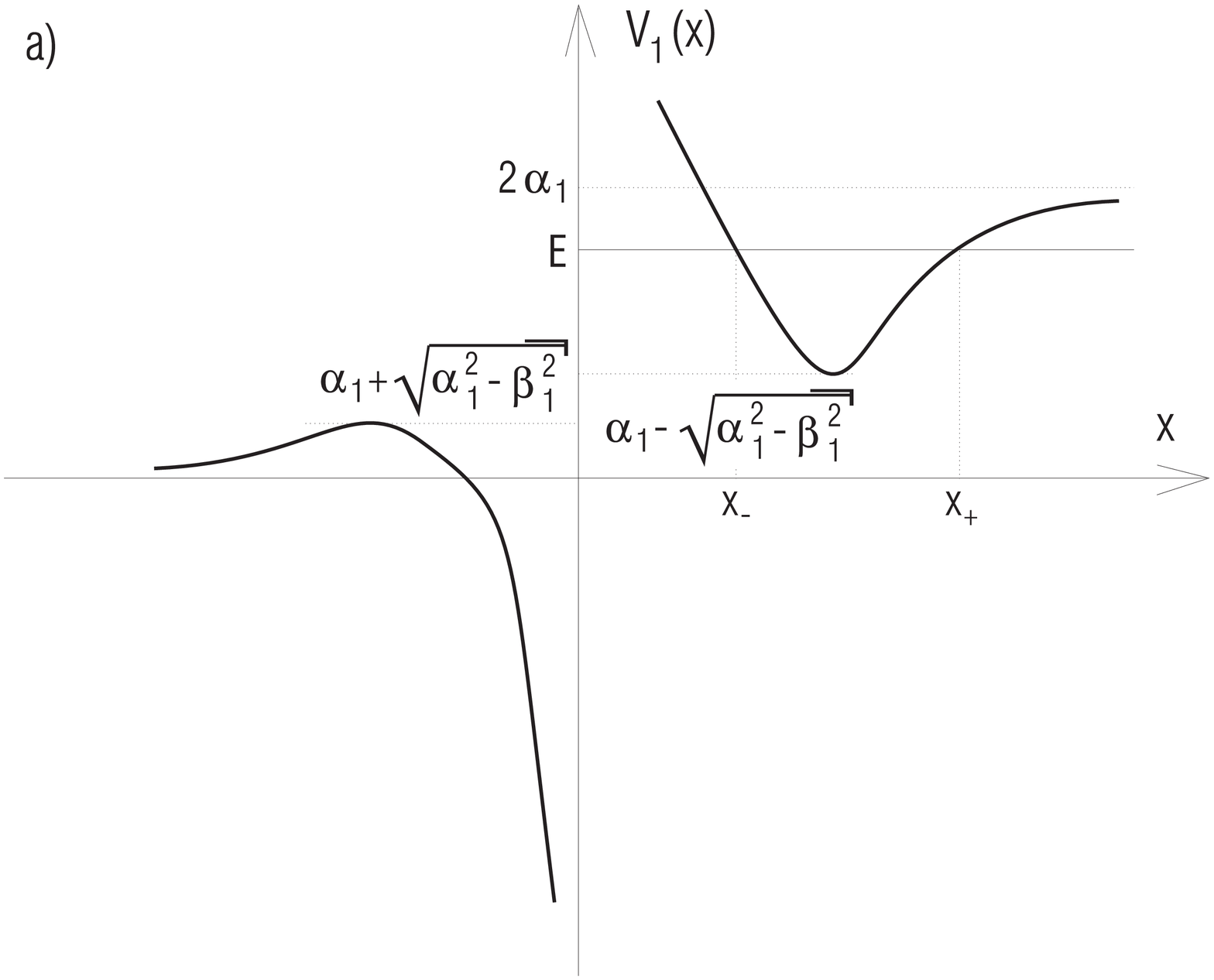,width=7cm} & \psfig{figure=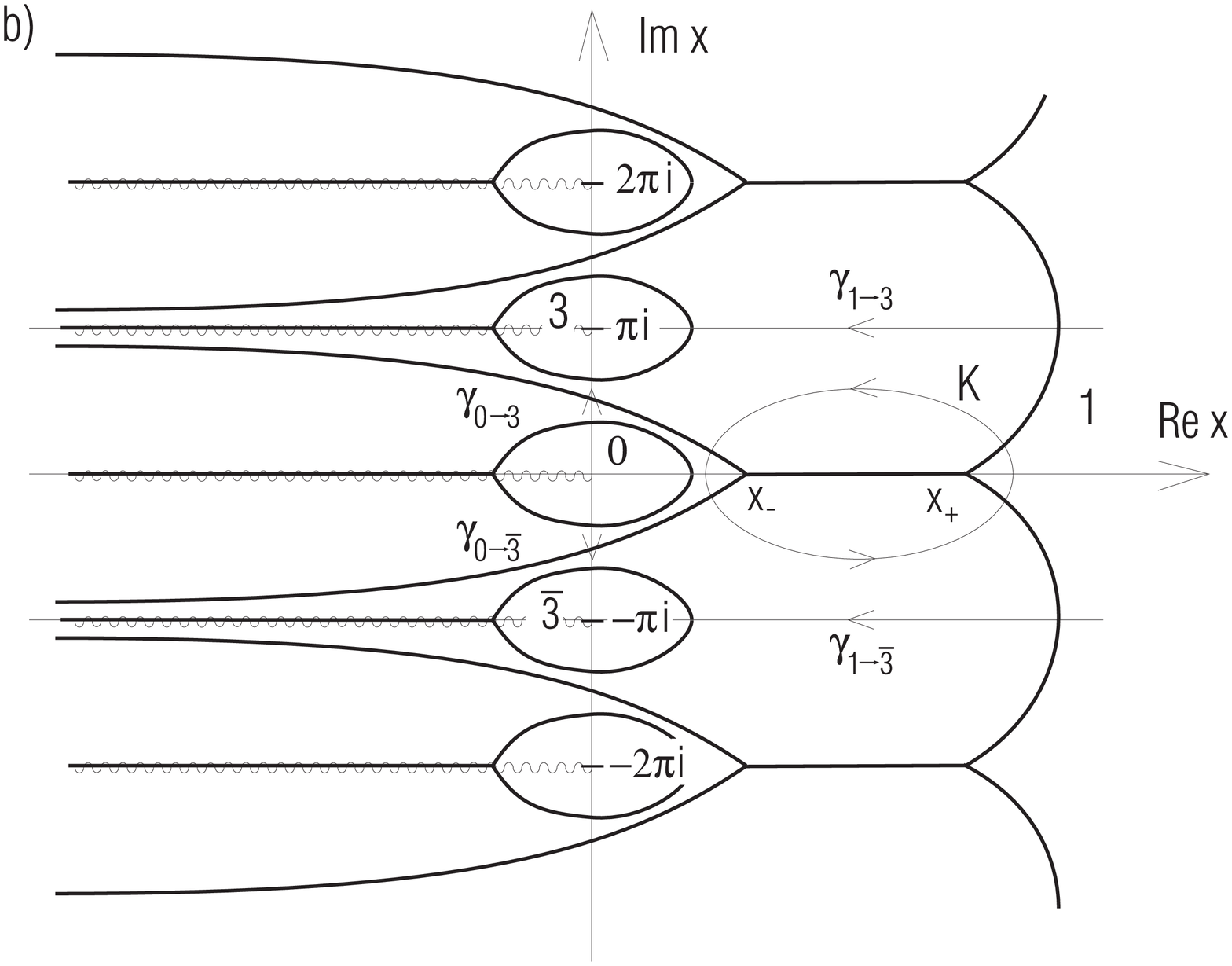,width=7cm} \\
\parbox{7cm}{Fig.9  The first of the potentials \mref{w3.22} and the 
corresponding SG} &
\end{tabular}

A shape of the potential is shown in Fig.9a. (Note that its
local maximum is {\it below} its local minimum). The
corresponding SG has to be modified because of the presence of
simple poles in the potential and because of the corresponding
quantization condition demanding for the wave function to vanish
at $x=0$. A 'minimal' choice for $\delta$\ is $\delta=(2\sinh
x)^{-2}$\ so that the corresponding SG looks now as in Fig.9b. 

The energy quantization demands now the solution $\Psi_1(x)$\
corresponding to Sector 1 and the solution $\Psi_0(x)$) from
Sector 0 to coincide. This provides us with the following
quantization condition:
\be\label{w3.24}
exp\ll[-\lambda\oint\limits_K\ll[\frac{\alpha_1e^x+\beta_1}{\sinh x}+
\frac{1}{4\lambda^2}\frac{1}{\sinh^2x}-E\r]^\fr dx\r]= \nn \\
=-\frac{\chi_{1\to \bar{3}}(E,\lambda)\chi_{0\to
3}(E,\lambda)}{\chi_{1\to3}(E,\lambda)\chi_{0\to\bar{3}}(E,\lambda)}
\ee
   
From \mref{w3.24} it follows, however, that although the
coefficients $\chi_{1\to3}$\ and $\chi_{1\to\bar{3}}$\ of the
formula mutually cancel (by periodicity) the remaining two
coefficients do not (i.e. the latter coefficients are not real)
and therefore the corresponding JWKB formula which follows of
\mref{w3.24} {\it cannot} be exact.

\begin{eqnarray*}
\psfig{figure=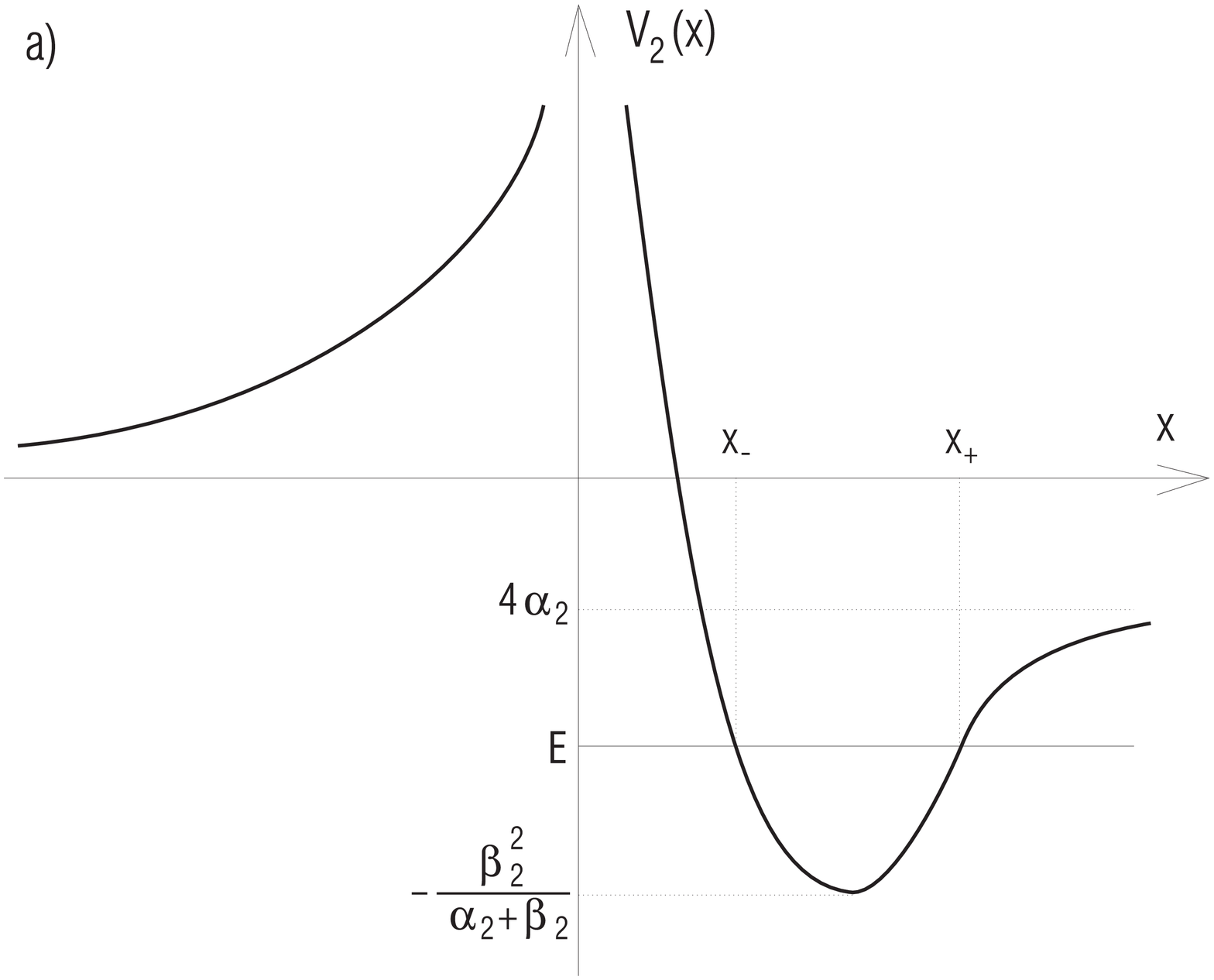,width=7cm} & \psfig{figure=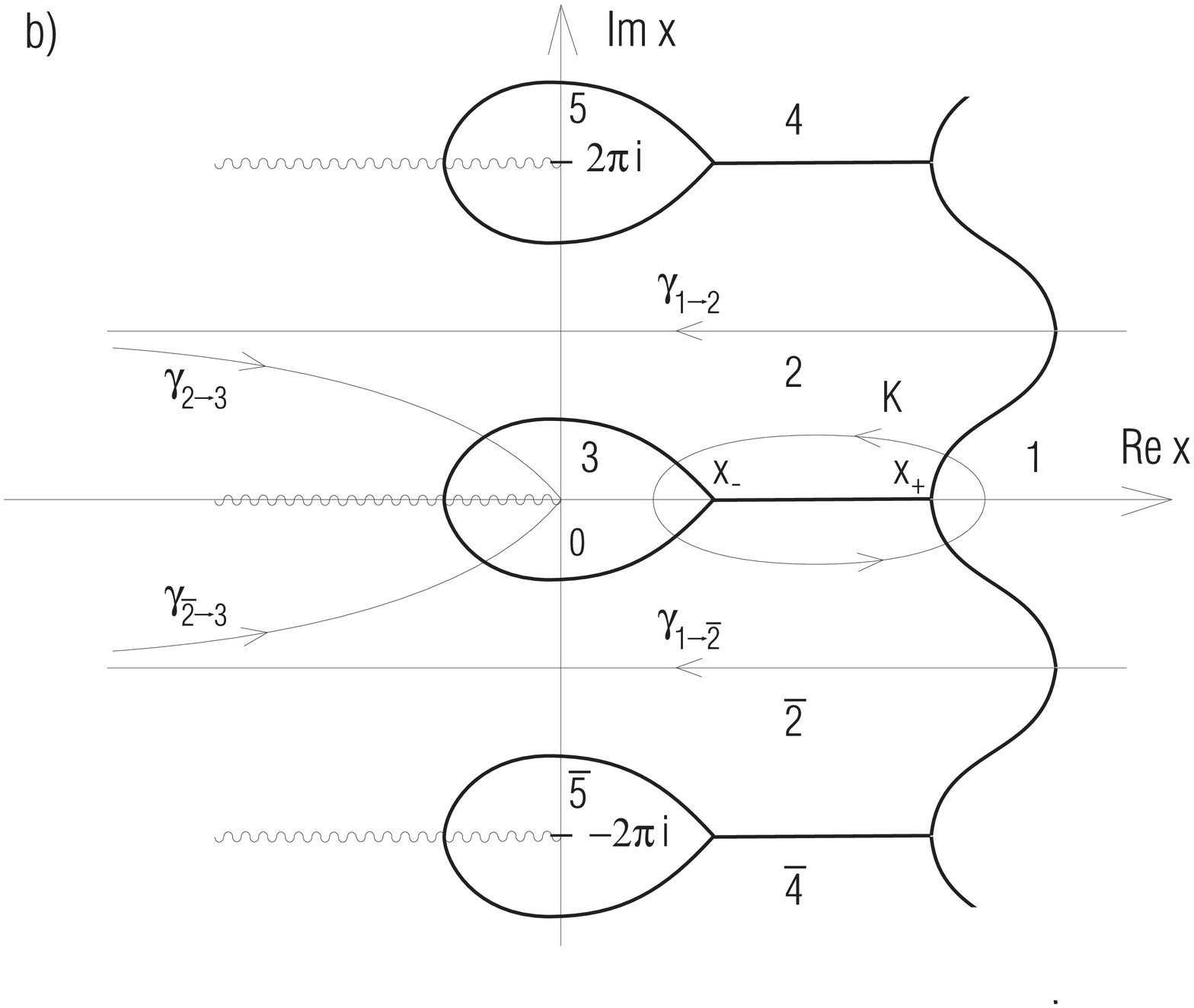,width=7.2cm} \\
\psfig{figure=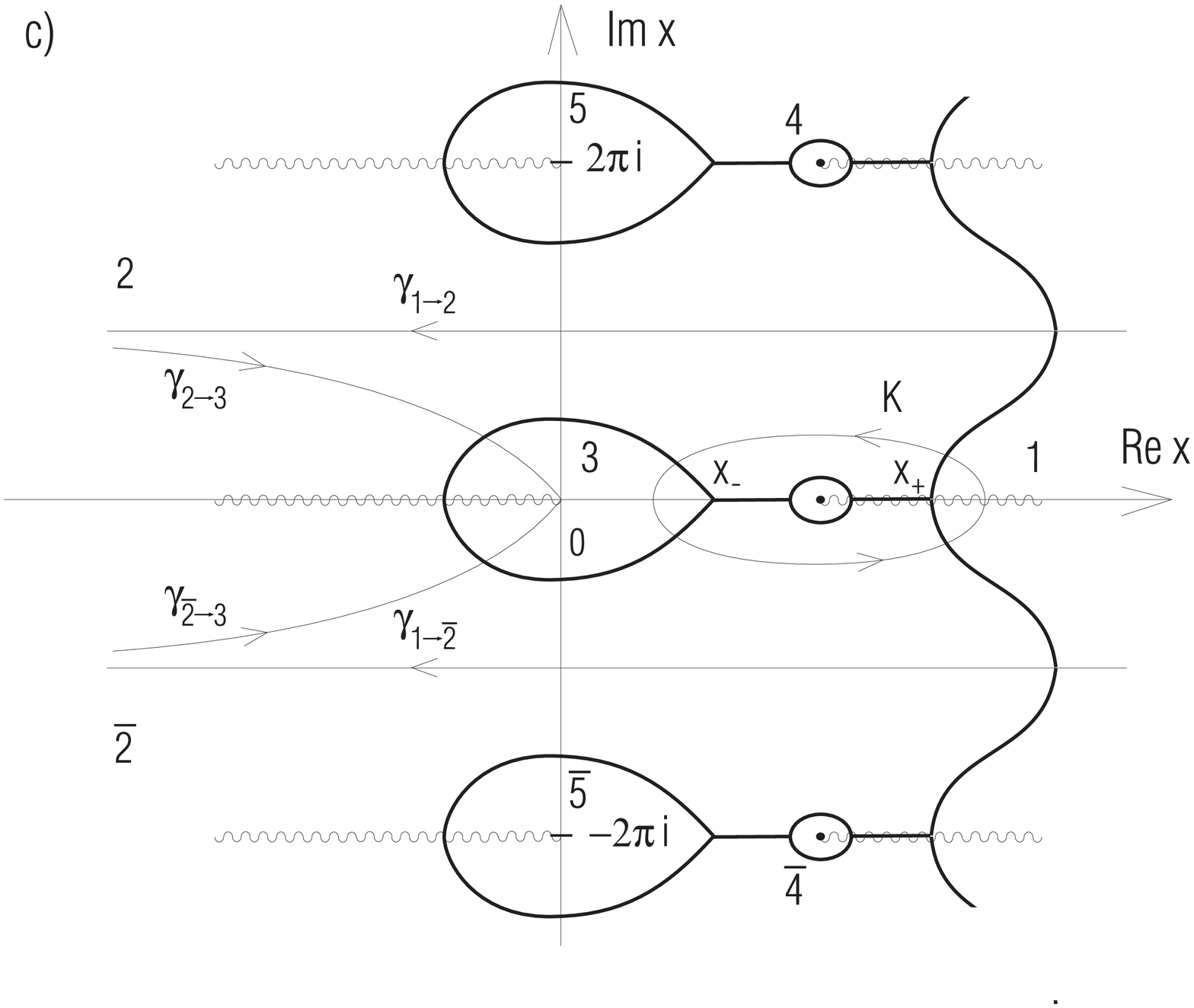,width=7cm} & 
\parbox[b]{7cm}{Fig.10  The second of potentials \mref{w3.22}  and 
the SG's  corresponding to the quantization 
formulae \mref{w3.20} (Fig.10a) and \mref{w3.27} (Fig.10b)}
\end{eqnarray*}

Not loosing a generality we can assume the parameters
$\alpha_2$\ and $\beta_2$\ of the second potential in
\mref{w3.22} to satisfy the following conditions:
\be\label{w3.25}
\beta_2,\alpha_2+\beta_2>0>2\alpha_2+\beta_2>\alpha_2\nn\\
-\frac{\beta^2_2}{\alpha_2+\beta_2}<E<4\alpha_2
\ee
to ensure an existence of bound states in the local potential
well. The shape of the potential is shown in Fig.10a. The
potential has to be modified by the 'standard' $\delta$--term:
$\delta=(4\sinh(x/2))^{-2}$\ to allow the construction of the FS
at $x=0$ what results with the change:
$\beta_2\to\beta_2+(4\lambda)^{-2}$\ in the potential. The
quantization condition corresponding to SG of Fig.10b reads
now:
\be\label{w3.26}
exp\ll[-\lambda\oint\limits_K\ll[\frac{\alpha_2e^x+\beta_2+
\frac{1}{16\lambda^2}}{\sinh^2\frac{x}{2}}-E\r]\fr dx\r]
=-\frac{\chi_{1\to\bar{2}}(E,\lambda)\chi_{3\to2}(E,\lambda)}
{\chi_{1\to2}(E,\lambda)\chi_{3\to\bar{2}}(E,\lambda)}
\ee

It follows from \mref{w3.26} and from Fig.10b that in this case
the coefficients $\chi_{1\to2}$\ and $\chi_{1\to\bar{2}}$\
cancel mutually by the periodicity arguments (the phase
difference produced by the nominator of
$\tilde{q}(x,E,\lambda)$\ and equal to $2\pi$\ for the two
integration paths $\gamma_{1\to2}$\ and $\gamma_{1\to\bar{2}}$\
is cancelled by its denominator $\sinh^2(x/2)$) whilst the
remaining two coefficients cancels by their reality (they are
real and complex conjugated to each other). It means that the
JWKB formula which follows from \mref{w3.26} is $exact$.

Again it is worth to note that the considered potential can be
modified by the $\delta$--function in a different way to
generate at least four zeros in the basic period strip of the
corresponding SG allowing the nominator of the modified
$\tilde{q}(x,E\lambda)$\ to change its phase by $4\pi$\ between
the earlier mentioned paths. This can achieved by putting
$\delta=(4\sinh(x/2)^{-2} + a\lambda^2\sinh^{-2}((x-x_0)/2))$\
with real and positive $a$ but sufficiently small for $x_0$\
satisfying $x_-<x_0<x_+$\ where $x_\pm$\ are the outer two (of
the total four) real turning points of the modified potential.
The corresponding SG is then shown in Fig.10c and for the
quantization condition we get:
\be\label{w3.27}
exp\ll[-\lambda\oint\limits_K\ll[\frac{\alpha_2e^x+\beta_2+
\frac{1}{16\lambda^2}}{\sinh^2\frac{x}{2}}+\frac{a}{\sinh^2\frac{x-x_0}{2}}
-E\r]^\fr dx\r]=\nn\\
=-\frac{\chi_{1\to\bar{2}}(E,\lambda)\chi_{3\to2}(E,\lambda)}{\chi_{1\to2}
(E,\lambda)\chi_{3\to\bar{2}}(E,\lambda)}
\ee
    
It follows from \mref{w3.27} and from Fig.10c that in this case
the coefficients $\chi_{1\to2}$\ and $\chi_{1\to\bar{2}}$\ again
cancel mutually by the periodicity arguments whilst the
remaining two again by their reality. It means that the JWKB
formula which follows from \mref{w3.27} is again exact and
coincides with the previous one when $a=0$.

The possibility of making the last modification enlarging the
number of roots in the basic period strip to four still suggests
to complete it differently, namely by adding the term coinciding
exactly with the second of the potential \mref{w3.15}. This of
course needs also to add the corresponding standard
$\delta$--term to the potential obtained in this way. The
potential we get in this way does not, however, satisfy the rule
of no more than two turning points in the period strip so that a
possibility of the exact JWKB quantization condition to appear
should mostly depend on symmetry properties of the relevant
$\chi$--coefficients. The considered potential can have bound
states for the following regime of its parameters (see the
formula below for the definition of the parameters):
$\alpha,\alpha'$\ real and sufficiently close to zero and
$\beta,\beta'>0$. Then the SG corresponding to the case is shown
in Fig.11 and the quantization condition related to it is:
\be\label{w3.28}
exp\ll[-\lambda\oint\limits_K\ll[\frac{\alpha
e^x+\beta+\frac{1}{16\lambda^2}}{\sinh^2\frac{x}{2}}+
\frac{\alpha'e^x-\beta'-\frac{1}{16\lambda^2}}{\cosh^2\frac{x}{2}}-E\r]^\fr
dx\r]=\nn\\
=-\frac{\chi_{1\to\bar{2}}(E,\lambda)\chi_{3\to2}(E,\lambda)}{\chi_{1\to2}
(E,\lambda)\chi_{3\to\bar{2}}(E,\lambda)}
\ee

\begin{tabular}{c}
\psfig{figure=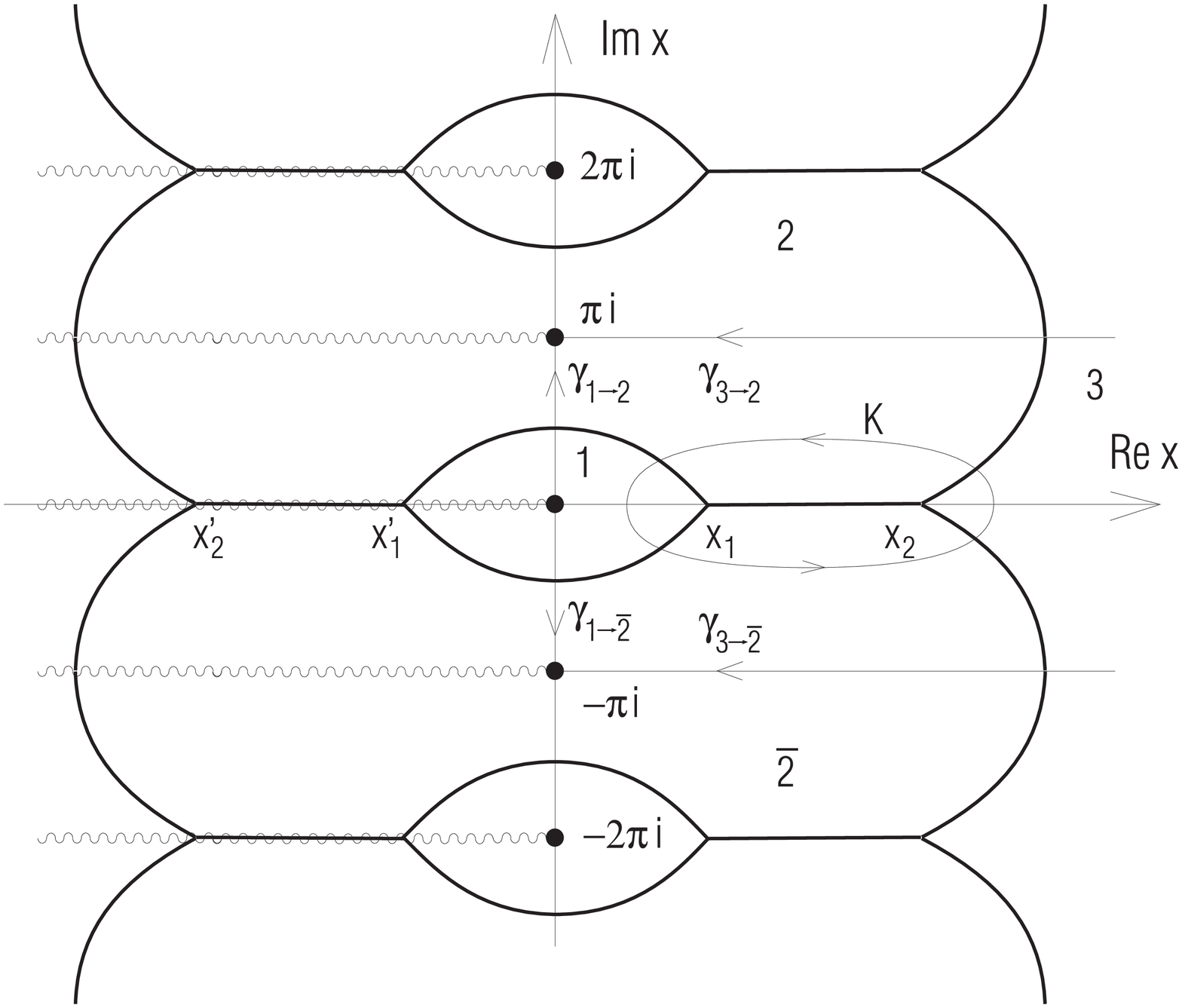,width=10cm} \\
Fig.11  The SG corresponding to the formula \mref{w3.29} \\
quantizing the potential of P\"oschl and Teller
\end{tabular}

It follows from the figure that the coefficients $\chi_{3\to2}$\
and $\chi_{3\to\bar{2}}$\ have to cancel mutually by periodicity
but not the remaining two: none symmetry (except the complex
conjugation) relates these two coefficients. However, when
$\alpha=\alpha'=0$\ the potential in \mref{w3.28} becomes
invariant under the reflection $x\to-x$\ and then the
coefficients $\chi_{1\to2}$\ and $\chi_{1\to\bar{2}}$\ are equal
just by the last symmetry.  (Note, however, the role played in
fulfilling this symmetry by $4\pi$\ of difference between the
arguments of $\tilde{q}(x,E,\lambda)$\ corresponding to the case
the latter takes on the paths $\gamma_{1\to2}$\ and
$\gamma_{1\to\bar{2}}$).

Therefore the following quantization condition:
\be\label{w3.29}
exp\ll[-\lambda\oint\limits_K\ll[\frac{\beta+\frac{1}{16\lambda^2}}
{\sinh^2\frac{x}{2}}-\frac{\beta'+\frac{1}{16\lambda^2}}
{\cosh^2\frac{x}{2}}-E\r]^\fr
dx\r]=-1
\ee
is {\it exact}. The potential in the above formula is of
P\"oschl and Teller \cite{22}.
\vskip 12pt
\underline{       case c.}
\vskip 12pt

The case contains the following four potentials:
\be\label{w3.30}
V_1(x,\lambda)=\frac{\alpha_1\sin x+\beta_1}{\cos
x}&&-\frac{\pi}{2}<x<\frac{\pi}{2}\nn\\
V_2(x,\lambda)=\frac{\alpha_2\sin
x+\beta_2}{\cos^2\frac{x}{2}}&&-\pi<x<\pi\\
V_3(x,\lambda)=\frac{\alpha_1'\sin x+\beta_1'}{\cos
x}+\frac{\alpha_1''\cos x+\beta_1''}{\sin
x}&&0<x<\frac{\pi}{2}\nn\\ V_4(x,\lambda)=\frac{\alpha_2'\sin
x+\beta_2'}{\cos^2\frac{x}{2}}+\frac{\alpha_2''\sin
x+\beta_2''}{\sin^2\frac{x}{2}}&&0<x<\pi\nn
\ee
the second of which is essentially another of P\"oschl-Teller
\cite{22}.

According to our earlier experience we cannot expect energy
levels of the first potential (where $\pm\alpha_1+\beta_1>0$) as
well as of the third one to be exactly quantized with its
corresponding JWKB formulae. This is because the points
$x=\pm\pi/2$\ which are singular for the first potential both
lie {\it inside} its basic period strip and thererefore since
the corresponding boundary conditions are formulated just for
these points the $\chi$--coefficients entering the relevant
quantization formula cannot mutually cancel i.e. the periodicity
argument should not work in this as well as in the third cases.

The second and the fourth potentials are more promising and as
it is well known the case $\alpha_2=0$\ of the first one is
quantized exactly by the JWKB formula.

Consider therefore the first of them. In order to have the
binding potential well we have to assume $\beta_2>0$ but the
choice of sign of $\alpha_2$\ is arbitrary since both the cases
are equivalent. So we shall put $\alpha_2>0$ for convenience.
Next we have to notice however that asymmetry introduced to the
potential by $\alpha_2\not=0$\ completely eliminates the
possibility of using the periodicity arguments.  Therefore we
shall put $\alpha_2=0$\ in \mref{w3.30}. Once more we have to
choose $\delta$\ taking it in its 'standard' form
$\delta=(4\cos(x/2))^{-2}$ and getting the SG of Fig.12a. It is
seen from the figure that the coefficients $\chi_{1\to-2}$\ and
$\chi_{1\to2}$\ as well as $\chi_{\bar{1}\to-2}$\ and
$\chi_{\bar{1}\to2}$\ are now equal by the periodicity
arguments. Because of that the following JWKB quantization
formula:
\be\label{w3.31}
exp\ll[-\lambda\oint\limits_K\ll[\frac{\beta_2+
\frac{1}{16\lambda^2}}{\cos^2\frac{x}{2}}-E\r]^\fr dx\r]=-1
\ee
is {\it exact}.

\begin{tabular}{cc}
\psfig{figure=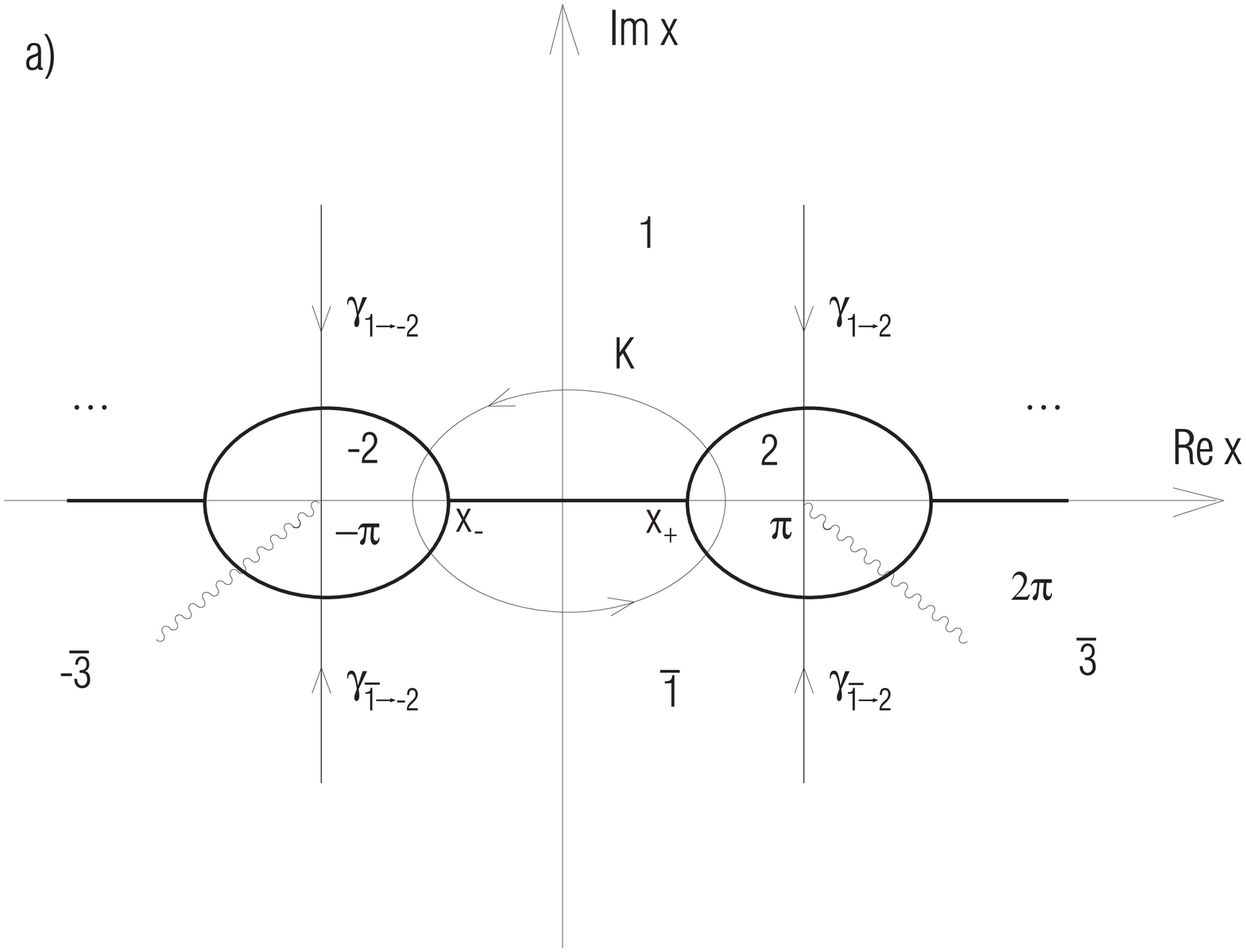,width=7.2cm} & \psfig{figure=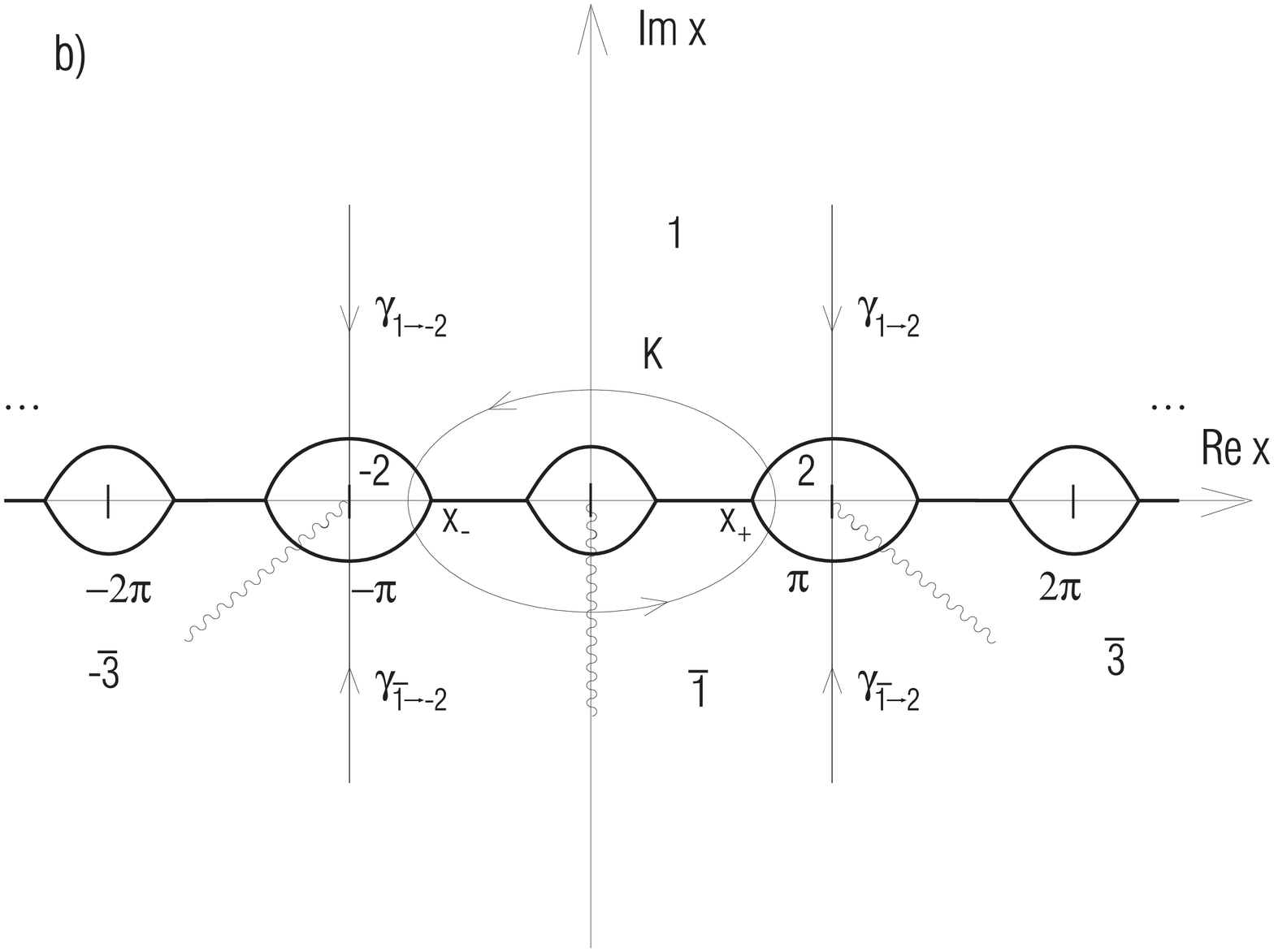,width=7.2cm} \\
\parbox{7cm}{Fig.12  The SG's corresponding to the two variants 
\mref{w3.31} (Fig.12a) and \mref{w3.32} (Fig.12b) of the JWKB formula 
for another P\"oshl--Teller potential} &
\end{tabular}

Similarly to the hyperbolic cosine case there is again
possibility to modify $\tilde{q}(x,E,\lambda)$\ differently by
putting $\delta=a\lambda^2\sin^{-2}(x/2)+(4\cos(x/2))^{-2}$. The
first term of $\delta$\ (with an arbitrary real a) introduces
two additional zeros in the basic period strip (necessary for
the corresponding $\chi$'s to have period $4\pi$). The standard
second term allows to construct the convergent solutions at the
points $x=\pm\pi$. Choosing for definitness $a>0$\ we obtain the
effective form of the SG corresponding to these modifications as
shown in Fig.12b. It is seen from the figure that the
coefficients $\chi_{1\to-2}$\ and $\chi_{1\to2}$\ as well as
$\chi_{\bar{1}\to-2}$\ and $\chi_{\bar{1}\to2}$\ are equal by
the periodicity arguments. Because of that the following JWKB
quantization formula:
\be\label{w3.32}
exp\ll[-\lambda\oint\limits_K\ll[\frac{\beta_2+\frac{1}{16\lambda^2}}
{\cos^2\frac{x}{2}}+\frac{a}{\sin^2\frac{x}{2}}-E\r]^\fr dx\r]=-1
\ee
is {\it exact} for any $a>0$\ and coincides with \mref{w3.31}
for $a=0$.

Considering finally the last of the potentials \mref{w3.30}
first we remove (for the same reason discussed earlier)
asymmetry in the latter putting $\alpha_2'=\alpha_2''=0$\ and
next we notice that in the basic period strip $-\pi<\Re
x\leq\pi$\ the number of the four turning points is sufficient
to make the relevant $\chi$'s periodic across the strip.
Therefore the only necessary modification of the potential is
the 'standard' choice for $\delta$\ i.e.
$\delta=(4\sin(x/2))^{-2}+(4\cos(x/2))^{-2}$\ what gives the SG
shown in Fig.13.  The following relations come then out from
the figure: $\chi_{1\to\bar{3}}=\chi_{-1\to3}=\chi_{1\to3}$\ and
$\chi_{0\to3}=\chi_{0\to\bar{3}}$. The first equalities in both
of these equality sequences follows from the parity invariance
of the potential considered whilst the second in the first one
is satisfied by the periodicity arguments. Therefore the
following JWKB quantization condition:
\be\label{w3.33}
exp\ll[-\lambda\oint\limits_K\ll[\frac{\beta_2'+\frac{1}{16\lambda^2}}
{\cos^2\frac{x}{2}}+\frac{\beta_2''+\frac{1}{16\lambda^2}}
{\sin^2\frac{x}{2}} - E\r]^\fr dx\r]=-1
\ee
is {\it exact}.

\begin{tabular}{c}
\psfig{figure=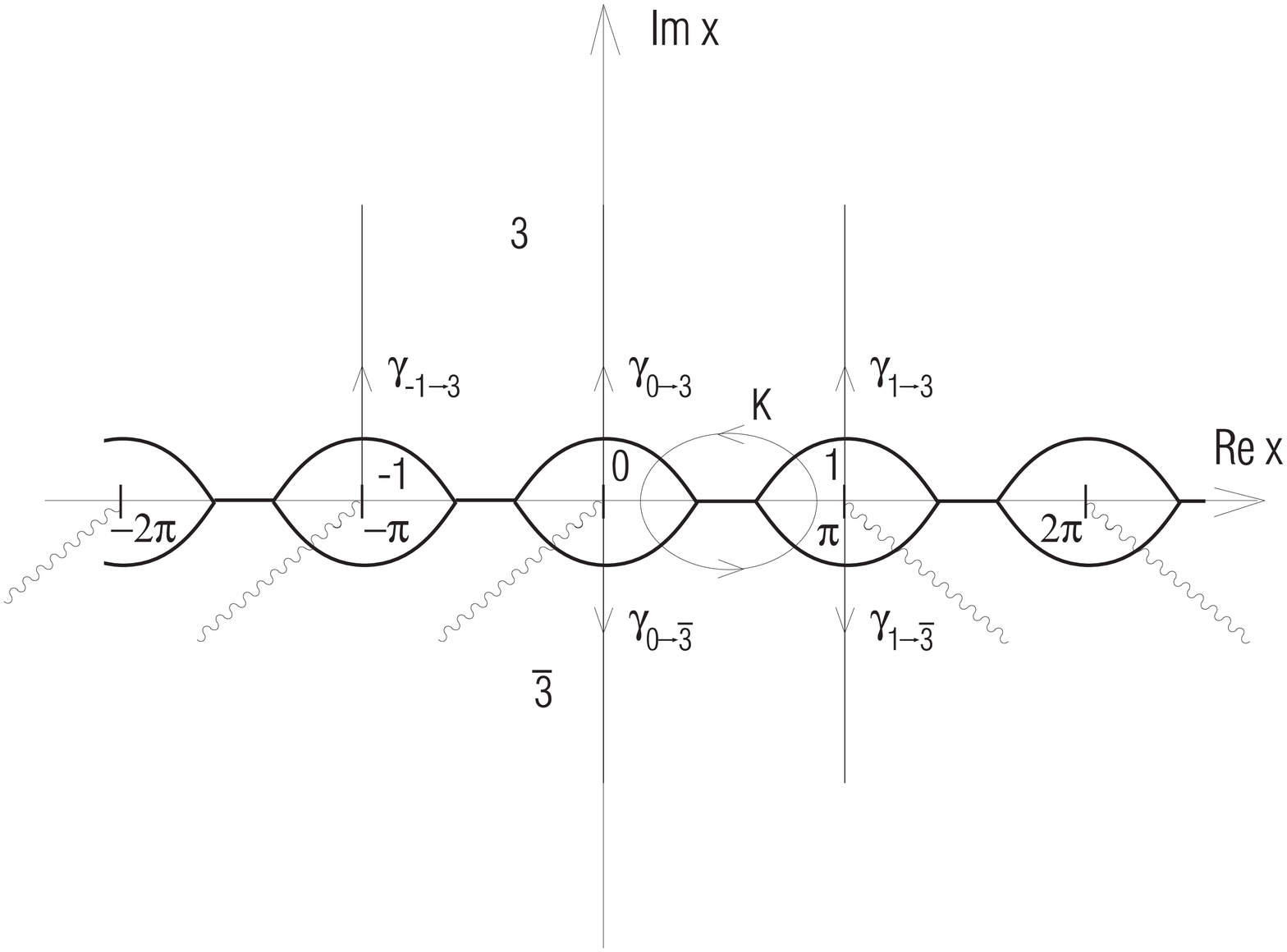,width=11cm} \\
Fig.13  The SG corresponding to the exact JWKB formula \mref{w3.33}
\end{tabular}

\underline{       case d.}

Examples of the case are provided by elliptic functions
\cite{19}. The simplest candidates are
the elliptic functions of the second order i.e. containing two
simple poles or a pole of the second order in the corresponding
basic parallelograms. In this case there are also two simple
roots or a double one in each such a parallelogram of periods.
However, in general, for a given elliptic function a number of
its roots in each parallelogram of periods is always equal to
the order of an elliptic function. Therefore, it is rather
hopeless to expect an elliptic functions with its order higher
than two to be a good candidate of a potential to produce the
corresponding exact JWKB quantization formulae. Considering,
however, $q(x,E,\lambda)$ to be a second order elliptic function
we have to assume that the two of its zeros are its real turning
points $x_\pm$\ placed between $0$ and $2\pi$\ (the latter being
the real period of $q(x,E,\lambda)$), i.e $0<x_-<x_+<2\pi$,
whilst one of its poles is at $x=0$. According to that we have the
following two possibilities:
\begin{description}
\item[a.] there are two real simple poles of $q(x,E,\lambda)$: one 
at $x_0=0$\ and the second at $x_1$,
$x_+<x_1<2\pi$;
\item[b.] there is one double pole of $q(x,E,\lambda)$\ (at $ x$=0).
\end{description}

Of course, in each of the above cases $q(x,E,\lambda)$\ has to
be completed to $\tilde{q}(x,E,\lambda)$\ by the corresponding
$\delta$--function. The latter, however, has to be a sum of the
second order elliptic functions with {\it double} poles at each
singular point of $q(x,E,\lambda)$. Therefore $\delta$\ can be
represented by a linear combination of two corresponding
Weierstrass elliptic functions (case a) or should be
proportional to such a function (case b). The periods of the
Weierstrass functions coincide in each case with those of
$q(x,E,\lambda)$. Therefore the SG's corresponding to the two
cases have to look as in Fig.14a,b respectively and the
quantization conditions which can be prescribed to each of the
graphs are the following:
\be
exp\ll[-\lambda\oint\limits_K\ll[F(x;2\pi,i\omega)+\frac{1}{4\lambda^2}
\wp(x;2\pi,i\omega)+\frac{1}{4\lambda^2}
\wp(x-x_1,2\pi,i\omega)-E\r]^\fr dx\r]= \nn \\
-\frac{\chi_{0\to2}(\lambda,E)\chi_{1\to\bar{2}}(\lambda,E)}
{\chi_{0\to\bar{2}}(\lambda,E)\chi_{1\to2}(\lambda,E)}\label{w3.34}\\
exp\ll[-\lambda\oint\limits_K\ll[(\alpha+\frac{1}{4\lambda^2})
\wp(x;2\pi,i\omega)-E\r]^\fr dx\r]=\nn\\
=-\frac{\chi_{0\to2}(\lambda,E)\chi_{1\to\bar{3}}(\lambda,E)}
{\chi_{0\to\bar{3}}(\lambda,E)\chi_{1\to2}(\lambda,E)}
\label{w3.35}
\ee

\begin{tabular}{cc}
\psfig{figure=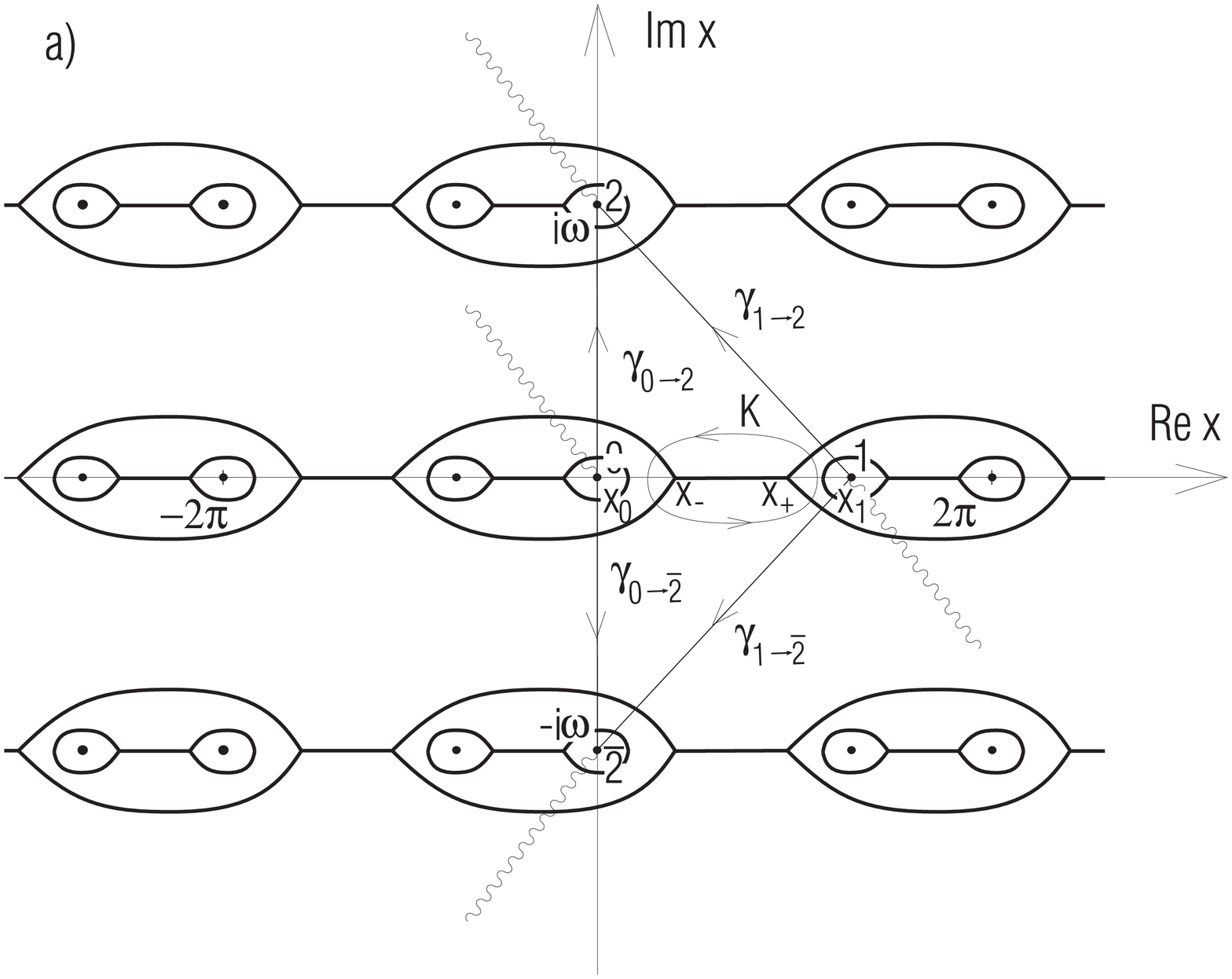,width=7.5cm} & \psfig{figure=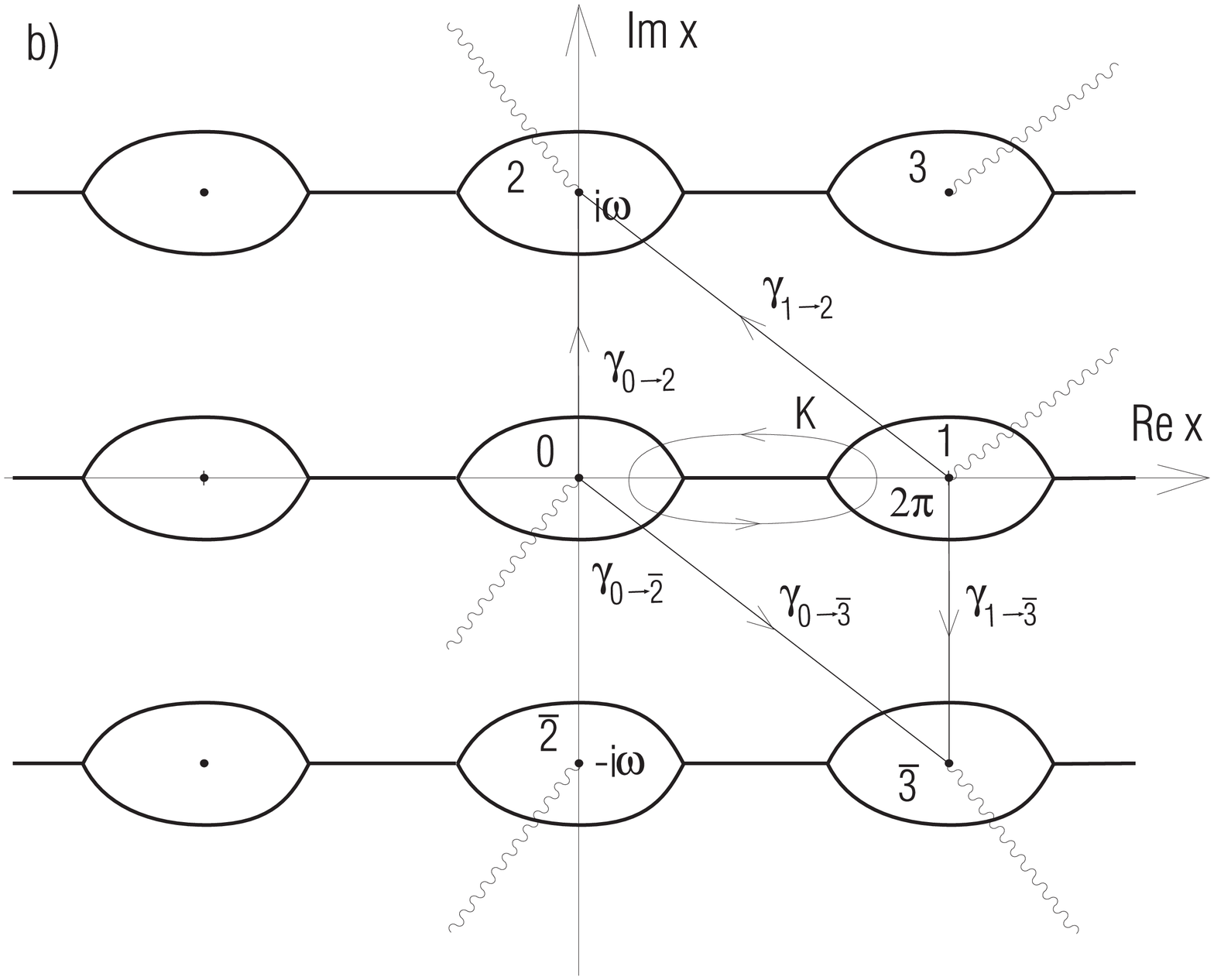,width=7.5cm} \\
\parbox{7cm}{Fig.14  The SG's corresponding to the elliptic function 
$F(x;2\pi,i\omega)$\ (Fig.14a)  and Weierstrass one $\wp(x;2\pi,i\omega)$
(Fig.14b) taken as potentials} &
\end{tabular}

Here $F(x;2\pi,i\omega)$\ is an elliptic function with two
simple poles at $x_0$\ and at $x_1$, whilst
$\wp(x;2\pi,i\omega)$\ is the Weierstrass elliptic function.

It follows from Fig.14a that in the first of the above
quantization formulae the periodicity arguments cannot work
($x_1$\ is not a shift of $x_0$\ by $2\pi$) nor in the form used
in \mref{w3.34} nor in any other of its mutations. 

In the condition \mref{w3.35} one can use periodicity and
reality arguments to show only that
$\chi_{1\to2}\chi_{0\to3}=
\chi_{0\to2}\chi_{1\to\bar{3}}exp(-2\delta_{0\to\bar{3}})$
(where $\delta_{0\to\bar{3}}$\ is the phase of
$\chi_{0\to\bar{3}}$) so that also in this case the RHS of this
condition is {\it not} -1 i.e. the corresponding JWKB formula is
{\it not} exact.

\section{ More general exactly JWKB quantized potentials\label{s4}}
\zero

\hskip+2em The periodic potentials considered in the previous section are
the simpliest ones of the potentials quantized exactly by the
JWKB formula. A generalization of their forms to the ones which
still can keep the exactness of the corresponding JWKB formula
can be done in the following way. 

Let $V(x)$ means any exactly JWKB quantized periodic potential of
the previous section.  Let $V(x,p)$ means a real parameter family
of periodic (with respect to $x$) potentials with the property
that in the limit $p\to0$, $V(x,p)$ approaches smoothly $V(x)$.
Then, for $p$ small enough the Stokes graph corresponding to
$V(x,p)$ has to resemble the Stokes graph corresponding to $V(x)$.
By such a resembling we mean the following:
\begin{enumerate}
\item To any singular point of SG of $V(x)$ there correspond a set 
of singular points of SG
of $V(x,p)$ which reduce to the former point in the limit $p\to0$.
Each such a set we shall call a singular point blob (SPB);
\item To any turning point of SG of $V(x)$ there correspond a set 
of turning points of $V(x,p)$
which reduce to the former point in the limit $p\to0$. We shall
call such a set a turning point blob (TPB);
\item There is one to one correspondence between the sectors of 
the two Stokes graphs such
that the sectors of SG of $V(x,p)$ reduce smoothly to the
corresponding sectors of SG of $V(x)$ when $p\to0$. In particular
the boundary conditions are formulated for both the potentials
$V(x,p)$ and $V(x)$ in sectors satisfying the correspondence just
describe;
\item Each set of Stokes lines which emerge from some SPB (TPB) can 
be mapped into a
definite set of Stokes lines of SG corresponding to $V(x)$
emerging from the point to which this SPB (TPB) reduces in the
limit $p\to0$. Each such a set can be divided into disjoint
subsets of SL's each of them transforming smoothly when $p\to0$\
into one particular SL emerging from the limiting point.
\end{enumerate}

Examples of $V(x,p)$ with the properties $1-4$ above with the
Planck constant $\hbar$\ as the parameter $p$ can be found in
\cite{1}. With these properties $V(x,p)$ provides us with the JWKB
formula quantizing exactly the energy levels of $V(x,p)$.

\section{ Some supersymmetric JWKB formula exactness\label{s5}}
\zero

\hskip+2em In connection with the supersymmetric formulation of quantum
mechanics the supersymmetric (SUSY) JWKB approximations have
been suggested some of which being different from the
conventional ones have appeared to be exact \cite{7,8}. 
It has been also noticed
however that their exactness have been parallel to the exactness
of the conventional ones
\cite{7,8,11}.

In the previous sections we have shown that the exactness of the
conventional (i.e. not SUSY) JWKB formulae was rather
exceptional and related to the periodicity properties of the
corresponding SG's. Since the SUSY QM quantization problems seem
to be governed by the same rules we can expect that the
exactness of the SUSY JWKB formulae have to follow in some way
from the traditional ones. We are going to show below that
indeed this is the case.

Let us note, however, that there is also a common conviction that
the SUSY JWKB exact quantization conditions are not only
independent of the conventional ones but also their exactness in
some cases of potentials is to be in contrast with the
approximate character in these cases of the conventional JWKB
formulae. As such potentials are considered the ones with the
shape invariance property \cite{30}. We would like to argue below that
also in these cases the parallelness of the exactness of both
the kind of the formulae seems to be still maintained.

Leaving the investigation of the latter relation for the later
discussion let us examine first the question how the SUSY JWKB
exact formulae follow from the conventional ones.
     
For this goal let us remind that if a potential $V(x,\lambda)$\
can be put in its SUSY form $V(x,\lambda)\equiv
V_-(x\lambda)=\phi^2(x,\lambda)-\phi'(x,\lambda)/\lambda+\epsilon_0$\
($\epsilon_0$\ is the energy of the fundamental level in $V(x)$ if
SUSY is exact) then the conventional JWKB quantization
condition:
\be\label{w5.1}
&-\lambda\oint\limits_K\sqrt{V(x,\lambda)+\frac{\delta(x,\lambda)}
{\lambda^2}-E}dx=(2m+1)\pi i&\\ &m=0,1,2,\ldots&\nn
\ee
for the exact SUSY is to be substituted by \cite{7,8}:
\be\label{w5.2}
&-\lambda\oint\limits_K\sqrt{\phi^2(x,\lambda)-(E-\epsilon_0)}dx=2\pi
im&\\ &m=0,1,2,\ldots&\nn
\ee   
If \mref{w5.1} is exact then as we have mentioned \mref{w5.2} is
very frequently also.

Let us analyze how this can happen. We shall perform our
analysis also for the cases of broken superpotentials $\phi$\
which can represent $V(x,\lambda)$. We shall show that in these
cases the condition \mref{w5.2} remains unchanged if it is exact
what is in contrast with its form representing the lowest JWKB
approximation only in which case its RHS coincides rather with
\mref{w5.1} \cite{27}.

At the beginning, let us note that because $\lambda$\ can vary
we can take it sufficiently large to expand the integrand in
\mref{w5.2} into a series with respect to $\phi'-\delta/\lambda$. We get:
\be\label{w5.3}
-\lambda\oint\limits_K\sqrt{\phi^2-\tilde{E}}dx-\sum\limits_{n\geq1}
\frac{1}{\lambda^{n-1}}\frac{\Gamma\ll[n-\fr\r]}{n!\Gamma\ll[-\fr\r]}
\oint\limits_K\frac{\ll[\phi'-\frac{\delta}{\lambda}\r]^n}
{(\phi^2-\tilde{E})^{n-\fr}}dx=(2m+1)\pi i
\ee
where $E=E-\epsilon_0$.

Making further a change of variable: $x\to\phi=\phi(x,\lambda)$\
in the integrands of the series in \mref{w5.3} and putting
$F_1(\phi,\lambda)\equiv \phi'(x(\phi,\lambda),\lambda) -
\delta(x(\phi,\lambda),\lambda)/\lambda$\ and $F_2(\phi,\lambda) \equiv 
\phi'(x(\phi,\lambda),\lambda)$\ we obtain:
\be\label{w5.4}
-\lambda\oint\limits_K\sqrt{\phi^2-\tilde{E}}dx-\sum\limits_{n\geq}\frac{1}
{\lambda^{n-1}}\frac{\Gamma\ll[n-\fr\r]}{n!\Gamma\ll[-\fr\r]}
\oint\limits_{K_\phi}\frac{F_1^n(\phi,\lambda)}{(\phi^2-\tilde{E})^{n-
\fr}}\frac{d\phi}{F_2(\phi,\lambda)}=(2m+1)\pi
\ee
where the integrations under the sum in \mref{w5.4} go now in
the $\phi$--plane.
       
Next let us observe that for all the exactly JWKB quantized
potentials considered above $F_{1,2}(\phi,\lambda)$\ are
holomorphic functions of $\phi$\ outside some circles of a
sufficiently large radius so that the circles contain the branch
points at $\phi=-\sqrt{E}$\ and $\phi=+\sqrt{E}$\ of the
integrand denominators in \mref{w5.4}. Moreover, both the functions 
$F_{1,2}$\ grow with the same powers of $\phi$\ but not faster than the
second ones. It follows then from \mref{w5.4} that all the
integrands of the series in \mref{w5.4} are also holomorphic
outside such circles. The integrals can all be easily calculated
then by taking the size of the contour $K$ large enough and
expanding their denominators:
\be\label{w5.5}
-\lambda\oint\limits_K\sqrt{\phi^2-\tilde{E}}dx-\sum\limits_{n\geq1}
\frac{1}{\lambda^{n-1}}\frac{\Gamma\ll[n-\fr\r]}{n!\Gamma\ll[-\fr\r]}
\times \nn\\
\sum\limits_{k\geq0}\tilde{E}^k\frac{\Gamma\ll[k+n-\fr\r]}{k!\Gamma
\ll[n-\fr\r]}\oint\limits_{K_\phi}\frac{F_1^n(\phi,\lambda)}{\phi^{2n+2k-1}}
\frac{d\phi}{F_2(\phi,\lambda)}=(2m+1)\pi i
\ee

A final result of the integrations in \mref{w5.5} depends now of
course on the particular forms of the expansions of
$F_{1,2}(\phi,\lambda)$\ into their corresponding Laurent
series.

It can be easily checked that the series in LHS of \mref{w5.5}
becomes energy independent only in the case when the Laurent
series expansions of $F_{1,2}$\ both abbreviate at least on the
second power of $\phi$. This is just the case of the potentials
considered.

Suppose therefore that
$F_{1,2}(\phi,\lambda)=\sum_{k\geq1}F_{1,2;k}(\lambda)\phi^{-k}
+ a_{1,2}(\lambda) + b_{1,2}(\lambda)\phi + c_{1,2}(\lambda)\phi^2$. 
Then from \mref{w5.5} we get:
\be\label{w5.6}
-\lambda\oint\limits_K\sqrt{\phi^2-\tilde{E}}dx+\pi
i\delta_{b,0}\delta_{c,0}+\pi i\delta_{c,0} - 2\pi
i\frac{\lambda}{c_2}\ll[\sqrt{1-\frac{c_1}{\lambda}}-1\r]=(2m+1)\pi i
\ee

Below we shall do an inspection of the JWKB-quantization exact
formulae of the previous section to calculate the corresponding
coefficients $a_{1,2},\; b_{1,2}$\ and $c_{1,2}$\ as well as to
show that all these formulae allow the quantization form
\mref{w5.2}\ {\it independently}\ of whether the supersymmetry
represented by $\phi$\ is exact or broken. This result, however,
does not contradict the one obtained recently by Inomata
\underline{et al} \cite{27} (see also \cite{29} for further 
references) who have modified the
Comtet \underline{et al} formula \mref{w5.2} with the aim to
cover also the cases when supersymmetric potentials $\phi$\
represent broken supersymmetry. They have argued that in such
cases the RHS of \mref{w5.2} had to be transformed again into
the conventional form of the RHS of \mref{w5.1}. The source of
the difference between both the conclusions is that our concerns
the exact result whilst this of Inomata \underline{et al} is
only the lowest semiclassical approximation of the {\it exact}
quantization condition.  Nevertheless, as we shall see further
that \mref{w5.2} having the same form gives, however, {\it
different} results for energy levels depending on whether the
supersymmetry is exact or broken in the latter case reproducing
effectively the result of Inomata \underline{et al}.

To follow further let us recapitulate all the potentials
$V_k(x)$\ and the corresponding
$\tilde{q}_k(x,E,\lambda)$--functions we have found in the
previous section to be quantized exactly by the corresponding
JWKB-formulae. They are:

\be
\tilde{q}_1(x,E,\lambda)=V_1(x)-E=\alpha^2e^{2x}-2\beta e^x-E,\nn\\
-\infty<x<+\infty,\; \beta>0>E \nn\\
\nn\\
\tilde{q}_2(x,E,\lambda)=V_2(x)+\frac{1}{4\lambda^2x^2}-E=-\frac{\alpha}{x}+
\frac{\beta+\frac{1}{4\lambda^2}}{x^2}-E,\nn\\
x,\alpha,\beta>0>E\nn\\
\nn\\
\tilde{q}_3(x,E,\lambda)=V_3(x)+\frac{1}{4\lambda^2x^2}-E=
\alpha^2x^2+ \frac{\beta+\frac{1}{4\lambda^2}}{x^2}-E ,\nn\\
x,\beta,E>0\nn\\
\nn\\
\tilde{q}_4(x,E,\lambda)=V_4(x)-\frac{1}{(4\lambda\cosh\frac{x}{2})^2}-E
=\nn\\
\frac{\alpha e^x-\beta-\frac{1}{16\lambda^2}}{\cosh^2\frac{x}{2}}-E, \nn\\
-\infty<x<+\infty,\;\beta>0,\; -\beta<2\alpha\nn\\
\nn\\
\tilde{q}_5(x,E,\lambda)=V_5(x)+ \frac{1}{(4\lambda\sinh\frac{x}{2})^2}
-E=\nn\\
\frac{\alpha e^x+\beta+\frac{1}{16\lambda^2}}{\sinh^2\frac{x}{2}}-E, \nn\\
0<x<+\infty,\; \beta,\alpha+\beta>0>2\alpha+\beta>\alpha, \nn\\
\nn\\
\tilde{q}_6(x,E,\lambda)=V_6(x)+\frac{1}{(4\lambda\sinh\frac{x}{2})^2} -
\frac{1}{(4\lambda\cosh\frac{x}{2})^2} -E=\nn\\
\frac{\beta+\frac{1}{16\lambda^2}}{\sinh^2\frac{x}{2}}-
\frac{\alpha +\frac{1}{16\lambda^2}}{\cosh^2\frac{x}{2}}-E, \nn\\
0<x<+\infty, \alpha,\beta>0, \nn\\
\nn\\
\tilde{q}_7(x,E,\lambda)=V_7(x)+\frac{1}{(4\lambda\cos\frac{x}{2})^2}-E=
\frac{\alpha +\frac{1}{16\lambda^2}}{\cos^2\frac{x}{2}}-E,\nn\\
-\pi<x<\pi,\; \alpha>0\nn\\
\nn\\
\tilde{q}_8(x,E,\lambda)=V_8(x)+\frac{1}{(4\lambda\cos\frac{x}{2})^2} +
\frac{1}{(4\lambda\sin\frac{x}{2})^2} -E=\nn\\
\frac{\alpha +\frac{1}{16\lambda^2}}{\cos^2\frac{x}{2}} +
\frac{\beta +\frac{1}{16\lambda^2}}{\sin^2\frac{x}{2}}-E,\nn\\
0<x<\pi,\; \alpha,\beta>0\nn
\ee

In order to represent the above potentials by their
supersymmetric ones one has in principle to solve non uniform
Riccati equations with their RHS given by the potentials listed.
In general such a task is rather difficult. For most of the
above potentials, however, it is possible to find these
representations just by a trivial guess. To each of the
potentials listed above one can guess several (at least two)
solutions one of which correspond to a superpotential realizing
the supersymmetry exactly and the remaining ones corresponding
to a broken supersymmetry.  The latter means that the
supersymmetry breaking can be realized in many ways. The ways
considered below take into account only the possibility to
define by a superpotential $\phi$\ the corresponding ground
state solution $\Psi_0$\ by the following representation:
\be\label{w5.7}
\Psi_0(x)=exp\ll[-\lambda\int\limits^x\phi(y)dy\r]\\
a<x<b\nn
\ee
where $a,b\; (a<b)$\ define boundaries of the corresponding
quantization problem. Note that $\Psi_0$\ as given by
\mref{w5.7} satisfies the SE \mref{w2.1} for $E=\epsilon_0$\ 
with the potentials $V(x,\lambda)(\equiv V_-(x,\lambda)$) listed above.
There are four possibilities:
\begin{description}
\item[$1^0$] $\Psi_0$\ vanishes at both the boundaries $a,b$ - 
the supersymmetry is exact and $\Psi_0$\ is the
ground state wave function;
\item[$2^0$]  and $3^0$\ $\Psi_0$\ vanishes at one of the 
boundaries only ($a$ or $b$ respectively) - the
supersymmetry {\it has} to be broken; and
\item[$4^0$] $\Psi_0$\ blows up at both the boundaries - the 
supersymmetry seems essentially to be
broken but there is still possibility that the ground state
$\Psi_0$\ has been constructed by the erroneous choice of
$\phi$\ --- there are infinitely many solutions satisfying the
SE considered with $E=\epsilon_0$\ but blowing up at both the
boundaries even if the corresponding ground state exists with
this energy.
\end{description}

The latter possibility cannot happen in the cases $2^0$\ and
$3^0$: blowing up of $\Psi_0$\ at one of the boundaries only
means that the ground state with $E=\epsilon_0$\ cannot exists
in these cases. One can expect therefore that resulting
relations between the energy spectra provided by the
quantization conditions defined by the allowed superpotentials
$\phi_k$, corresponding to each of the potentials $V_k,\;
k=1,\ldots,8$, listed earlier, and the original spectra of the
latter potentials can depend on the way the supersymmetry is
broken by each particular $\phi_k$.

Below we have enumerated all the allowed superpotentials
$\phi_k$\ corresponding to each of the potentials $V_k,\;
k=1,\ldots,8$, with the properties $1^0-4^0$\ just discussed
(attaching to each of them the corresponding category) together
with their $F_{1,2}$--functions:

\hskip-3em $1^0\;\;$
$$\phi_1(x,\lambda)=|\alpha|e^x-\frac{\beta}{|\alpha|}
+\frac{1}{2\lambda},$$
$$  F_1(\phi_1)=F_2(\phi_1)=\phi_1+\frac{\beta}{|\alpha|}
+\frac{1}{2\lambda}, $$
$$ b_1=1;\;\;\epsilon_0=-(\frac{\beta}{|\alpha|}
-\frac{1}{2\lambda})^2;\\$$
\vskip 12pt

\hskip-3em $4^0\;\;$
$$\phi_1(x,\lambda)=-|\alpha|e^x+\frac{\beta}{|\alpha|}
\frac{1}{2\lambda},$$
$$ F_1(\phi_1)=F_2(\phi_1)=\phi_1-\frac{\beta}{|\alpha|}
+\frac{1}{2\lambda}, $$
$$  b_1=1;\;\;\epsilon_0=-(\frac{\beta}{|\alpha|}
+\frac{1}{2\lambda})^2;\\$$
\vskip 12pt

\hskip-3em $1^0\;\;$
 $$  \phi_2(x,\lambda)=-\frac{|2l+1|+1}{2\lambda
x} +\frac{ \lambda\alpha}{|2l+1|+1}, $$
$$ F_1(\phi_2)=\lambda(2|2l+1|+1)\frac{(\phi_2-\frac{\alpha\lambda}
{|2l+1|+1})^2}{(|2l+1|+1)^2}, $$
$$ F_2(\phi_2)=2\lambda\frac{(\phi_2-\frac{\alpha\lambda}
{|2l+1|+1})^2}{|2l+1|+1}, $$
$$c_1=\lambda\frac{2|2l+1|+1}{(|2l+1|+1)^2},\;\; 
c_2=\frac{2\lambda}{|2l+1|+1};\;\; $$
$$ \epsilon_0=-\frac{(\lambda\alpha)^2}{(|2l+1|+1)^2},$$
$$\beta=\frac{l(l+1)}{\lambda^2},\;\;\;\;\;\; l<-1, \;\; l>0; \\$$
\vskip 12pt

\hskip-3em $4^0\;\;$
$$ \phi_2(x,\lambda)=\frac{|2l+1|-1}{2\lambda x} -
\frac{\lambda\alpha}{|2l+1|-1},$$
$$ F_1(\phi_2)=-\lambda(2|2l+1|-1)\frac{(\phi_2+\frac{\alpha\lambda}
{|2l+1|-1})^2}{(|2l+1|-1)^2}, $$
$$    F_2(\phi_2)=-2\lambda\frac{(\phi_2+\frac{\alpha\lambda}
{|2l+1|-1})^2}{|2l+1|-1}, $$
$$    c_1=-\lambda\frac{2|2l+1|-1}{(|2l+1|-1)^2},\;\; 
c_2=\frac{2\lambda}{|2l+1|-1};\;\; $$
$$   \epsilon_0=-\frac{(\lambda\alpha)^2}{(|2l+1|-1)^2},$$
$$    \beta=\frac{l(l+1)}{\lambda^2}, \;\;\;\;\;\; l<-1, \;\; l>0; \\$$

\hskip-3em $1^0\;\;$
$$\phi_3(x,\lambda)=|\alpha|x-\frac{|2l+1|+1}{2\lambda x}, $$
$$  F_1(\phi_3)=\lambda(2|2l+1|+1)\left[\phi_3^2 +2|\alpha|
\frac{|2l+1|+1}{\lambda} \right]^\frac{1}{2}
\frac{\phi_3+(\phi_3^2+2|\alpha|\frac{|2l+1|+1}{\lambda})^\fr}
{2(|2l+1|+1)^2} + \frac{|\alpha|}{2|2l+1|+2}, $$
$$ F_2(\phi_3)=\lambda\left[ \phi_3^2 
+2|\alpha|\frac{|2l+1|+1}{\lambda} 
\right]^\fr \frac{\phi_3+(\phi_3^2+2|\alpha|\frac{|2l+1|+1}{\lambda})^\fr}
{|2l+1|+1} , $$
$$ c_1^0=\lambda\frac{2|2l+1|+1}{(|2l+1|+1)^2},\;\; 
c_2^0=\frac{2\lambda}{|2l+1|+1};\;\;$$
$$  \epsilon_0=(|2l+1|+2)\frac{|\alpha|}{\lambda},\;\; \;\;
\beta=\frac{l(l+1)}{\lambda^2}, \;\;\;\;\; l<-1,\;\;\; l>0;\\ $$

\hskip-3em $2^0$\ \ we get the case from $1^0$\ substituting there $|2l+1|$
by $-|2l+1|$;

\hskip-3em $3^0$\ \ we get the case from $1^0$\ by the substitution 
$|\alpha| \to -|\alpha|$;

\hskip-3em $4^0$\ \  we get the case from $1^0$\ substituting there $|2l+1|$
by $-|2l+1|$\ and $|\alpha|$\ by $-|\alpha|$;
\vskip 12pt
\hskip-3em $1^0\;\;$
$$  \phi_4=\frac{|2l+1|-1}{4\lambda}\tanh\frac{x}{2}+
\frac{4\lambda\alpha}{|2l+1|-1},$$
$$    F_1(\phi_4)=-\lambda(2|2l+1|-1)\ll[{\frac{\phi_4-\frac{4\lambda\alpha}
{|2l+1|-1})^2}{|2l+1|-1}}\right]^{\fr} +\frac{2|2l+1|-1}{16\lambda},$$
$$    F_2(\phi_4)=-2\lambda\frac{(\phi_4-\frac{4\lambda\alpha}
{|2l+1|-1})^2}{|2l+1|-1}+\frac{|2l+1|-1}{8\lambda}, $$
$$    c_1=-\lambda\frac{2|2l+1|-1}{(|2l+1|-1)^2},\;\; 
c_2=-\frac{2\lambda}{|2l+1|-1};\;\; $$
$$ \epsilon_0=-\left[\frac{|2l+1|-1}{2\lambda}-\frac{2\lambda\alpha}
{|2l+1|-1} \right]^2; $$
$$   \alpha+\beta=\frac{l(l+1)}{(2\lambda)^2},\;\;\;\;\; l<-1,\;\;\l; l>0; $$

\hskip-3em $4^0$\ we get the case from $1^0$\ by the substitution 
$|2l+1|\to-|2l+1|$\
\vskip 12pt

\hskip-3em $1^0\;\;$
$$ \phi_5=-\frac{|2l+1|+1}{4\lambda}\coth\frac{x}{2}-
\frac{4\lambda\alpha}{|2l+1|+1},$$
$$    F_1(\phi_5)=\lambda(2|2l+1|+1)\frac{\phi_5+\frac{(4\lambda\alpha}
{|2l+1|+1})^2}{(|2l+1|+1)^2}-\frac{2|2l+1|+1}{16\lambda}, $$
$$    F_2(\phi_5)=+2\lambda\frac{(\phi_5+\frac{4\lambda\alpha}
{|2l+1|+1})^2}{|2l+1|+1}-\frac{|2l+1|+1}{8\lambda}, $$
$$    c_1=\lambda\frac{2|2l+1|+1}{(|2l+1|+1)^2},\;\; 
c_2=\frac{2\lambda}{|2l+1|+1};\;\; $$
$$ a_1^{\infty} = a_2^{\infty} = |\alpha|; \;\;\;
\epsilon_0=-\left[-\frac{|2l+1|+1}{2\lambda}+\frac{2\lambda\alpha}
{|2l+1|+1} \right]^2; $$
$$   \alpha+\beta=\frac{l(l+1)}{(2\lambda)^2},\;\;\;\; l<-1,\;\;\; l>0; \\ $$

$\hskip-3em 4^0$ we get the case from $1^0$\ by the substitution 
$|2l+1|\to -|2l+1|$\

\hskip-3em $1^0\;\;$
$$ \phi_6=\frac{|2l+1|-1}{4\lambda}\tanh\frac{x}{2}-
\frac{|2l'+1|+1}{4\lambda}\coth\frac{x}{2} ,$$
$$    F_1(\phi_6)=-\frac{\lambda}{2}(2|2l+1|-1)\left[\phi_6^2+(2|2l+1|-1)
\frac{|2l'+1|+1}{(2\lambda)^2}\right]^\fr 
\frac{\phi_6+ \left[\phi_6^2+(2|2l+1|-1)
\frac{|2l'+1|+1}{(2\lambda)^2}\right]^\fr}{(|2l+1|-1)^2}+ $$
$$ -\frac{\lambda}{2}(2|2l'+1|-1)\left[\phi_6^2+(2|2l+1|-1)
\frac{|2l'+1|+1}{(2\lambda)^2}\right]^\fr $$
$$\frac{\phi_6- \left[\phi_6^2+(|2l+1|-1)
\frac{|2l'+1|+1}{(2\lambda)^2}\right]^\fr }{(|2l'+1|+1)^2}, $$
$$    F_2(\phi_6)=-\lambda\left[\phi_6^2+(|2l+1|-1)\frac{|2l'+1|+1}
{(2\lambda)^2}\right]^\fr \frac{\phi_6+ \left[\phi_6^2+(|2l+1|-1)
\frac{|2l'+1|+1}{(2\lambda)^2}\right]^\fr}{(|2l+1|-1)^2}+ $$
$$ -\lambda\left[\phi_6^2+(|2l+1|-1)\frac{|2l'+1|+1}{(2\lambda)^2}
\right]^\fr \frac{\phi_6- \left[\phi_6^2+(|2l+1|-1)
\frac{|2l'+1|+1}{(2\lambda)^2}\right]^\fr }{(|2l'+1|+1)^2}, $$
$$    c_1^0 =\lambda\frac{2|2l'+1|+1}{(|2l'+1|+1)^2},\;\;\;\; 
c_1^{\infty}=\lambda\frac{2|2l+1|-1}{(|2l+1|-1)^2};\;\; $$
$$    c_2^0 =\frac{2\lambda}{|2l'+1|+1},\;\;\;\; 
c_2^{\infty}=-\frac{2\lambda}{|2l+1|-1};\;\; $$
$$ \epsilon_0=\frac{(l-l' -1)^2}{(2\lambda)^2}, \;\;\; \;\; 
\alpha=\frac{l(l+1)}
{(2\lambda)^2}, \;\;\;\;\; \beta=\frac{l'(l'+1)}{(2\lambda)^2}, \;\; $$
$$ |2l+1|-|2l'+1|>2, \;\;\;\;\; l,l'<-1, \;\;\;\;\; l,l' >0; $$

\hskip-3em $2^0$\ we get the case from $1^0$\ taking $l,l'$\ satisfying 
$|2l+1|-|2l'+1|<2$\ or substituting $|2l+1|$\ by $-|2l+1|$\
there;

\hskip-3em $3^0$\ we get the case from $1^0$\ substituting there $|2l'+1|$ 
by $-|2l'+1|$\ and next taking $l,l'$\ satisfying
$|2l'+1|\pm|2l+1|>2$;

\hskip-3em $4^0$\  we get the case from $1^0$\ substituting there $|2l'+1|$
by $-|2l'+1|$\ and next taking $l,l'$\ satisfying
$|2l'+1|\pm|2l+1|<2$;

\hskip-3em $1^0\;\;$ 
$$ \phi_7=-\frac{|2l+1|-1}{4\lambda}\tan\frac{x}{2},$$
$$    F_1(\phi_7)=-\lambda(2|2l+1|-1)\frac{\phi_7^2}{(|2l+1|-1)^2}-
\frac{2|2l+1|-1}{16\lambda},  $$
$$    F_2(\phi_7)=-2\lambda\frac{\phi_7^2}{|2l+1|-1}-\frac{|2l+1|-1}
{8\lambda}, $$
$$    c_1=-\lambda\frac{2|2l+1|-1}{(|2l+1|-1)^2},\;\;\;\; 
 c_2=-\frac{2\lambda}{|2l+1|-1} \;\; $$
$$ \epsilon_0=\left(\frac{|2l+1|-1}{4\lambda} \right)^2 $$
$$   \alpha=\frac{l(l+1)}{(2\lambda)^2},\;\;\;\;\;\; l<-1, \;\;\;\; l>0; $$

\hskip-3em $4^0$\ we get the case from $1^0$\ substituting there $|2l+1|$\ 
by $-|2l+1|$;

$$\hskip-3em 1^0\;\;$$ $$ \phi_8=\frac{|2l+1|+1}{4\lambda}\tan\frac{x}{2}-
\frac{|2l'+1|-1}{4\lambda}\cot\frac{x}{2} , $$
$$    F_1(\phi_8)=\frac{\lambda}{2}(2|2l+1|+1)\left[\phi_8^2+(|2l'+1|-1)
\frac{|2l+1|+1}{(4\lambda)^2}\right]^\fr 
\frac{\phi_8+ \left[\phi_8^2+(|2l'+1|-1)
\frac{|2l+1|+1}{(4\lambda)^2}\right]^\fr }{(|2l+1|+1)^2}+ $$
$$ -\frac{\lambda}{2}(2|2l'+1|-3)\left[\phi_8^2 
+\frac{l'(l+1)}{\lambda^2}\right]^\fr 
\frac{\phi_8- \left[\phi_8^2+(|2l'+1|-1)
\frac{|2l+1|+1}{(4\lambda)^2}\right]^\fr }{(|2l'+1|-1)^2},$$
$$  F_2(\phi_8)=\lambda\left[\phi_8^2+(|2l'+1|-1)\frac{|2l+1|+1}
{(4\lambda)^2}\right]^\fr \frac{\phi_8+
\left[\phi_8^2+(|2l'+1|-1)
\frac{|2l+1|+1}{(4\lambda)^2}\right]^\fr }{|2l+1|+1}+ $$
$$ -\lambda\left[\phi_8^2+(2|l' +1|-1)\frac{|2l+1|+1}{(4\lambda)^2}
\right]^\fr 
\frac{\phi_8- \left[\phi_8^2+(|2l' +1|-1)
\frac{|2l+1|+1}{(4\lambda)^2}\right]^\fr }{(|2l'+1|-1)^2},$$
$$    c_1^0 =\lambda\frac{4l' -1}{(2l')^2}, \;\;\;\; 
c_1^{\infty}=\lambda\frac{4l+3}{(2l+2)^2}, \;\;\;\; 
c_2^0=\frac{\lambda}{l'},\;\;\;\; c_2^{\infty}=\frac{\lambda}{l+1}, \;\;$$
$$ \epsilon_0=\frac{(l-l' +1)^2}{(2\lambda)^2}, \;\;\;\;\;\;\;\;
\alpha=\frac{l(l+1)}
{(2\lambda)^2}, \;\;\;\;\; \beta=\frac{l'(l' +1)}{(2\lambda)^2}, \;\;\;\;\;\;
l,l'<-1, \;\;\;\;\;\; l,l' >0; $$

\hskip-3em $2^0$\ we get the case from $1^0$\ substituting there $|2l+1|$\ 
by $-|2l+1|$;

\hskip-3em $3^0$\ we get the case from $1^0$\ substituting there $|2l'+1|$\ 
by $-|2l'+1|$;

\hskip-3em $4^0$\ we get the case from $1^0$\ substituting there $|2l+1|$\ 
by $-|2l+1|$\ as well as $|2l'+1|$\ by $-|2l'+1|$.

In the above calculations only the non vanishing coefficients
$a_{1,2},\; b_{1,2}$\ and $c_{1,2}$\ at the {\it highest} power
of $\phi$\ have been given. Now we can use them to calculate the
three pieces of the LHS of \mref{w5.6} and to convince ourselves
that in all the cases considered all the pieces contribute the
total value $\pi i\lambda$\ only so that \mref{w5.5} {\it
always} reduces to \mref{w5.2}. 

A comment to these calculations is necessary in the cases of the
$V_3,V_6$\ and $V_8$\ potentials.  The necessary integrations
that lead to \mref{w5.6} are performed here on the two sheeted
Riemann surfaces of the variable $\phi$\ on which the
corresponding functions $F_{1,2}(\phi)$\ have different
asymptotic properties for $\phi\to\infty$. The superscripts $'0'$
and $'\infty'$\ at the coefficients $a_{1,2},\; b_{1,2}$\ and
$c_{1,2}$\ indicate that the integrations contributed to them
have been performed on the two different sheets. In such cases
the coefficients with both supercripts contribute to \mref{w5.6}
but these contribution should be multiplied by $1/2$ each (see
Appendix 2 for an example of such calculations).

Let us finally note that compairing the energy levels obtained
by the formula \mref{w5.1} with those obtained by \mref{w5.2}
using in the latter the respective superpotentials of the cases
$1^0-4^0$\ we get the result that the energy levels given by
\mref{w5.1} are reproduced by \mref{w5.2}
\begin{description}
\item[i.] exactly in the case $1^0$\ of the superpotentials;
\item[ii.] being {\it shifted up} by a {\it half} of a unit 
used to enumerate the levels in the cases $2^0-3^0$\ of
the superpotentials;
\item[iii.] being {\it shifted up} by a {\it whole} unit used to enumerate 
the levels in the cases $4^0$\ of the superpotentials;
\end{description}

It is clear that the above differences follow as a result of the
different enumeration of energy levels in the compaired spectra
($m$ in \mref{w5.1} starts from zero whilst in \mref{w5.2} from
unity) as well as due to different choices of the energy levels
$\epsilon_0$ with respect to which the levels of the spectra are
measured in every of the cases $1^0-4^0$.

\subsection{SUSY and 
conventional JWKB quantization of shape invariant potentials \label{s5.1}}

\hskip+2em We have shown in the previous section that all the exactly JWKB
quantized cases of potentials are also quantized exactly by
their SJWKB quantization formulae. However, the latter property
of the considered potentials has been established also as being
closely related to their common property of being shape
invariant \cite{30}. The latter means that each
$V_k(x,\lambda)\equiv V_{-,k}(x,\lambda),\; k=1,\ldots,8$,
depends additionally on some parameter $a$ so that for its SUSY
partner $V_{+,k}(x,\lambda,a)$\ we have:
\be\label{w5.8}
&V_{k,+}(x,\lambda,a)=V_{,-k}(x,\lambda,a)+R_k(a_1)&\\
&k=1,\ldots,8&\nn
\ee
with $a_1=f_k(a)$. In the case of the considered potentials each
$f_k$\ is simply a translation of the parameter $a$.

The exactness of \mref{w5.2} following from \mref{w5.8} has been
suggested by Dutt \underline{et al} \cite{31} and established
by Barclay \underline{et al} \cite{32}. It was argued also (see
Cooper \underline{et al} and \cite{29}, for example) that the
exactness of SJWKB formulae \mref{w5.2} takes place even when the
exactness of the conventional ones fails.  The latter claim, however, 
needs not be necesserilly true and we would like to show below that in the
case of the translational shape invariance all the known cases
of the potentials are JWKB quantized exactly, too. To this aim
let us note that on the list of the eight of them cited above
there are two still lacking on the list when compaired with the
corresponding list of Cooper \underline{et al} \cite{29}. These
two are:

\be\label{w5.9}
V_9(x,\lambda)=\frac{\alpha+\beta\sin
x}{\cos^2x},\;\;-\frac{\pi}{2}<x<+\frac{\pi}{2},\;\alpha>\beta>0\nn\\
V_9^{min}=\fr(\sqrt{\alpha^2-\beta^2}+\alpha)\\
V_{10}(x,\lambda)=\frac{\alpha+\beta\sinh
x}{\cosh^2x},\;\;\;-\infty<x<+\infty,\;\;\beta>0\nn\\
V_{10}^{min}=-\fr(\sqrt{\alpha^2+\beta^2}-\alpha)\nn
\ee

One can easily convince oneself, however, that completed by the
'standard' $\delta$--terms ($(2\cos x)^{-2}$\ for the first
potential and $-(2\cosh x)^{-2}$\ for the second one) the
potentials are {\it exactly} JWKB quantized no doubt reflecting
the fact that this exactness follows in some although not
obvious way from the underlying periodicity of their SG's shown
in Fig.15.

\begin{tabular}{cc}
\psfig{figure=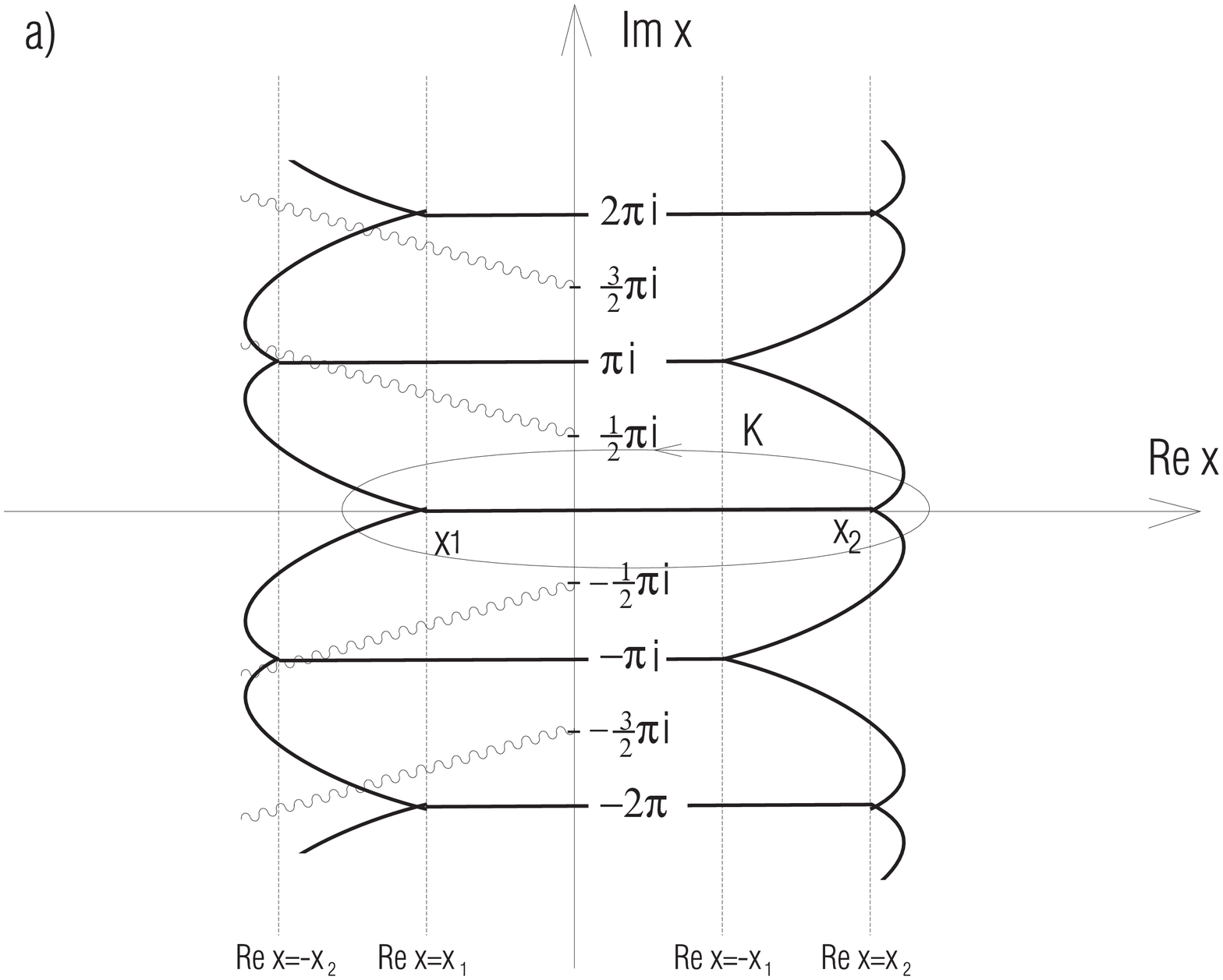,width=7.2cm} & \psfig{figure=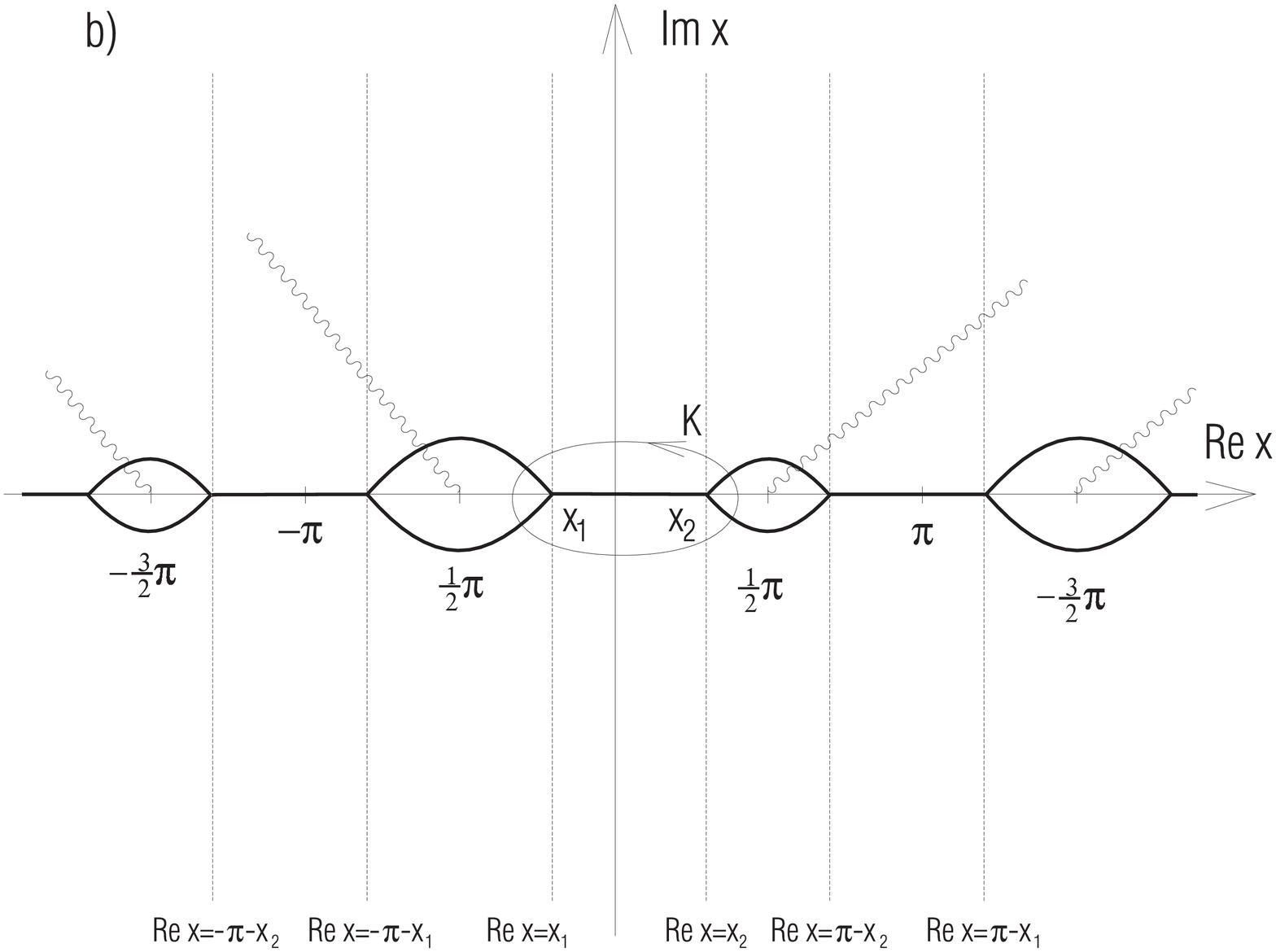,width=7.2cm} \\
\parbox{7cm}{Fig.15  The SG's corresponding to the potentials 
$V_9(x,\lambda)$ (Fig.15a) and $V_{10}(x,\lambda)$ (Fig.15b) given by the 
formulae \mref{w5.9}} &
\end{tabular}

This exactness can be, however, concluded also applying back the
procedure described in the previous section i.e. the exact JWKB
formulae for the considered potentials {\it follow} from the
exactness of the corresponding SJWKB ones. To see this let us
find the superpotentials $\phi_{9,10}$\ and their
$F_{1,2}$--functions corresponding to the potentials
\mref{w5.9}. They are:

\hskip-3em $1^0\;\;$ 
$$\phi_9=\frac{|2l+1|+|2l'+1|+2}{4\lambda}\tan x+
\frac{|2l+1|-|2l'+1|}{4\lambda\cos x}$$
$$F_1(\phi_9)=(\phi_9^2+a^2)\frac{(\phi_9^2+a^2)a'-ab^2+
b\phi_9\sqrt{\phi_9^2+a^2-b^2}}{(b^2+a^2)\phi_9^2+a^2(a^2-b^2)+2ab\phi_9
\sqrt{\phi_9^2+a^2-b^2}}$$
$$F_2(\phi_9)=(\phi_9^2+a^2)\frac{(\phi_9^2+a^2)a-ab^2+b\phi_9
\sqrt{\phi_9^2+a^2-b^2}}{(b^2+a^2)\phi^2_9+a^2(a^2-b^2)+2ab\phi_9
\sqrt{\phi_9^2+a^2-b^2}}$$
$$c_1^{\infty_1}=\frac{1}{a+b}-\frac{1}{4\lambda}\frac{1}{(a+b)^2},\;\;\;
c_1^{\infty_2}=\frac{1}{a-b}-\frac{1}{4\lambda}\frac{1}{(a-b)^2};$$
$$c_2^{\infty_1}=\frac{1}{a+b},\;\;\;c_2^{\infty_2}=\frac{1}{a-b}$$
$$  \alpha+\beta=\frac{l(l+1)}{\lambda^2},\;\;\; 
\alpha-\beta=\frac{l'(l'+1)}{\lambda^2}, $$
$$ a=\frac{|2l+1|+|2l'+1}{4\lambda}+\frac{1}{2\lambda},\; \;\;\;\;\;
a'=a-\frac{1}{4\lambda}, $$
$$b=\frac{|2l+1|-|2l'+1|}{4\lambda}, \;\;\; l,l'<-1,\;\;\;\; l,l'>0$$;

\hskip-3em $2^0$\ we get the case substituting in $1^0\;\; |2l+1|$\ 
by $-|2l+1|$;

\hskip-3em $3^0$\ we get the case substituting in $1^0\;\; |2l'+1|$\ 
by $-|2l'+1|$;

\hskip-3em $4^0$\ we get the case substituting in $1^0\;\; |2l+1|$\ 
by $-|2l+1|$\ and $|2l'+1|$\ by $-|2l'+1|$;

\hskip-3em $1^0\;\;$
$$\phi_{10}=\frac{|2l+1|-1}{2\lambda}\tanh x +\frac{b}{\cosh x}$$
\be F_1(\phi_{10})=\frac{1}{4\lambda}\frac{(\phi_{10}^2-a^2)^2+4\lambda
(\phi_{10}^2-a^2)\ll(a(\phi_{10}^2-a^2-b^2)-\imath b\phi_{10}
\sqrt{\phi_{10}^2-a^2-b^2}\r)}
{(b^2-a^2)(\phi_{10}^2-a^2)+2a^2b^2+2\imath ab\phi_{10}
\sqrt{\phi_{10}^2-a^2-b^2}}\nn\ee
$$F_2(\phi_{10})=(\phi_{10}^2-a^2)\frac{a(\phi_{10}^2-a^2-b^2)-
\imath b\phi_{10}\sqrt{\phi_{10}^2-a^2-b^2}}{(b^2-a^2)(\phi_{10}^2-a^2)+
2a^2b^2+2\imath ab \phi_{10}\sqrt{\phi_{10}^2-a^2-b^2}}$$
$$c_1^{\infty_1}=-\frac{1}{a-\imath b}-\frac{1}{4\lambda}\frac{1}{(a-\imath
b)^2},\;c^{\infty_2}_1=-\frac{1}{a+ib}-
\frac{1}{4\lambda}\frac{1}{(a+ib)^2};$$
$$c_2^{\infty_1}=-\frac{1}{(a-ib},\;\;c_2^{\infty_2}=-\frac{1}{a+ib}$$
$$ \alpha=b^2-\frac{l(l+1)}{\lambda^2},\;\;\;\; \beta=|b||2l+1|,\;\;\;\;
a=\frac{|2l+1|-1}{2\lambda}>0,$$ 
$$ b>0,\;\;\;\; l>0,\;\;\;\;\; l<-1;$$

\hskip-3em $4^0$\ we get the case substituting in $1^0\; |2l+1|$\ by 
$-|2l+1|$\ or allowing $l$ to vary in $1^0$\ in the
segment $-1<l<0$\ (the allowed $b$ is then negative in both the cases).

Since the behaviour of the functions $F_{1,2}$\ in both of the
above cases corresponds exactly to the assumptions we have done
about them earlier to relate the conventional and supersymetric
JWKB formulae these relations have to be maintained also in both
the cases considered. A detailed demonstration of this is
performed in Appendix 2. The latter illustrates a typical way of
getting such relations.

All the results we have obtained in this section are the good
illustrations of theorems we are going to formulate and to prove
below.

Let $V(x,\lambda)\equiv V_-(x,\lambda)$\ satisfies the following
assumptions:
\vskip-1ex
\begin{description}
\vskip-2ex\item[$1^0$] $V_-(x,\lambda)$\ being shape invariant is 
meromorphic 
on the $x$--plane having there finite second
order poles and diverging at infinity quadratically or
exponentially with $x$ (the two latter possibilities excludes each
other of course);
\item[$2^0$] The map: $x\to\phi=\phi(x)$\ defines a finitely sheeted Riemann 
surface $R_\phi$\ of the $\phi$--variable;
\item[$3^0$] All turning points of 
$q(x)=V_\pm(x,\lambda)+\frac{\delta(x)}{\lambda^2}-E$\ 
are transformed into $R_\phi$\ pairwise i.e.
one pair into one sheet of $R_\phi$\ and each pair can be used
to define equivalently the contour $K$ in the quantization
condition \mref{w5.1} (that is each such a contour surrounds the
corresponding turning point pair);
\item[$4^0$] The functions $F_1(\phi)$\ and $F_2(\phi)$\ defined on 
$R_\phi$\ 
are holomorphic on $R_\phi$\ outside some
circle of sufficiently large radius not branching at infinities
of any sheet.
\end{description}

The above assumptions allow us for the following conclusions:
\begin{description}
\vskip-3ex\item[a.] $\phi(x)$\ maps the contours $K$ of the assumption 
$3^0$\ 
into respective contours $K_\phi$\ on $R_\phi$\ 
which surround the corresponding maps of the turning point pairs
on $R_\phi$;
\item[b.] $R_\phi$\  is cut with the branch points $\phi(x_k)$\ satisfying: 
$\phi'(x_k)=0$\ where $x_k$\ are finite regular
points of $\phi(x)$. All these points lie outside all the
contours $K_\phi$\ and are the square root branch points for
$F_1$\ and $F_2$. (The latter type of branching follows from the
equalities: if $\phi'(x_k)=0$\ then close to $x_k$
$\phi'(x)\approx \alpha(x-x_k)$\ and $\phi\approx\alpha(x-x_k)^2/2)$;
\item[c.] If $\phi'4$ vanishes on $R_\phi$\ at some of its regular point 
$\phi_0$ linearly i.e. $\phi'(x)\approx a(\phi-\phi_0)$\ then
the point is a map of an essential singularity of $\phi(x)$\
lying at infinity i.e. $\phi$\ approaches $\phi_0$\
exponentially: $\phi(x)\approx\phi_0+Ce^{\alpha'x}$;
\item[d.] Second order poles of $V_\pm(x,\lambda)$\ as well as their 
infinite singular ones are transformed
into infinities of different sheets of $R_\phi$. The divergence
to infinity of the $F_{1,2}$--functions is the following:
\begin{description}              
\item[i.] For $V_-(x,\lambda)$\ diverging as $e^x$\ for $x\to\infty\;\; 
F_{1,2}$\ diverge linearly with $\phi$\ when $\phi\to\infty$\ 
on a given sheet;
\item[ii.] For $V_-(x,\lambda)$\ diverging as $x^2$\ for $x\to\infty\;\; 
F_{1,2}$\ approaches constant values when
$\phi\to\infty$;
\item[iii.] For $x$ close to a second order pole of $V_-(x,\lambda)\;\; 
F_{1,2}$\ diverge to infinity as $\phi^2$\ 
when $\phi\to\infty$\ on a sheet which vicinity of its infinity
is a map of the corresponding vicinity of the pole $x_0$;
\end{description}
\end{description} 

Let us add yet to the four above the following one more assumption:

\hskip-3em $5^0$\ The integrand of the following integrals:
\be\label{w5.10}
\oint\limits_{K_\phi}\ll[\sqrt{\phi^2\pm\frac{1}{\lambda^2}\phi'+
\frac{\delta}{\lambda^2}-\tilde{E}}-\sqrt{\phi^2-\tilde{E}}\r]
\frac{d\phi}{\phi'}
\ee
where $K_\phi$\ is any of the contours of the remark {\bf a.} above do
not possess outside the contours $K_\phi$\ singularities
different than those described in the conclusion {\bf b.}.

From the assumptions $1^0-5^0$\ and from the remarks {\bf a}.-{\bf d}. the
following two theorems come out:
\newtheorem{moje}{Theorem}
\begin{moje}
The SJWKB formulae with the superpotentials $\phi(x,\lambda)$\
corresponding to $V_\pm(x,\lambda)$\ are exact independently of
whether the supersymmetry is exact or broken.
\end{moje}
\begin{moje}
The conventional JWKB formulae for $V_\pm(x,\lambda)$\ are
exact.
\end{moje}

{\it Proof of the Theorem 1.}

The theorem follows from the following sequence of equalities:
\be\label{w5.11}
\oint\limits_K(\phi^2(a)-\tilde{E})^\fr
dx=\oint\limits_K(\phi^2(a_1)-\tilde{E}+R(a_1))^\fr dx +&\nn\\
\mbox{}+\oint\limits_K(f(F^-_1(a_1),\tilde{E}-R(a_1))-f(F_1^+ (a),
\tilde{E}))dx=\ldots&\nn\\
=\oint\limits_K(\phi^2(a_m)-\tilde{E}+R(a_1)+\ldots+R(a_m))^\fr
dx+&\\
\mbox{}+\sum\limits^m_{p=1}\oint\limits_K(f(F^-_1(a_p),\tilde{E}-
R(a_1)-\ldots-R(a_p))-&\nn\\
\mbox{}-f(F^+_1(a_{p-1}),\tilde{E}-R(a_1)-\ldots
R(a_{p-1})))dx,&\nn\\ 
a_0=a,\;\;\;\;\;\;R(a_0)=0&\nn
\ee
where $f(F_1^\pm,\tilde{E})$\ is defined by:
\be\label{w5.12}
\oint\limits_K\ll[\phi^2(a)\pm\frac{1}{\lambda}\phi'(a)+\frac{\delta}
{\lambda^2}-\tilde{E}\r]^\fr dx=\oint\limits_K(\phi^2(a)-\tilde{E})^\fr dx
+\oint\limits_Kf(F_1^\pm(a),\tilde{E})dx
\ee
with $F_1^\pm=\pm\phi'+\delta/\lambda$.

From assumption $3^0$\ it follows that every of the contour
integrals in the sum of the RHS of \mref{w5.11} when rewritten
to be taken on some sheet of $R_\phi$\ can be taken on {\it
each} sheet of $R_\phi$\ in the following way:
\be\label{w5.13}
\oint\limits_Kf(F^\pm(x,a),\tilde{E})dx=\frac{1}{n}\sum\limits^n_{r=1}
\oint\limits_{K_{\phi,r}}f(F^\pm_1(\phi,a),\tilde{E})\frac{d\phi}{F_2(\phi)}
\ee
where $F_2(\phi)\equiv \phi'(x(\phi))$.

Now it follows further from assumption $5^0$\ that every contour
$K_{\phi,r},\, r=1,\dots,n$, can be deformed on a sheet which it
is defined on to a circle of sufficiently large radius and to
pieces of this contour which cancel mutually with analogous
pieces of other contours. The net result of these deformations
are the integrations performed on every sheet along the circle
with sufficiently large radius. Outside the circle the
integrated $f$'s (divided by $F_2$) are holomorphic and diverging
to infinity not faster than the second power of $\phi$. This
guarantees that all these integrals can be calculated in the way
similar to that used by us earlier. It is easy to check that
independently of the type of the divergencies listed in the
points {\bf i.-iii.} above each infinity contributes the same to the
sum \mref{w5.13} namely $\mp i\pi/\lambda$\ for the
$F_1^\pm$--cases respectively. Therefore, the total value of the
integral in the LHS of \mref{w5.13} is also $\mp i\pi/\lambda$\
accordingly. Finally the formula \mref{w5.11} becomes:
\be\label{w5.14}
&\oint\limits_K(\phi^2(a)-\tilde{E})^\fr
dx=\oint\limits_K(\phi^2(a_m)-\tilde{E}+R(a_1)+\ldots+R(a_m))^\fr
dx+2\pi im&\nn\\ &a_0=a,\;\;\;\;R(a_0)=0
\ee
       
Putting now in \mref{w5.14} $E=R(a_1)+\ldots+R(a_m)\equiv
\tilde{E}_m$\ we get the result \mref{w5.2} where for the
broken supersymmetry the integer $m$ starts rather from $m=1$. QED.

{\it Proof of Theorem 2.}

The claim of the theorem follows immediately from the formula
\mref{w5.12} and from the
above proof of the theorem 1. Namely, from \mref{w5.12} we get:
\be\label{w5.15}
\oint_K\ll[\phi^2(a)\pm\frac{1}{\lambda}\phi'(a)+\frac{\delta}
{\lambda^2}-R(a_1)-\ldots-R(a_m)\r]^\fr dx=(2m\mp1)\pi \imath
\ee
QED.

Some remarks are in order.
   
First if $F_{1,2}(\phi)$\ diverged to infinity faster than
$\phi^2$\ then every integral of $f(F_{1,2}^\pm)$\ in
\mref{w5.11}
would contain $E$--dependent infinite series not reducing of
course to simple values $\pm i\pi$\ i.e. the relation
\mref{w5.14} as well as \mref{w5.15} could not be valid any longer. 

Second one can easily check that if $F_{1,2}(\phi)$\ do not
diverge to infinity faster than $\phi^2$\ and the potentials
$V_\pm(x,\lambda)$\ are holomorphic then they have to satisfy
the assumptions $1^0-5^0$\ above.

Therefore we can draw a conclusion that our assumptions about
the potentials $V_\pm$\ and the functions $F_{1,2}$\ fit in some
way in with the property of $V_\pm$\ being shape invariant.
But as we have checked they are not determined in some unique way by the
shape invariance condition \mref{w5.8}. (For example, others than the second 
power rate of growth of  $F_{1,2}(\phi)$ with $\phi$ are allowed by 
\mref{w5.8}, see Appendix 3).

\section{ Discussion and conclusions\label{s6}}

\hskip+2em In this thesis we have demonstrated that there are 
two basic symmetries, a
reflection: $x\to-x$\ and a translation: $x\to x+a$, of
potentials and of their corresponding Stokes graphs which decide
whether the JWKB quantization formulae are only approximations
to the exact formulae \mref{w2.10} or they are exact by
themselves.

We have established also that despite the above two symmetries
for the latter case to happen an additional property of the
considered potentials and the corresponding Stokes graphs has to
be present. Namely, this is the simplicity of SG's generated by
the original potential expressing itself in no more than two
turning points and in no more than two singular points in the
basic period strips to appear. In the opposite case a
proliferation of additional sectors in the basic period strip
prevents the periodicity properties of the corresponding
quantization conditions \mref{w2.10} to be used to reduce 
the conditions to the pure JWKB ones. The possible relaxation of
these conditions has been described in Sec.\ref{s4}. and the
corresponding examples were given in  \cite{1}.

Altogether, the above two symmetries and the simplicity
condition reduce effectively a number of exactly JWKB-quantized
potentials to only eight of them. {\it All} of them have long
been known. But due to our investigations we have given them the
property of being rather exceptional.

We have also shown that the SUSY JWKB exact quantization
formulae seem to be only different formulations of the exact
conventional ones at least in the case of the translationally
shape invariant potentials. We have supported the validity of this
conclusion showing the exactness of the JWKB formulae for the
two cases of the shape invariant potentials ($V_9$\ and $V_{10}$\
of Sec.\ref{s5.1}) not found by our earlier analysis of
Sec.\ref{s3}. Additionally, our two theorems of Sec.\ref{s5.1}
suggest also that there is close relation between the property
of being translationally shape invariant and the meromorphic
structure of the considered potentials on the $x$--plane which is
constrained to contain a limited number of second order poles
(in the whole $x$--plane or in the basic period strip if the
potential is periodic) and to have a particular behaviour at the
infinity.

We would like also to stress that the earlier proofs of the
exactness of some JWKB quantization formulae as done by
Rosenzweig and Krieger \cite{4,5} and in the case of
their SUSY forms by Crescimanno \cite{11} are incorrect by
erroneous calculations of necessary phases.

We have to note also that the results obtained by Inomata
\underline{et al} \cite{27} for the form of the
SUSY JWKB formulae in the cases of the broken supersymmetric
potentials do not contradict ours since the latter concern their
exact, not approximated forms which appear to coincide rather
with those of Comtet \underline{et al} \cite{7}.

\section*{ Appendix 1}
\renewcommand{\theequation}{A1.\arabic{equation}}
\zero

\hskip+2em Here we show that for the following holomorphic $2\pi
i$--periodic function:
\be\label{A1.1}
q(x,E,\lambda)=\sum\limits^k_{n=l}q_n(E,\lambda)e^{nx}
\ee
with {\it even k-l} and having only simple zeros its Weierstrass
product representation is the following:
\be\label{A1.2}
q(x,E,\lambda)=Ce^{\frac{k+l}{2}} x \prod\limits_{n\geq1}\ll[
1-\frac{x}{x_n}\r]
\ee

where $C=q(x,E,\lambda)/x|_{x=0}$\ or $C=q(0,E,\lambda)$\ if $x=0$
is not a root of $q(x,E,\lambda)$.

\hskip-2em The above formula follows from the observation 
that $Q(x,E,\lambda)= q(x,E,\lambda)\cdot\\
\cdot exp(-k/2 - l/2)$\ 
is also $2\pi i$--periodic and holomorphic with the same roots
as $q(x,E,\lambda)$\ and therefore its WP representation should be:
\be\label{A1.3}
Q(x,E,\lambda)=Ce^{\alpha x} x\prod\limits_{n\geq1}\ll[1-\frac{x}{x_n}\r]
\ee
where $\alpha$\ is an integer by periodicity of $Q$. On the other
hand the representation \mref{A1.3} depends analytically on the
coefficients $q_n$\ of \mref{A1.1} and we can always choose them
in such a way to make $Q$ symmetric under the reflection:
$x\to-x$. This operation does not change in \mref{A1.3} the
product itself (the distribution of roots are then invariant
under the operation) but changes $e^{\alpha x}$\ into
$e^{-\alpha x}$. However, $\alpha$\ being integer cannot change
with analytic continuation od $q_n$'s and therefore it has to be
zero from the very beginning.

As an example consider $q(x,E,\lambda)$\ given by \mref{w3.4}
for which its distribution of roots is shown in Fig.3. We have
for it:
\be\label{A1.4}
\alpha e^{2x}-2\beta e^x+\gamma=(\alpha-2\beta+\gamma)e^x
\prod\limits_{n\geq1}\ll[1-\frac{x}{x_n}\r]
\ee

We want to calculate with the help of \mref{A1.4} a change of
phase of $q(x,E,\lambda)$\ when transporting it from a point
$x_0$\ of the line $\Im x=\pi$\ to the point $x_0-2\pi i$\ of
the line $\Im x=-\pi$. To this goal we note that as it follows
from \mref{A1.4} the roots of $q(x,E,\lambda)$\ lying in large
distances from the points considered almost do not contribute to
the values of $q(x,E,\lambda)$\ in the considered strip (their
product in \mref{A1.4} is close to $1$). Therefore we can take a
sufficiently large but finite number of roots around the
considered points to perform the calculations needed (eventually
we can take the limit of the infinite number of roots). 

Starting from the point $x_0$\ we can consider $n$ pairs of roots
lying above the line $\Im x=\pi$\ ($n$ is large) and $n$ pairs of
roots lying below the line. The arguments of $x_0-x_k$\ we take
to be positive for $x_k$\ lying below the line $\Im x=\pi$\ and
negative in the opposite case. It is clear that the net result
of summing the corresponding arguments of the product in
\mref{A1.4} is zero. But there
is still non zero contribution to the argument of
$q(x_0,E,\lambda)$\ coming from the factor $e^x$\ of
\mref{A1.4}.
It amounts of course to $\pi$\ and this is the total argument of
$q(x_0,E,\lambda)$.

At the points $x_0-\pi i$\ our calculations are similar. Keeping
the {\it same} set of roots as chosen previously we see that to
the total phase of the product at $x_0-\pi i$\ contribute only
the two most distant pairs of roots lying {\it above} the line
$\Im x=\pi$\ so according to our convention this contribution
amounts to $4(-\pi/2)= -2\pi$\ (in the limit of the root number
going to infinity). Together with the argument $-\pi$\ provided
by the factor $e^x$\ we get the argument of $q(x_0-\pi
i,E,\lambda)$\ to be equal to $-3\pi$.  Therefore the total
change of the argument of $q(x,E,\lambda)$ between the lines
considered is equal to $-4\pi$.

\section*{Appendix 2}
\renewcommand{\theequation}{A2.\arabic{equation}}
\zero

\hskip+2em We demonstrate here particularities of our statement that the
exactness of the conventional JWKB formula for the potential
$V_9(x,\lambda)\equiv V_{9,-}(x,\lambda)$, the first one of
those in \mref{w5.9}, follows from its SJWKB one. For the
potential $V_{10}(x,\lambda)$\ our considerations would be
similar.

\begin{tabular}{cc}
\psfig{figure=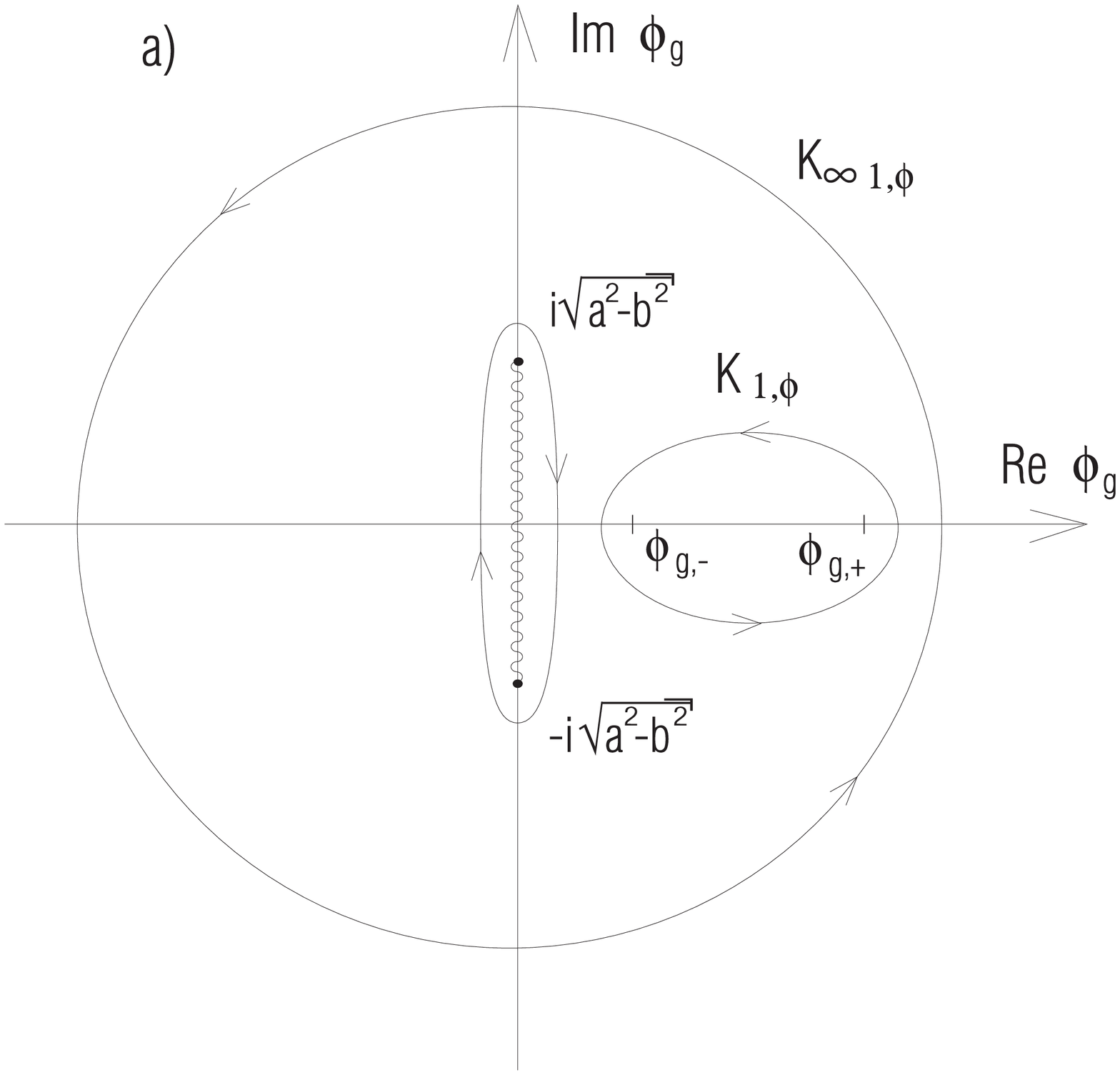,width=7cm} & \psfig{figure=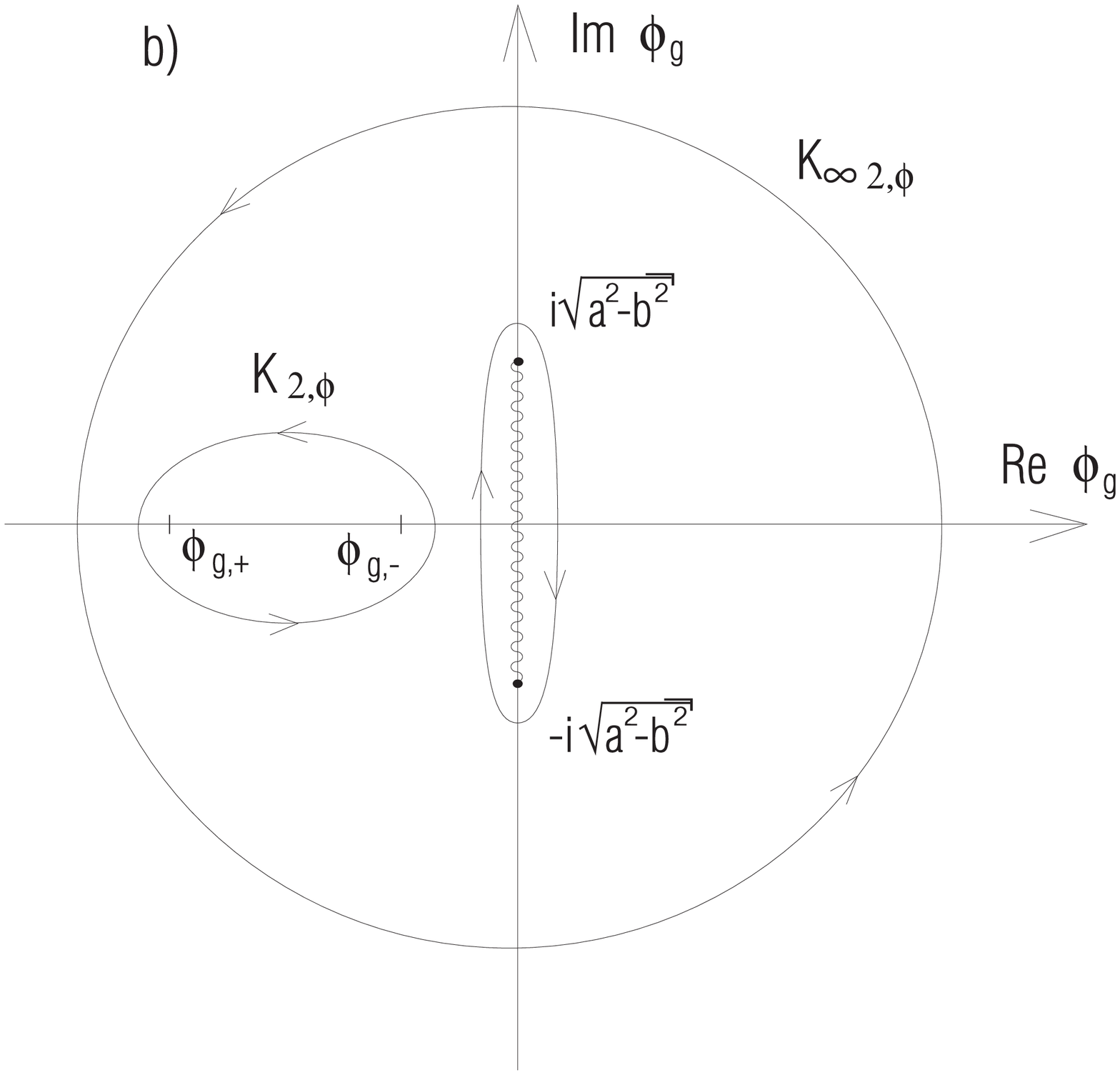,width=7cm} \\
\parbox{7cm}{Fig.16  The two sheeted $\phi_9$--Riemann surface for the 
unbroken superpotential $\phi_9$\ (the case $1^0$)} &
\end{tabular}

To this end let us consider the relation \mref{w5.12} using the
superpotentials $\phi_9(x,\lambda)$\ given above. First consider
the case $1^0$\ of the exact supersymmetry. The corresponding
Riemann surface $R_{\phi_9}$\ is depicted on Fig.16. This is
two sheeted surface with the branch points at $\phi_9=\pm
i(a^2-b^2)^\fr$. The latter are the unique singularities of the
integrand of the following integral:
\be\label{A2.1}
\oint\limits_K{\phi_9}\ll[\sqrt{\phi_9^2-\frac{1}{\lambda}
F_1(\phi_9)-\tilde{E}}-\sqrt{\phi_9^2-
\tilde{E}}\r]\frac{d\phi_9}{F_2(\phi_9)}
\ee
(since roots of $F_2$\ at $\phi=\pm ia$\ are also the roots of
$F_1$).

$R_{\phi_9}$\ is, as it can be easily noticed, a map of the basic
period strip $-\pi\leq x\leq\pi$\ of the $x$-plane (see Fig.15),
so that the four turning points of $q_{9,-}(x,\lambda,E)$\ from
this strip are mapped pairwise into $R_{\phi_9}$: the two from
the segment $(-\pi/2,\pi/2)$\ into the sheet a) of Fig.16 and
the other two into the second one. It is also easy to note that
in the quantization formulae \mref{w5.1} and \mref{w5.2} the
contour $K_1$\ on Fig.15 can be substituted by the contour
$K_2$\ of the figure by the periodicity. The contours are mapped
into $R_{\phi_9}$\ as $K_{1,\phi}$\ and $K_{2,\phi}$\
respectively, the latter surrounding the respective pairs of the
turning point pictures on $R_{\phi_9}$\ (see Fig.16). Therefore
for the quantization formulae
\mref{w5.1} and \mref{w5.2} we can write:
\be\label{A2.2}
\hskip-18em -\lambda\oint\limits_{K_1}\sqrt{\phi_9^2-\frac{1}{\lambda}\phi_9'
+\frac{\delta}{\lambda^2}-\tilde{E}}dx=-\frac{\lambda}{2}
\ll[\oint\limits_{K_1}+\oint\limits_{K_2}\r]\sqrt{\phi_9^2-
\frac{1}{\lambda}\phi_9'+\frac{\delta}{\lambda^2}-\tilde{E}}dx=  \\
=-\lambda\oint\limits_{K_1}\sqrt{\phi_9^2-\tilde{E}}dx-\frac{\lambda}{2}
\ll[\oint\limits_{K_{1,\phi}}+\oint\limits_{K_{2,\phi}}\r]
\ll[\sqrt{\phi_9^2-\frac{1}{\lambda}F_1(\phi_9)-\tilde{E}}-
\sqrt{\phi_9^2-\tilde{E}}\r]\frac{d\phi_9}{F_2(\phi_9)}= \nn \\ 
=-\lambda\oint\limits_{K_1}\sqrt{\phi_9^2-\tilde{E}}dx-\frac{\lambda}{2}
\ll[\oint\limits_{K_{\infty_1,\phi}}+\oint\limits_{K_{\infty_2,\phi}}\r]
\ll[\sqrt{\phi_9^2-\frac{1}{\lambda}F_1(\phi_9)-\tilde{E}}-
\sqrt{\phi_9^2-\tilde{E}}\r]\frac{d\phi_9}{F_2(\phi_9)} \nn
\ee
where $K_{\infty_1,\phi}\;{\rm and}\; K_{\infty_2,\phi}$\ are
the contours obtained by an obvious deformations of the contours
$K_{1,\phi}$\ and $K_{2,\phi}$\ which contain all the
singularities of $F_{1,2}(\phi_9)$. Making use of the explicite
forms of $F_{1,2}(\phi_9)$\ as given in Sec.\ref{s5} we can
calculate the last integral in \mref{A2.2} getting for it the
value $+i\pi$. Altogether with \mref{w5.2} this gives the result
\mref{w5.1}.

Consider now the broken case $2^0$\ of the superpotential
$\phi_9$. The corresponding basic period strip of
$q_{9,-}(x,\lambda,E)$\ and the quantization contours $K_1$\ and
$K_2$\ transform into $R_{\phi_9}$\ as it is shown in Fig.17.
Once again we can write the sequence analogous to \mref{A2.2}
deforming the contorus $_{K1,\phi}$\ and $K_{2,\phi}$\ of
Fig.17 into $K_{\infty_1,\phi}$\ and $K_{\infty_2,\phi}$\
respectively to perform the final integration getting
\underline{again} $+i\pi$\ and consequently the exact formula \mref{w5.1}. 
Of course, the starting value of $m$ can be now zero.

\begin{tabular}{cc}
\psfig{figure=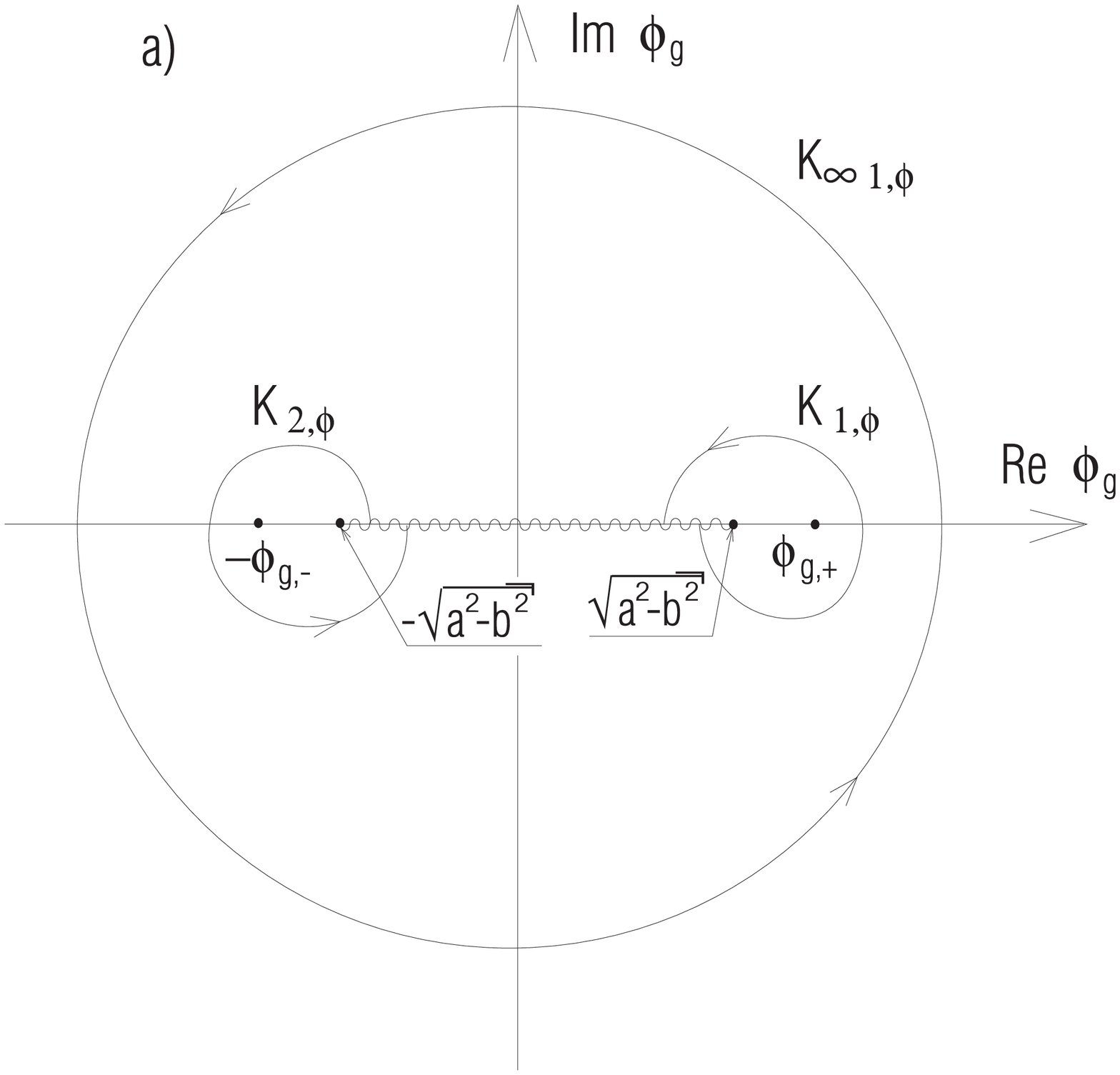,width=7cm} & \psfig{figure=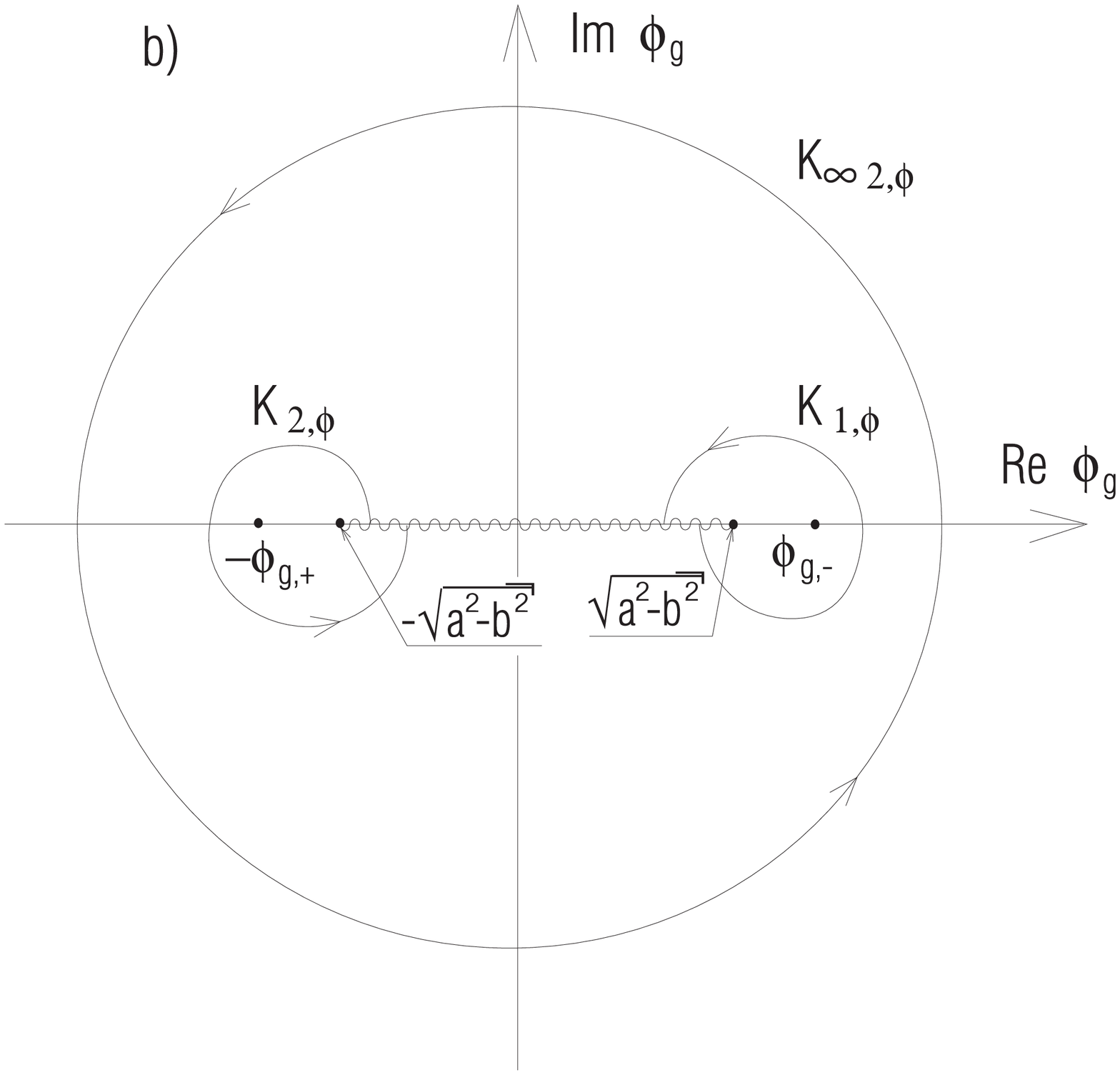,width=7cm} \\
Fig.17  The two sheeted $\phi_9$-Riemann surface & for broken 
superpotential $\phi_9$\ (the case $2^0$)
\end{tabular}

\section*{Appendix 3}
\renewcommand{\theequation}{A3.\arabic{equation}}
\zero

\hskip+2em We shall show here that the shape invariance 
condition \mref{w5.8} does not
prevent in some obvious way for $F_{1,2}(\phi)$ to diverge with any
power of $\phi$ when $\phi \to \infty$. To this end let us
rewrite (\ref{w5.8}) in terms of superpotentials. We get:
\be\label{A3.1}
\phi^2(x,\lambda,a)+\frac{1}{\lambda}\phi^{\prime}(x,\lambda,a)=
\phi^2(x,\lambda,a_1)-\frac{1}{\lambda}\phi^{\prime}(x,\lambda,a_1)+
R(a_1) \\
a_1=f(a) \nn
\ee

Introducing farther to (\ref{A3.1}) the function $F_2(\phi,a)$
($\equiv \phi^{\prime}(x(\phi,a),a)$)
we obtain:
\be\label{A3.2}
F_2(\phi,a)=\lambda\frac{2\phi\Delta(\phi,a)+\Delta^2(\phi,a)+R(f(a))}
{2+\Delta^{\prime}_{\phi}(\phi,a)}
\ee
where $\Delta(\phi,a)$ is defined as:
\be\label{A3.3}
\Delta(\phi,a) \equiv \tilde{\Delta}(x(\phi,a),a) \\
\phi(x,a_1)=\phi(x,a)+\tilde{\Delta}(x,a) \nn
\ee

It follows from (\ref{A3.2}) that the behaviour of
$F_2(\phi,a)$ when $\phi \to \infty$ comes out from the corresponding
behaviour of $\Delta(\phi,a)$. The latter, however, under the
assumption that $\phi = \infty$ is at most a pole for it has to be
following:
\be\label{A3.4}
\Delta(\phi,a) = \sum_{k\geq 0} b_k (a)\phi^{-k+1}
\ee
i.e. this pole has to be simple at most. 

The last equation is a conclusion of the condition:
\be\label{A3.5}
x(\phi+\Delta(\phi,a),a_1) = x(\phi,a)
\ee
under which the shape invariance property (\ref{A3.1}) is satisfied.
Note also that due to equality:
$x_{\phi}^{\prime}(\phi,a)=1/F_2(\phi,a)$, the following relation is coming
out from (\ref{A3.5}): 
\be\label{A3.6}
F_2(x(\phi+\Delta(\phi,a),f(a)) = (1+\Delta_{\phi}^{\prime}(\phi,a))
F_2(\phi,a)
\ee

It is now easy to conclude from (\ref{A3.2}) that if $b_0 \neq 0,-2$ 
then $F_2(\phi,a)$ grows as $\phi^2$ when $\phi \to \infty$.
But for example if $b_0 = -2$ and $b_1, b_n \neq $0
with $b_2, ..., b_{n-1}=0$, $n \geq 2$, then $F_2(\phi,a)$ has to grow as
$\phi^{n+1}$ when $\phi \to \infty$. 

Of course, whether $\Delta(\phi,a)$ can really behave in the above ways 
depends totally on the properties of the superpotentials considered which on 
their own are constrained by (\ref{A3.6}).


\begin{thebibliography}{99}
\bibitem{1} Giller S., {\it J. Phys. A: Math. Gen.} {\bf 21} (1988) 909 

\bibitem{2} Fr\"oman N. and Fr\"oman P.O. , 
{\it JWKB Approximation. Contribution to the Theory}, North-Holland, 
Amsterdam 1965 

\bibitem{3} Bailey P.B., J. Math. Phys. 5 (1964) 1293

\bibitem{4} Rosenzweig C. and Krieger J.B., J. Math. Phys. 9 (1968) 849

\bibitem{5} Krieger J.B., J. Math. Phys. 10 (1969) 1455

\bibitem{6} Bruev A.S., Phys. Lett. A 161 (1992) 407

\bibitem{7} Comtet A., Bandrauk A.D. and Campbell D.K., 
Phys. Lett. B 150 (1985) 159

\bibitem{8} Khare A., Phye. Lett. B 161 (1985) 131

\bibitem{9} Eckhardt B. Phys. Lett. B 168 (1986) 245

\bibitem{10} Delaney D. and Nieto M.M., LANL preprint LA-UR-90-1708

\bibitem{11} Crescimanno M., J. Math. Phys 31 (1990) 2946

\bibitem{12} Landau L.D., Lifshitz E.M., 
{\it Quantum mechanics (nonrelativistic theory)}, 3rd Ed.  Pergamon
Press, Oxford, New York, 1977

\bibitem{13} Giller S. , {\it J, Phys. A: Math. Gen.} {\bf 22} (1989) 2965 

\bibitem{14} Giller S., {\it Acta Phys. Pol.} {\bf B21} (1990) 675-709

\bibitem{15} Giller S. and Milczarski P. {\it Borel summable solutions 
to Schr\"odinger equation} {\bf quant-ph/9801031}, to be published 

\bibitem{16} Giller S., Acta Phys. Pol. B 23 (1992) 457-511

\bibitem{17} Fedoryuk M.V. , {\it Asymptotic Methods For Linear Ordinary}

\bibitem{18} Lee T.D. , {\it Mathematical Methods in Physics},
Columbia University, New York 1964

\bibitem{19} Whittaker E.T. and Watson G.N., {\it A Course of Modern 
Analysis}
4th Ed. CUP Cambridge 1963

\bibitem{20} Morse P.M., Phys. Rev. 34 (1929) 57

\bibitem{21} Rosen N. and Morse P.M., Phys. Rev. 42 (1932) 210

\bibitem{22} P\"oschl G. and Teller E., Z. Phys. 83 (1933) 143

\bibitem{23} Langer R.E., Phys. Rev. 51 (1937) 669

\bibitem{24} Giller S., Milczarski P., 
{\it Change of variable as Borel resummation},  {\bf quant-ph/9712039}, 
to be published

\bibitem{25} Berry M.V. and Mount K.E., Reps. Prog. Phys. 35 (1972) 315

\bibitem{26} Maslov V.P. and Fedoriuk M.V., 
{\it Semi-classical Approximation in Quantum Mechanics}, 
Dodrecht, Boston, London: D. Reidel Pub. C. 1981

\bibitem{27} Inomata A., Junker G. and Suparmi A., J. Phys. A 26 (1993) 2261

\bibitem{28} Bose A.K., Nuo. Cim. 32 (1964) 679

\bibitem{29} Cooper F., Khare A., Sukhatme U., Phys. Rep. 251 (1995) 267

\bibitem{30} Gendenshtein L., JETP Lett. 38 (1983) 356

\bibitem{31} Dutt R., Khare A. and Sukhatme U.P., Phys. Lett. B 181 (1986) 
295

\bibitem{32} Barclay D.T. and Maxwell C.J., Phys. Lett. A 157 (1991) 357

\end{thebibliography}
\end{document}